\documentclass[aps,prl,reprint,groupedaddress]{revtex4-1}
\pdfoutput=1
\usepackage{braket}
\usepackage{amsmath, amssymb, graphicx}
\usepackage{breqn}
\usepackage{verbatim}

\newcommand{\Dgs}{D_{\mathrm{gs}}}

\newcommand{\Swe}{\tilde{S}\left(\omega\right)}

\newcommand{\Tone}{$T_1$}
\newcommand{\us}{$\mu$s}
\newcommand{\um}{$\mu$m}

\newcommand{\nit}[1]{$^{#1}$N}
\newcommand{\C}[1]{$^{#1}$C}
\newcommand{\Tpm}{$T_{1}^{(-1,+1)}$}
\newcommand{\Tonez}{$T_{1}^{(0)}$}
\newcommand{\Ttwomax}{$T_{2}^{\rm{SQmax}}$}
\newcommand{\Sw}{$\tilde{S}\left(\omega\right)$}

\begin{document}


\title{Double-Quantum Spin-Relaxation Limits to Coherence of Near-Surface Nitrogen-Vacancy Centers}


\author{B. A. Myers, A. Ariyaratne, A. C. Bleszynski Jayich}
\makeatletter
\let\cat@comma@active\@empty
\makeatother
\affiliation{Department of Physics, University of California, Santa Barbara, California 93106 USA}


\date{\today}

\begin{abstract}
We probe the relaxation dynamics of the full three-level spin system of near-surface nitrogen-vacancy (NV) centers in diamond to define a $T_{1}$ relaxation time that helps resolve the $T_{2} \leq 2T_{1}$ coherence limit of the NV's subset qubit superpositions. We find that double-quantum spin relaxation via electric field noise dominates $T_{1}$ of near-surface NVs at low applied magnetic fields. Furthermore, we differentiate $1/f^{\alpha}$ spectra of electric and magnetic field noise using a novel noise-spectroscopy technique, with broad applications in probing surface-induced decoherence at material interfaces.

\end{abstract}


\maketitle
Nitrogen-vacancy (NV) centers in diamond excel as room-temperature quantum sensors and quantum bits, where long-lived spin coherence and population are critical to an NV's functionality in these roles. In particular, the coherent manipulation of near-surface NVs has been used to detect few to single electronic spins \cite{Grinolds2013} and nuclear spins \cite{Loretz2014,SushkovProton2014,Muller2014} and to perform nanoscale magnetic resonance imaging \cite{Grinolds2013,Grinolds2014,Rugar2015,Haberle2015}. The placement of these NVs just nanometers from the diamond surface is vital to strongly couple to external degrees of freedom \cite{Cai2013} and achieve nanoscale spatial resolution in imaging \cite{Taylor2008}. However, surface-related noise and its effect on coherence is an incomplete puzzle and understanding this noise remains a grand challenge \cite{Okai2012,Rosskopf2014, Myers2014,Romach2015,Kim2015, Wang2016} to improving sensitivity and spatial resolution. In a broader context, identifying surface noise on the nanoscale is useful to the study of a variety of quantum technologies, such as trapped ions \cite{Safavi2011,Enoise2013}, mechanical resonators \cite{Stipe2001}, and superconducting circuits \cite{Wellstood1987, deSousa2007}, whose performance is known to be limited by pervasive surface-related decoherence.

The NV spin levels that display long coherence reside in the orbital ground state, a three-level spin $S=1$ system \cite{Doherty2013}. Any two of the levels may constitute a qubit for coherent quantum sensing, and although the sensor's functionality resides in the coherence of the qubit \cite{Maze2008, Rondin2014}, this functionality is compromised by the coupling of all three levels to the environment. For a two-level system, coherence time $T_{2}$ is known to be ultimately limited by spin relaxation time $T_{1}$ as $T_{2} \leq 2T_{1}$ \cite{Slichter,Bylander2011}, and much attention has been paid to this theoretical $T_{1}$ limit for NVs \cite{Bar-Gill2013,Knowles2014}. However, for NVs in bulk diamond a saturating $T_{2} = 0.53(2)T_{1}$ has been reported \cite{Bar-Gill2013}, and for shallow NVs, those within $\sim25$ nm of the surface, the discrepancy is more striking with $T_{2} \lesssim 0.1T_{1}$ \cite{Myers2014,Romach2015}. These prior results suggest a decoherence channel that has not been accounted for.

\begin{figure}
\includegraphics[width = 1.0\columnwidth]{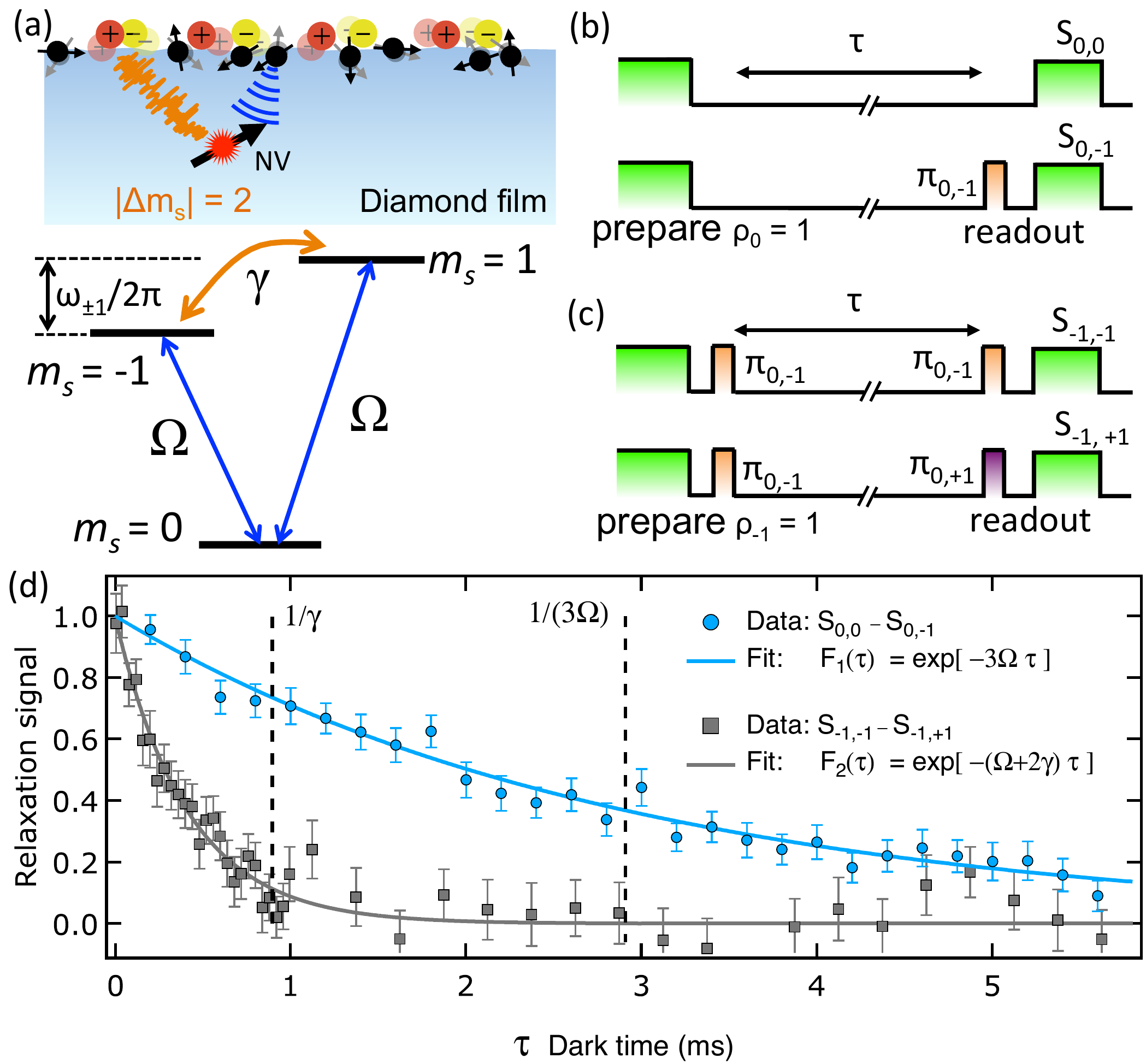}
 \caption{\label{fig:Figure1} (a) Surface-noise spectroscopy with a shallow NV center showing relaxation channels in its triplet ground state. The double-quantum (DQ) relaxation (orange, $\gamma$) is sensitive to electric field noise, and the single-quantum (SQ) channel (blue, $\Omega$) is sensitive to magnetic field noise. An applied dc magnetic field tunes the frequency $\omega_{\pm1}/2\pi$ of the DQ transition. (b,c) Measurement  sequences to extract the relaxation rates $\Omega$ and $\gamma$. The spin is initialized into (b) $\ket{0}$ or (c) $\ket{-1}$ by a green laser pulse and, for $\ket{-1}$, a microwave $\pi_{0,-1}$ pulse. After a dark time $\tau$, any of the three spin state populations $\rho_{j}$ can be read out by a choice of $\pi_{0,\pm1}$ pulse before photoluminescence detection, giving signal $S_{i,j}$. (d) Population decay data with three-level relaxation model fits as solid lines.}
 \end{figure}

The NV qutrit is rendered a powerful and versatile sensor by the different frequency scales and selection rules of its spin transitions. And for precisely the same reasons -- the double-edged sword of sensitivity -- the NV is also highly susceptible to environmental noise of various origins. The NV has both single- $(\Delta m_{s}=\pm1)$ and double- $(\Delta m_{s}=\pm2)$ quantum transitions \cite{Manson1990} tunable in the MHz to GHz frequency range, as shown in Fig. \ref{fig:Figure1}(a). This full capacity of probing noise has not been utilized until now, in particular concerning the direct relaxation rate between the $m_{s}=\pm1$ spin states of the qutrit, which we will refer to as double-quantum (DQ) relaxometry. Here, we measure both single-quantum (SQ) and DQ relaxation rates of the three-level system and find that shallow NVs exhibit particularly fast DQ relaxation, accounting for decoherence that has not been directly observed before. We then use multipulse dynamical decoupling to show that $T_{2}$ of the $m_{s}=0,-1$ qubit can exceed a properly defined $T_{1}$, where DQ relaxation dominates this limit when the $m_{s}=\pm1$ transition splitting is below $\sim 100$ MHz. Furthermore, because the DQ relaxation channel is a magnetic-dipole-forbidden transition, it can be used to selectively probe electric fields \cite{Klimov2014} and strain \cite{MacQuarrie2013,Barfuss2015,Ovartchaiyapong2014}. We combine spectroscopic DQ relaxometry with standard SQ dephasing spectroscopy \cite{Bar-Gill2012,Myers2014,Romach2015} to quantitatively map the spectral character of noise sources responsible for decoherence of near-surface NVs, and this technique enables us to distinguish electric and magnetic contributions to the noise spectrum.

The ground-state spin Hamiltonian \cite{Doherty2012,Dolde2011} of the NV center indicates how magnetic, electric, and strain fields contribute to dephasing and spin relaxation, with the corresponding energy level diagram shown in Fig. \ref{fig:Figure1}(a).
\begin{multline}
H_{\mathrm{NV}} = \left(h\Dgs + d_{\parallel}\Pi_{\parallel}\right)S^2_{z}  + g\mu_{B} \textbf{B}\cdot\textbf{S}\\
- \frac{d_{\perp}\Pi_{\perp}}{2}\left(S^{2}_{+} +S^{2}_{-}\right)
\label{eq:hamiltonian}
\end{multline}
where $\textbf{S}$ is the spin-1 operator, $h$ is Planck's constant, $g\mu_{B}/h= 2.8$ MHz$/$G is the gyromagnetic ratio, $\Dgs{}=2.87$ GHz is the crystal-field splitting, $d_{\parallel}/h=0.35$ Hz$\cdot$cm$/$V  and $d_{\perp}/h=17$ Hz$\cdot$cm$/$V are the components of the NV's electric dipole moment parallel and perpendicular to its symmetry axis \cite{VanOort1990}, and $\Pi_{\parallel}$ and $\Pi_{\perp}$ are the corresponding total effective electric field components \cite{Doherty2012,Dolde2011}. $\boldsymbol{\Pi} = \left(\textbf{E} + \pmb{\sigma}\right)$ contains electric field $\textbf{E}$ and scaled strain $\pmb{\sigma}$ terms. We attribute the $\boldsymbol{\Pi}$ noise identified in our experimental results with $\textbf{E}$ electric fields, as discussed in the supplemental information \cite{MyersSOM2016}. 

The form of Eq. \ref{eq:hamiltonian} \cite{Ovartchaiyapong2014} makes apparent the $m_{s}=\pm1$ DQ coupling: the quadratic spin raising and lowering operators in the last term couple the $\ket{1}$ and $\ket{-1}$ states, a route for electric noise-induced $\left|\Delta m_{s}\right|=2$ spin relaxation with a rate $\gamma$ (Fig. \ref{fig:Figure1}(a)). For the $\{\ket{0},\ket{-1}\}$ qubit, the $d_{\parallel}E_{\parallel}$ term describes electric-field-induced energy shifts \cite{Dolde2011,Kim2015} that contribute to the dephasing rate $\Gamma_{d}^{(-10)}$. The second (Zeeman) term accounts for magnetic fields that cause additional dephasing \cite{Taylor2008,deLange2010,Myers2014} and SQ relaxation \cite{Tetienne2013, Rosskopf2014, Pelliccione2014,Yacoby2015,Hall2016} between $\ket{0}$ and $\ket{\pm1}$ with rates $\Omega_{0,\pm1}$ ($\Omega_{0,+1} = \Omega_{0,-1} \equiv \Omega$, as we verified experimentally \cite{MyersSOM2016}). The energy splitting between the $\ket{\pm 1}$ levels, $\hbar\omega_{\pm1} = 2g\mu_{B}B_{z}$, is tunable via a dc magnetic field $B_{z}$, enabling us to probe the noise spectral density that affects $\gamma$ \cite{spectrometer2003}. We consider the regime where dc strain, dc electric field, and dc transverse magnetic field are small compared to the applied $B_{z}$, so the eigenstates are approximately $\{\ket{0},\ket{-1},\ket{1}\}$ of the $S_{z}$ operator \cite{Dolde2011}. Temperature changes can also lead to dephasing through $\Dgs{}$ \cite{Fang2013}, though we neglect them here as our measurements are insensitive to slow (sub-kHz) fluctuations.

For the NV qutrit in Fig. \ref{fig:Figure1}(a), the total $T_{1}$ relaxation time that limits $T_{2}$ of the qubit is built from the relaxation rates between the three $\ket{m_{s}}$ spin states. However, the most prevalent definition of $T_{1}$ in the NV literature \cite{Jarmola2012,Bar-Gill2012,Tetienne2013,Steinert2013,Rosskopf2014,Myers2014,Romach2015,Knowles2014} considers only $\Omega$.  We label that quantity \Tonez{}, the time constant for an NV prepared in density matrix $\rho= \ket{0}\bra{0}$ to depolarize into a fully-mixed state $\rho = I_{3\times 3}/3$. In using the definition $T_{1}=T_{1}^{(0)}$, past experiments have implicitly assumed the DQ relaxation rate $\gamma=0$. To understand what limits $T_{2}$ of the $\{\ket{0},\ket{-1}\}$ qubit the correct definition should be
\begin{equation}
\frac{1}{T_{1}} = \frac{1}{T_{1}^{(0)}} + \gamma = 3\Omega + \gamma.
\label{eq:T1sq}
\end{equation}
SQ coherence $\rho_{-10}$ initialized between the $\ket{0}$ and $\ket{-1}$ states will decay at a total rate $1/T_{2}$ due to the sum of pure dephasing  $\Gamma_{d}^{(-10)}$ and spin relaxation rates \cite{threelevel2004}, so that in the zero-dephasing limit $T_{2}^{\rm {SQmax}}= 2T_{1} = 2\left(3\Omega + \gamma\right)^{-1}$ \cite{MyersSOM2016}. Hence, to evaluate the revised decoherence limit $T_{2} \leq 2T_{1}$ we used SQ and DQ relaxometry to extract $\Omega$ and $\gamma$.

The experimental setup consists of a homebuilt, room-temperature confocal microscope with a 532-nm excitation laser and single-photon counters to collect sideband photoluminescence (PL) \cite{Gruber1997}. A single-crystal diamond film was grown via chemical vapor deposition (CVD) using isotopically purified methane ($99.99\%$ \C{12}) to minimize NV decoherence due to a \C{13} nuclear spin bath \cite{Balasubramanian2009}. The sample contains NVs at a mean depth of 7 nm, formed via 4-keV nitrogen implantation \cite{Rabeau2006,Okai2012,MyersSOM2016}. A microwave stripline was used for coherent $\ket{0}\leftrightarrow\ket{\pm1}$ spin rotations, namely spin inversion $\pi_{0,\pm1}$ pulses \cite{Jelezko2004}. 

The transition rates $\Omega$ and $\gamma$ can be experimentally determined by measuring the decay of each diagonal element of the density matrix through pulsed optically detected magnetic resonance (ODMR) measurements  \cite{Jarmola2012}. The population dynamics of the pure $S_{z}$ eigenstates are given by solutions to three differential equations relating the transition rates and spin populations \cite{Steinert2013,Tetienne2013,MyersSOM2016}
\begin{equation}
\rho_{0}\left(\tau\right)  = \frac{1}{3} + \left(\rho_{0}\left(0\right) - \frac{1}{3}\right)e^{-3\Omega \tau}
\label{eq:omegatP0}
\end{equation}
\begin{equation}
\rho_{-1}\left(\tau\right)  = \frac{1}{3} - \frac{1}{2}\Delta \rho_{\pm1}\left(0\right)e^{-(\Omega+2\gamma) \tau} -\frac{1}{2} \left(\rho_{0}\left(0\right) - \frac{1}{3}\right) e^{-3\Omega \tau}
\label{eq:omegatPm1}
\end{equation}
where $\tau$ is the time between initialization and readout, $\rho_{m_{s}}$ are the $\ket{m_{s}}$ state populations with condition $\rho_{+1}= 1-\rho_{-1}-\rho_{0}$, and the initial conditions determine cofficients $\rho_{0}\left(0\right)$ and $\Delta \rho_{\pm1}\left(0\right)=\left[\rho_{+1}\left(0\right) - \rho_{-1}\left(0\right)\right]$.

We performed two sets of pulse sequences that directly probe the spin populations under initial conditions $\rho_{0}(0)=1$ (Fig. \ref{fig:Figure1}(b)) and $\rho_{-1}(0)=1$ (Fig. \ref{fig:Figure1}(c)). The final $\pi_{0,m_{s}}$ pulse before PL readout determines which $m_{s}$ population is probed, and the subscripts of each relaxation signal $S_{i,j}\left(\tau\right)$ indicate the initialized state $\ket{i}$ and population to read out $\rho_{j}$. Figure \ref{fig:Figure1}(b) shows a standard method to measure \Tonez{}$=\left(3\Omega\right)^{-1}$ \cite{Jarmola2012}, and applying this sequence to Eqs. \ref{eq:omegatP0} and \ref{eq:omegatPm1} yields a fit function \cite{MyersSOM2016}
\begin{equation}
F_{1}\left(\tau\right) = S_{0,0}\left(\tau\right) - S_{0,-1}\left(\tau\right) = r e^{-3\Omega \tau}
\label{eq:fit0t0}
\end{equation}
where fit parameters are PL contrast $r$ and rate $\Omega$. The second set of sequences (Fig. \ref{fig:Figure1}(c)) accounts for the indistinguishability of PL from the $\ket{\pm1}$ states by selectively swapping either $\rho_{-1}$ or $\rho_{+1}$ with the distinguishable $\rho_{0}$ population before PL readout. The fit function \cite{MyersSOM2016} is 
\begin{equation}
F_{2}\left(\tau\right) = S_{-1,-1}\left(\tau\right) - S_{-1,+1}\left(\tau\right) = r e^{-\left(\Omega + 2\gamma\right)\tau}.
\label{eq:fitptp2Tone}
\end{equation}
Figure \ref{fig:Figure1}(d) shows data for shallow NV A1 fitted to Eqs. \ref{eq:fit0t0} (blue circles data) and \ref{eq:fitptp2Tone} (gray squares data), revealing a slow SQ rate $\Omega = 0.115(4)$ kHz and faster DQ rate $\gamma = 1.11(5)$ kHz. Hence the traditional relaxation time $T_{1}^{(0)}= 2.90(3) $ ms overestimates by $4\times$ the full $T_{1}= 0.69(7)$ ms from Eq. \ref{eq:T1sq}, due to significant DQ relaxation.

\begin{figure}
\includegraphics[width = 1.0\columnwidth]{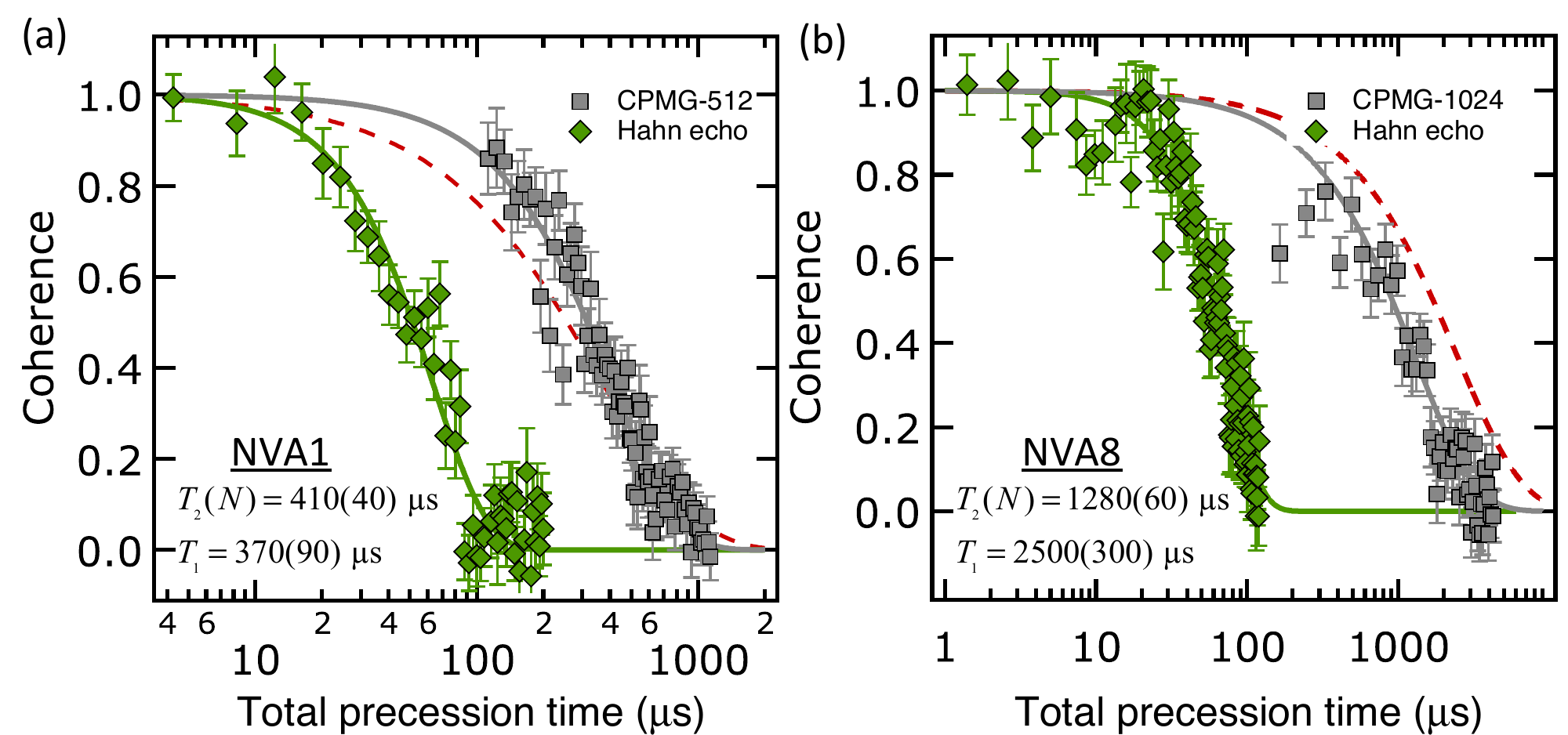}
 \caption{\label{fig:Figure2} Enhancement of SQ coherence time using CPMG-$N$ for shallow NVs under conditions of (a) large $\gamma$ at $\omega_{\pm1}$$/2\pi=37.1$ MHz and (b) small $\gamma$ at $\omega_{\pm1}$$/2\pi=1376$ MHz. Data shown are Hahn echo (green diamonds) and CPMG-$N$ (gray squares) where $N$ is the number of $\pi$ pulses, and solid lines are fits to $\exp\left[-(T/T_{2})^{n}\right]$. Dashed red lines are reference plots of $\exp\left(-T/T_{1}\right)$ using the measured $T_{1}=\left(3\Omega+\gamma\right)^{-1}$.}
 \end{figure}

The complete $T_{1}$ enables evaluation of the coherence time limit $T_{2} \leq 2T_{1}$, for which we reduced $\Gamma_{d}^{(-10)}$ via CPMG-$N$ dynamical decoupling (DD) measurements \cite{CPMG1958}, where $N$ is the number of $\pi_{y}$ pulses. Figure \ref{fig:Figure2} shows Hahn echo and CPMG-$N$ measurements for two shallow NVs. The coherence time $T_{2} = T_{2}\left(N\right)$ is extracted from a stretched-exponential fit $\exp\left[-(T/T_{2})^{n}\right]$ to data $C(T)$, where $T$ is total precession time. Figure \ref{fig:Figure2}(a) shows that for sufficiently large $N=512$, and at $\omega_{\pm1}/2\pi=37.1$ MHz, $T_{2}$ saturates at $1.2(3)T_{1}$ \cite{MyersSOM2016}, in clear contrast to the incomplete comparison $T_{2} = 0.14(1) T_{1}^{(0)}$. This demonstration of $T_{2} \gtrsim T_{1}$ for shallow NVs at room temperature also exceeds the ratio previously reported for bulk NVs, $T_{2} \approx 0.53(2) T_{1}^{(0)}$ \cite{Bar-Gill2013}. At a much larger $\omega_{\pm1}/2\pi=1376$ MHz (Fig. \ref{fig:Figure2}(b)), $T_{2}(N=1024)$ saturates at only $0.52(7)T_{1}$, while $T_{2}$ and $T_{1}$ both increase. The explanation for these changes at higher $\omega_{\pm1}$ lies in the frequency dependence of $\gamma$, as we discuss next.

\begin{figure}
\includegraphics[width = 1.0\columnwidth]{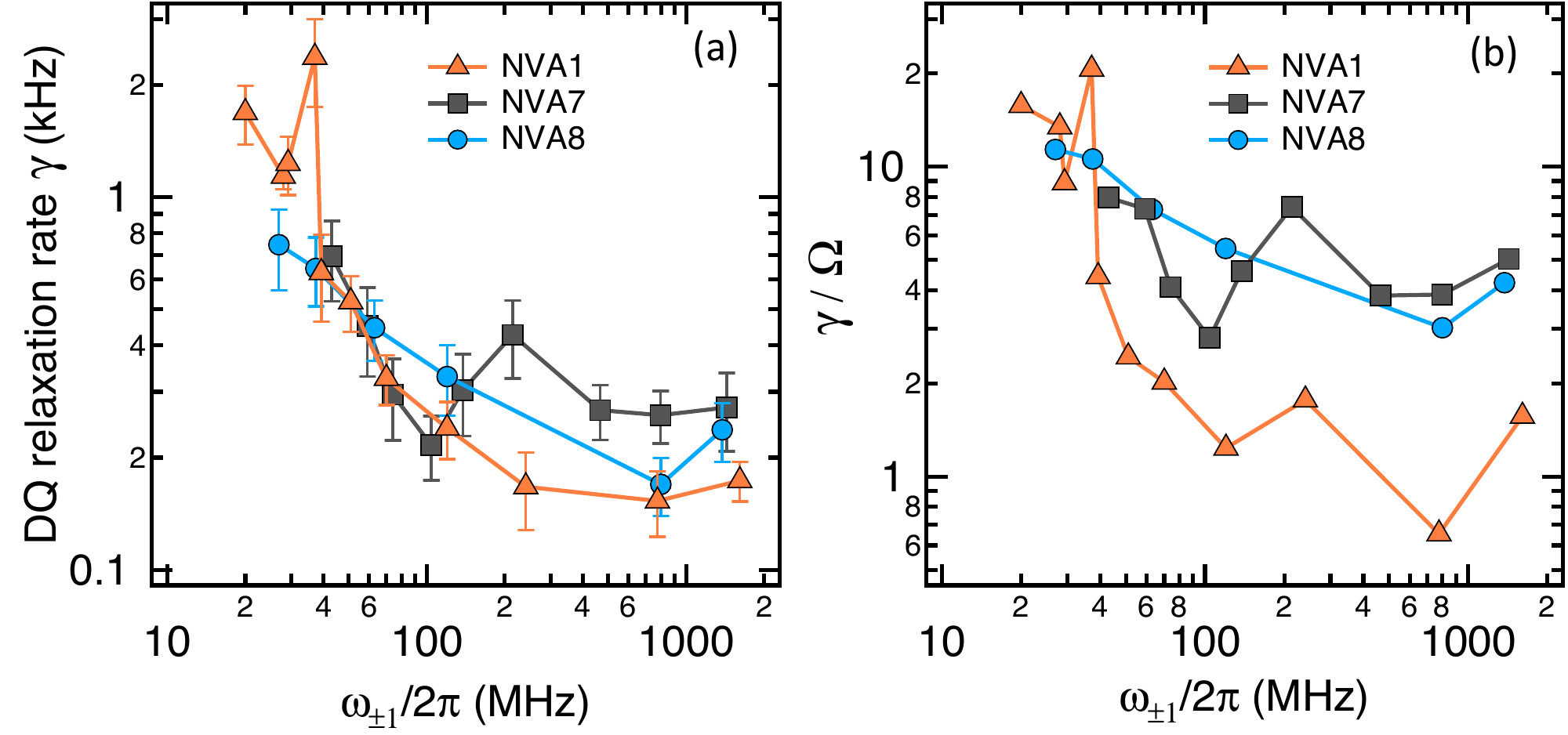}
 \caption{\label{fig:Figure3} (a) Measured DQ relaxation rates $\gamma$ for three shallow NV centers versus DQ frequency splitting $f=\omega_{\pm1}/2\pi$. Each symbol type refers to one NV. The $1/f^{\alpha}$-type dependence is attributed to surface-related electric field noise and saturation at large $\omega_{\pm1}$ is attributed to bulk effects. (b) Ratio $\gamma/\Omega$ plotted for the same NVs as in (a).}
 \end{figure}
\begin{table*}[t]
\begin{tabular}{|l|l|l|} 
\hline
  & DQ relaxation spectroscopy & SQ dephasing spectroscopy \\ \hline
Measurement  & Relaxation between $m_{s}=\pm1$ populations & CPMG multipulse on NV superposition \\ \hline
Filter frequency tuning & Applied $B_{z}$: $\omega_{\pm1}/2\pi \approx 2g\mu_{B}B_{z}/h$  & Number $N$ and spacing $T/N$ of $\pi$ pulses \\ \hline
Frequencies probed & 10 MHz - few GHz & kHz - few MHz \\ \hline
Primary noise probed & Transverse electric/strain & Parallel magnetic/electric/strain \\ \hline
Coupling power [Hz$^{2}/$Hz] & $S\left(\omega_{\pm1}\right)=\gamma$& $S\left(\omega=\pi N/T\right) \approx -\pi\ln{C\left(T\right)}/T$ \\ \hline
Assumption for validity & $g\mu_{B} B_{z}/h \gg \Pi_{\perp}d_{\perp}/h$ $\implies$ eigenstates $\ket{m_s}$  & $B_{z} \gtrsim 100$ G  $\implies$ small $\gamma$: $T_{2}\ll T_{1}$ \\
\hline
\end{tabular}
\caption{Comparison and complementarity of relaxation and dephasing for classical-noise spectroscopy with NV centers.}
\label{tab:spectroscopy}
\end{table*}
Figure \ref{fig:Figure3}(a) shows a strong dependence of $\gamma$ on frequency $f = \omega_{\pm1}/2\pi$ for shallow NVs, with two implications: 1) $T_{1}$ greatly decreases at lower magnetic fields, in contrast to \Tonez{}, and 2) double-quantum relaxation spectroscopy gives new insights about noise sources affecting $\gamma$. As $B_{z}$ tunes $\omega_{\pm1}/2\pi$ from 1612 MHz to 20 MHz, $\gamma$ increases by up to an order of magnitude, showing a $1/f^{\alpha} + \gamma_{\infty}$ type of dependence with $\alpha=1-2$. We observe the $1/f^{\alpha}$ part only for shallow NVs, and thus we identify its origin as surface-related electric field noise \cite{MyersSOM2016}. We attribute $\gamma_{\infty}$ relaxation to bulk effects \cite{MyersSOM2016, Taminiau2014}. In contrast to $\gamma$, we find $\Omega$ to be independent of magnetic field over the studied range of $B_{z} \approx 4-290$ G. The ratio $\gamma/\Omega$ plotted in Fig. \ref{fig:Figure3}(b) demonstrates that the DQ relaxation contributes substantially to the total decoherence rate regardless of applied magnetic field; $\gamma/\Omega \gg 1$ at low $B_{z}$ and $ \gamma/\Omega \gtrsim 1$ at higher $B_{z}$. The suppression of shallow-NV decoherence via the DQ channel at large $\omega_{\pm1}$ gives a practical reason for magnetometry experiments to operate at $B_{z} > 100$ G, and it also explains our observation in Fig. \ref{fig:Figure2}: $T_{1}$ relaxation slows down as $\omega_{\pm1}$ increases, and dephasing takes over as the dominant decoherence channel. This dephasing cannot be eliminated completely because experimental limitations  to $\pi$ pulse duration and spacing restrict the maximum CPMG filter frequency $f_{\mathrm{max}}$ to a few MHz. The noise spectrum that causes dephasing, although decaying in frequency, has a finite value at $f_{\mathrm{max}}$, which explains why we do not reach $T_{2} = 2T_{1}$ even for large $\gamma$ (short $T_{1}$). The $T_{2}/T_{1}$ ratio is reduced at small $\gamma$ because higher $N$ is required to make $\Gamma_{d}^{(-10)} \ll 1/T_{1}$. 

\begin{figure}
\includegraphics[width = 1.0\columnwidth]{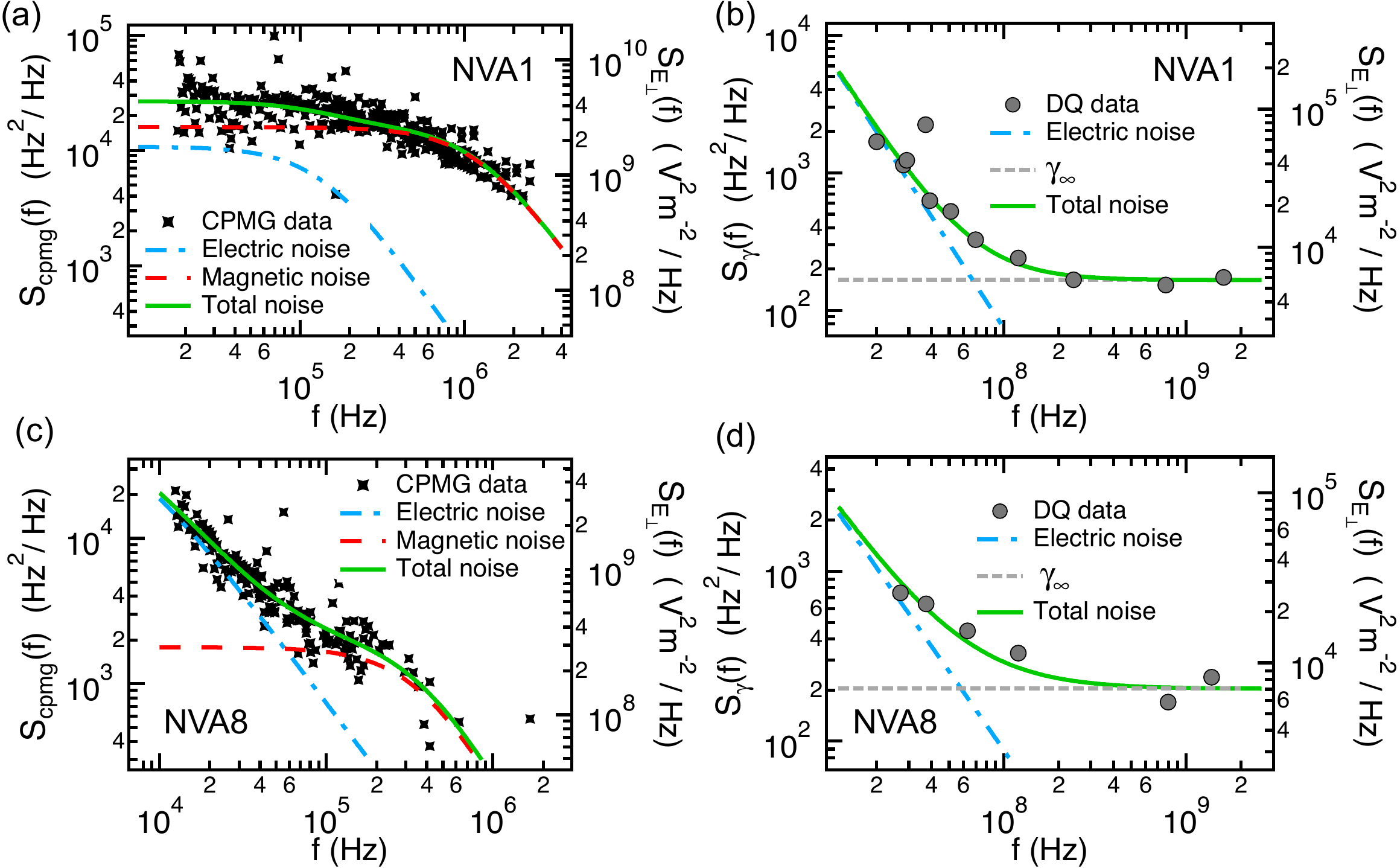}
 \caption{\label{fig:Figure4} Measured noise spectra in terms of coupling power (Hz$^{2}/$Hz) and transverse electric field power (V$^{2}$m$^{-2}/$Hz) for shallow NVs A1 (a,b) and A8 (c,d) using dephasing spectroscopy (left plots) and DQ relaxation spectroscopy (right plots). Each NV data set is jointly fit to a noise model (green solid line) of three parts: $1/f^{\alpha}$-like electric fields (blue dash-dot line), magnetic fields (red dashed line), and a minimum relaxation rate $\gamma_{\infty}$ due to bulk effects (horizontal dashed line). }
 \end{figure} 

Finally, we identify the spectra of surface electric and magnetic field noise over a broad frequency range by employing a combination of SQ dephasing spectroscopy \cite{dasSarma2008,deSousa2009,Bylander2011,Bar-Gill2012,Romach2015,Kim2015} and DQ relaxometry. These complementary techniques are summarized in Table \ref{tab:spectroscopy}. SQ dephasing spectra $S_{\mathrm{cpmg}}\left(f\right)$ and DQ relaxation spectra $S_{\gamma}\left(f\right)$, in units of coupling power Hz$^{2}/$Hz, were generated from measurements like those presented in Figs. \ref{fig:Figure2} and \ref{fig:Figure3}, respectively. The two spectral densities each have distinct noise origins (Table \ref{tab:spectroscopy}), and hence ``coupling power'' has different meanings for dephasing and relaxation. Therefore, to directly compare $S_{\mathrm{cpmg}}\left(f\right)$ and $S_{\gamma}\left(f\right)$ we scale each from a coupling rate to a shared effective transverse electric field noise power spectrum:
\begin{equation}
S_{E_{\perp}}^{\mathrm{cpmg}}\left(f\right) = 2\frac{S_{\mathrm{cpmg}}\left(f\right)}{d_{\parallel}^{2}/h^{2}}; \mathrm{  } S_{E_{\perp}}^{\gamma}\left(f\right) = \frac{S_{\gamma}\left(f\right)}{d_{\perp}^{2}/h^{2}}.
\label{eq:scaleSpectra}
\end{equation}
Equation \ref{eq:scaleSpectra} enables us to jointly model the dephasing and relaxation spectra (see supplementary information \cite{MyersSOM2016}), and the results are shown in Fig. \ref{fig:Figure4}, where the left axis of each plot is coupling power and the right axis is transverse electric noise power. To fit the $S_{E_{\perp}}^{\mathrm{cpmg}}\left(f\right)$ and $S_{E_{\perp}}^{\gamma}\left(f\right)$ data we assume a stationary Gauss-Markov process for electric and magnetic field sources  \cite{deSousa2009}. A double-Lorentzian is the sum of two such processes with different total noise power and frequency cutoffs. The fit results show that the electric noise (blue dash-dot line) is Lorentzian and has a lower-frequency cutoff than the magnetic noise (red dashed line): for NVA1 $\tau_{e}\approx1$ \us{} and $\tau_{m}\approx100$ ns. For NVA8 $\tau_{m}\approx400$ ns and its electric noise curve  actually fits best as $1/f^{\alpha}$ with $\alpha=1.5$. This $\alpha<2$ frequency dependence can be constructed from a sum of many discrete Lorentzians with a range of correlation times, as postulated for noise from charge traps \cite{McWhorter1957} or fluctuating electric dipoles \cite{Safavi2011}.

Our spectroscopy results help tie together prior work \cite{Rosskopf2014,Myers2014,Romach2015,Kim2015} on decoherence of near-surface NVs, which primarily focused on magnetic noise. Kim \textit{et. al.} \cite{Kim2015} gave evidence for shallow-NV dephasing via $1/f$-like $E_{\parallel}$ electric field noise by showing that 1) dephasing noise is reduced when a high-dielectric-constant liquid is placed on the diamond surface, and 2) coherences of SQ and DQ qubits exhibit a ratio that cannot be explained by purely magnetic noise. Our addition of DQ relaxometry to the surface-noise-spectroscopy toolbox enables us to differentiate magnetic and electric noise sources, and importantly, our two-bath model identifies the lower-frequency noise component to be electric, in contrast to previous experiments \cite{Myers2014, Romach2015}. Together with previous depth-resolved work that identified a $1/d^{3.6(4)}$ dependence of $S_{\mathrm{cpmg}}\left(f\right)$ \cite{Myers2014,Romach2015}, we suggest that electric field noise from fluctuating electric dipole moments, such as modeled on metal electrodes in ion traps \cite{Safavi2011,Enoise2013}, could explain the observed results.  Furthermore, we note that the magnitudes of our observed electric field noise are quantitatively consistent with those reported in experiments on ion-trap heating rates \cite{ion2008,MyersSOM2016}. Looking forward, the DQ relaxation technique can be readily combined with single-NV scanning probe microscopy \cite{Cole2009,Pelliccione2014,Luan2015} to shed further light on the microscopic origins of noise from various surfaces.

In conclusion, we have highlighted the importance of considering the coupling between all three levels of the NV ground state for understanding NV decoherence, especially for near-surface NVs. We find the double-quantum (DQ) spin relaxation rate $\gamma$ to be a major, and even dominant, contributor to the limit of qubit coherence time $T_{2}$. We have also used shallow NVs to perform combined dephasing and DQ relaxation spectroscopies of diamond surfaces and furthermore demonstrated a method to distinguish electric and magnetic field noise. To gain more insight into diamond-surface-related electric field noise, several experiments could be revisited with $\gamma$ measured in tandem with $T_{2}$. Since DQ relaxation should be even faster for ultra-shallow NVs at depths of 2-5 nanometers \cite{Loretz2014,reporters2014,Lovchinsky2016}, one could sensitively probe the effects of, for example, annealing with thermal oxidative etching \cite{Kim2014, Wang2016} oxygen plasma etching \cite{Cui2015,O2etch2015}, surface termination \cite{Cui2013, Ohashi2013, SF62015}, chemical treatments \cite{Lovchinsky2016}, temperature \cite{Rosskopf2014,Jarmola2012}, and variations in the work function \cite{Pakes2009}. $\gamma$ measurements might also be combined with optical studies such as the effects of photoinduced electric fields \cite{Bassett2011}. The DQ relaxometry technique we have presented will also be a useful asset for understanding the coupling of general spin $S > 1/2$ solid-state defects to interfaces in hybrid systems.

\begin{acknowledgments}
The authors thank K. Lee, D. Rugar, and J. Mamin for helpful discussions. P. Ovartchaiyapong performed the electron beam lithography for nanopillars and provided helpful knowledge on diamond fabrication. The authors thank D. D. Awschalom and K. Ohno for providing diamond growth equipment and training. This work was supported by the DARPA QuASAR program and the AFOSR YIP. B.A.M. acknowledges support from the Department of Defense (NDSEG) and the IBM Ph.D. Fellowship Program. 

\end{acknowledgments}

\bibliography{dqbibliography}

\section{Supplemental material}

\subsection{Diamond samples}
We prepared a $(001)$ single-crystal diamond film, labeled Sample A, with near-surface NV centers for the experiments presented in the main text is discussed. Table \ref{tab:samples} lists the main features in comparison with a Sample B used for supplemental measurements.

\begin{table*}[t]
\begin{tabular}{|l|l|l|} 
\hline
Properties &    Sample A     &       Sample B \cite{Myers2014} \\ \hline
substrate &  E6 electronic & E6 electronic \\ \hline
thickness & 150 \um{} & 30 \um{} \\ \hline
pre-growth & polished, ArCl$_{2}$ etch & polished, no etch \\ \hline
CVD growth & 50 nm \C{12} ($99.99\%$) & 150 nm \C{12} ($99.999\%$) \\ \hline
N incorporation & \nit{14} implanted 4.0 keV & \nit{15} delta doped, multi-layer \\ \hline
vacancies & from implantation & 2-MeV electron irradiation \\ \hline
annealing & 850 $^{\circ}$C vacuum 2.5 h & 850 $^{\circ}$C H$_{2}$/Ar 2 h \\ \hline
fabrication & on-chip waveguide, nanopillars & on-chip metal coordinates \\  \hline
NV depths & not measured & measured with MRI + NMR \\ \hline

\hline
\end{tabular}
\caption{Diamond samples with implanted and delta-doped NVs measured in this work.}
\label{tab:samples}

\end{table*}

Sample A started as a polished electronic grade (Element Six) substrate of original dimensions $2\times2\times0.5$ mm$^{3}$. The substrate was sliced into two plates and polished from the cut side down to a measured thickness of 150 \um{}. AFM measurements indicated a surface roughness of 160 pm with a step flow pattern. The diamond was etched with ArCl$_{2}$ plasma (Ar 25 sccm, Cl$_2$ 40 sccm, ICP 500 W, bias 200 W, 0.7 Pa) for 20 minutes to mitigate polishing damage. After cleaning in boiling acid H$_{2}$NO$_{3}$:H$_{2}$SO$_{4}$ 2:3 for 40 minutes we grew 40-50 nm of \C{12}-enriched diamond with plasma-enhanced chemical vapor deposition (CVD) at 800 $^{\circ}$C, 750 W, 0.1 sccm \C{12}H$_{4}$ (99.99$\%$), and 400 sccm H$_{2}$. The sample was implanted with 4 keV \nit{14} ions of dose $5.2\times10^{10}$cm$^{-2}$ at a tilt angle of $7^{\circ}$. This was followed by annealing in vacuum $(P<10^{-9}$ Torr at max temperature) at 850 $^{\circ}$C for 2.5 hours with a 40-minute temperature ramp. The sample was cleaned in HClO$_{4}$:H$_{2}$NO$_{3}$:H$_{2}$SO$_{4}$ 1:1:1 for 1 hour at 230-240$^{\circ}$C. 

Standard photolithography and deposition of Ti/Au 6 nm/350 nm was used to pattern a microstrip for microwave control.
Diamond nanopillars were patterned on sample A to increase the collection efficiency of the NV PL \cite{nanopillar2011,nanopillar2015}, which significantly reduces the required long averaging time of relaxation measurements. Tapered diamond nanopillars of 400-nm diameter were patterned with e-beam lithography and etched in O$_2$ plasma to a height of 500 nm. The small height was chosen to limit the amount of time of exposure of sidewalls to the plasma.

The diamond was glued to a thin metal sheet with a hole for optical high-NA access through the backside. The microstrip was wirebonded off-chip to a PCB waveguide to interface with a microwave amplifier circuit. A microwave source was gated by two in-series fast switches for extra isolation during long spin relaxation measurements. Two-tone measurements (Fig. \ref{fig:Figure1}(c)) for $\gamma$ were done by combining individually gated $f_{0,-1}$ and $f_{0,+1}$ carrier signals from two microwave sources. The total signal was amplified and delivered to the on-diamond waveguide. The use of the on-diamond waveguide mitigates drift of the Rabi frequency during long large-$N$ dynamical decoupling measurements in comparison to a free wirebond loop or off-chip waveguide.

Measurements on sample B spins NVB1 and NVB2 were carried out on the same shallow NV ``k7'' and deep NV ``k26'' discussed in our prior work relating NV depth and decoherence rates \cite{Myers2014}. Details of the diamond preparation are included in the supplementary online material of that work. No diamond nanopillars were fabricated on sample B and NV depths were measured.

\begin{figure}
\includegraphics[width = 0.6\columnwidth]{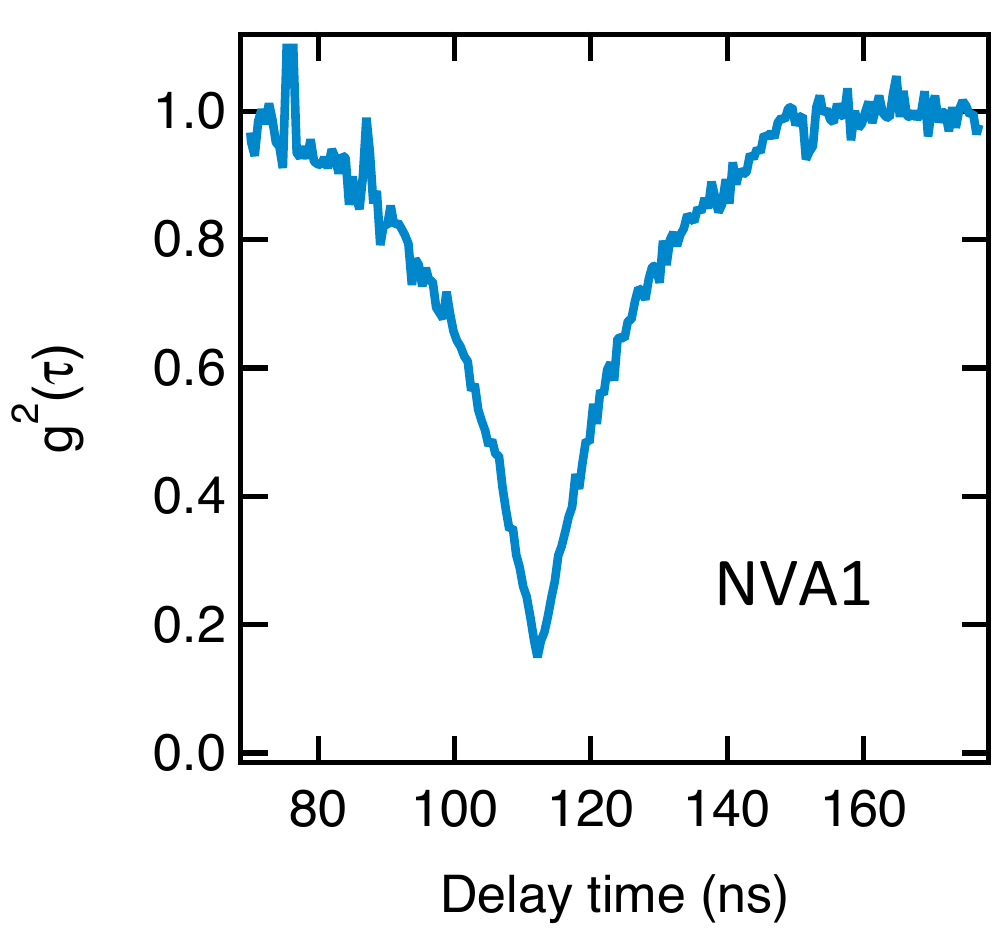}
 \caption{\label{fig:FigureSOMg2} Correlation photon count measurement of pillar containing NVA1, showing $g^{2}\left(\tau=0\right)<0.5$, indicating a single-spin emitter.}
 \end{figure}

Figure \ref{fig:FigureSOMg2} is a normalized $g^{2}\left(\tau\right)$ correlation measurement of emitted photons from the pillar containing NVA1, showing a dip in coincidence counts at zero delay time (offset). The high signal-to-background in the nanopillar causes the dip to be well below the $g^{2}\left(\tau=0\right)<0.5$ threshold. A histogram of time-tagged counts was collected from two APDs connected to a fiber beamsplitter.

\subsection{Derivations of population dynamics for equal SQ rates}
For $k_{b}T \gg \hbar \omega_{i,j}$ the relaxation is bidirectional, \textit{i.e.} $\Omega_{i,j} = \Omega_{j,i}$, as depicted by the double-headed arrows between levels in Figure \ref{fig:Figure1}.
The equality of $\Omega_{+}=\Omega_{-}$ is found in several measurements in past work where \Tonez appears relatively constant with tuning of the applied magnetic field over 100s of Gauss \cite{Jarmola2012,Kolkowitz2015} in the absence of cross-relaxation. Our CVD-grown diamond samples have a relatively low concentration of P1 centers and NV centers such that cross-relaxation is not observed.
In some other cases, like NV coupling to ferromagnetic materials \cite{Yacoby2015}, this equality does not always hold for all applied fields. In this section we look at this simple case discussed in the main text. The rate of population change $d\rho_{00}/dt$ of state $\ket{0}$ is a sum of rates into and out of the state each weighted by the current populations. Therefore, abbreviating $\rho_{ii}$ as $\rho_{i}$, the system of equations is 
\begin{equation}
\frac{d}{dt}
\left( \begin{array}{c} \rho_{0} \\
\rho_{-1}\\
\rho_{1}\end{array}\right)
= \left( \begin{array}{ccc}  -2\Omega & \Omega & \Omega \\
\Omega & -\Omega-\gamma & \gamma \\
\Omega & \gamma & -\Omega-\gamma \end{array}\right)
\left( \begin{array}{c} \rho_{0} \\
\rho_{-1}\\
\rho_{1}\end{array}\right)
\label{eq:differentialEquationRepeat}
\end{equation}
as compared to the more commonly treated situation in NV center literature \cite{Bar-Gill2013,Kolkowitz2015} of $\gamma=0$
\begin{equation}
\frac{d}{dt}
\left( \begin{array}{c} \rho_{0} \\
\rho_{-1}\\
\rho_{1}\end{array}\right)
= \left( \begin{array}{ccc}  -2\Omega & \Omega & \Omega \\
\Omega & -\Omega & 0 \\
\Omega & 0 & -\Omega \end{array}\right)
\left( \begin{array}{c} \rho_{0} \\
\rho_{-1}\\
\rho_{1}\end{array}\right)
\label{eq:differentialEquationRepeat}
\end{equation}

\begin{figure}
\includegraphics[width = 1.0\columnwidth]{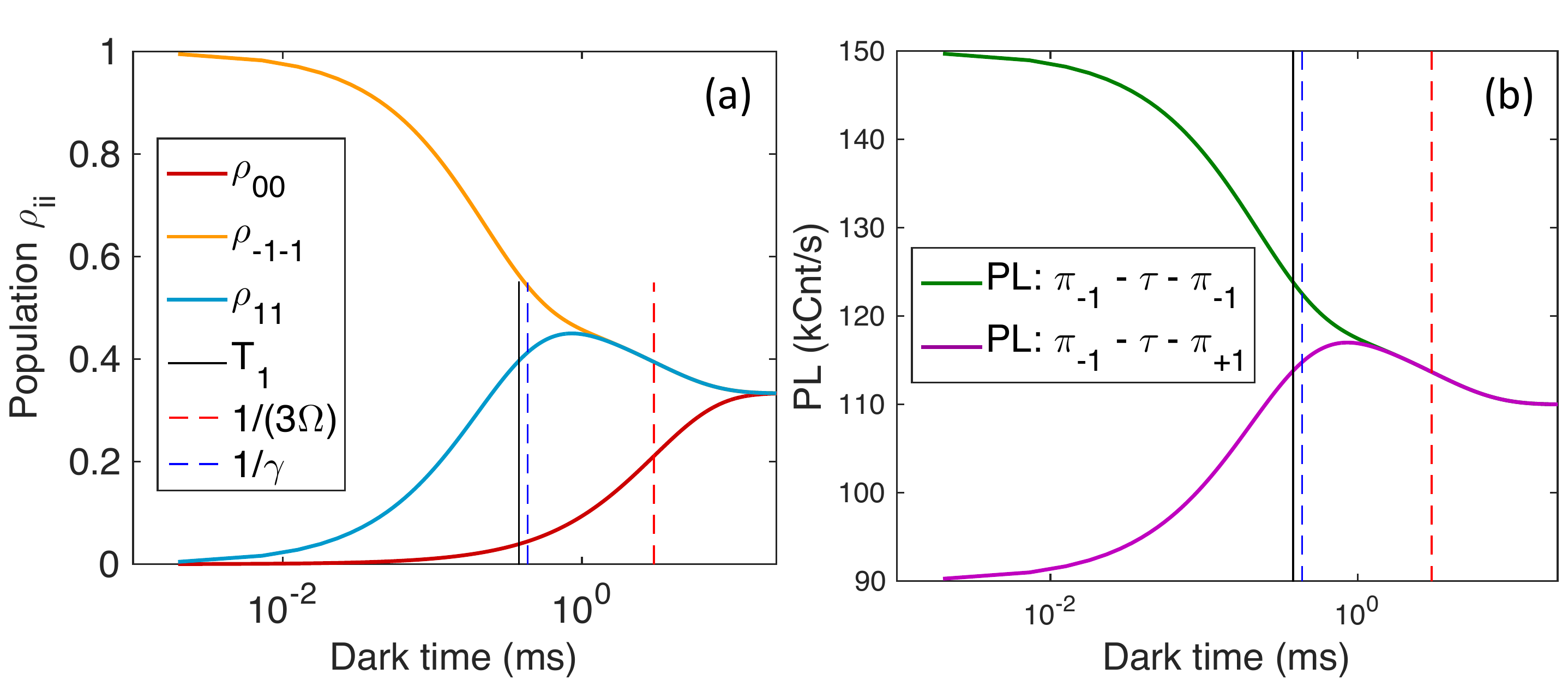}
 \caption{\label{fig:SOMCalcPopulationsPL} (a) Calculated NV ground state spin level populations as a function of time between initialization and readout. The case shown is for the parameters $\gamma=2.3$ kHz and $\Omega=0.11$ with initial state $\rho\left(0\right) = \ket{-1}\bra{-1}$. The dashed vertical lines shown in the legend mark time constants of relaxation relevant to SQ decoherence, and the solid vertical line indicates $T_{1}=1/\left(\gamma+3\Omega\right)$. (b) Calculated PL from these populations given typical PL rates $a_{0}=150$ kCnt/s and $a_{1}=0.6a_{0}$. The green line shows the measurement with two $\pi_{-1,0}$ pulses. The purple line is the result for a final $\pi_{+1,0}$ pulse. This results in effectively measuring the population of the initially-empty $\ket{1}$ state, seen as a non-monotonic change of PL that peaks between $1/\gamma$ and $1/(3\Omega)$. The difference of these two PL curves gives the single-exponential signal $F_{2}$.}
 \end{figure}

We plot in Figure \ref{fig:SOMCalcPopulationsPL}(a) the calculated populations for the case that $\gamma/\Omega \gg 1$, specifically using values for $\gamma$ and $\Omega$ similar to those of NVA1 shown in Fig. \ref{fig:Figure2} and an initial density matrix of $\rho\left(0\right) = \ket{-1}\bra{-1}$ as would be the case initially before the dark time for a measurement of $\gamma$. For this case the $\rho_{00}$ population changes little initially but population rapidly leaks from $\rho_{-1-1}$ to $\rho_{11}$ until the two equilibrate. Figure \ref{fig:SOMCalcPopulationsPL}(b) shows the PL calculated from these populations given two different types of $\pi$ pulses at the end before readout, to effectively either measure the population of the $\ket{-1}$ or $\ket{1}$ states. The purple line in particular shows the non-monotonic curve of PL that would result due to a change of sign in one of the two exponential terms using a final $\pi_{+1}$ pulse, leading to a competition between two decay processes. Note the PL curves in Fig. \ref{fig:SOMCalcPopulationsPL}(b) though are not yet normalized with the subtraction procedure in the data analysis, though the curves demonstrate the direct correspondence between PL and populations of the $\ket{m_{s}}$ states.

\subsection{Derivations of population dynamics for unequal SQ rates}
In the more general case the transition rates between the $S=1$ ground state levels are $\Omega_{+} \neq \Omega_{-} \neq \gamma$, where we use the abbreviation $\Omega_{\pm}\equiv\Omega_{\pm1}$ in this section. The system of differential equations is

\begin{multline}
\frac{d}{dt}
\left( \begin{array}{c} \rho_{0} \\
\rho_{-1}\\
\rho_{1}\end{array}\right)
=\\
 \left( \begin{array}{ccc}  -\Omega_{+}-\Omega_{-} & \Omega_{-} & \Omega_{+} \\
\Omega_{-} & -\Omega_{-}-\gamma & \gamma \\
\Omega_{+} & \gamma & -\Omega_{+}-\gamma \end{array}\right)
\left( \begin{array}{c} \rho_{0} \\
\rho_{-1}\\
\rho_{1}\end{array}\right)
\label{eq:differentialEquationGeneral}
\end{multline}
Substituting for $\rho_{1}$ (where $\rho_{1}= 1-\rho_{0} - \rho_{-1}$) in the equations of $\rho_{0}$ and $\rho_{-1}$ leads to the differential equation system
\begin{multline} 
\frac {d}{dt}
\left( \begin{array}{ccc}
\rho_{0}\\
\rho_{-1}\\
 \end{array} \right)
= \left( \begin{array}{ccc}
-\Omega_{-} - 2\Omega_{+} & -\Omega_{+}+\Omega_{-}\\
\Omega_{-} - \gamma & -\Omega_{-}-2\gamma\\
 \end{array} \right)
\left( \begin{array}{ccc}
\rho_{0}\\
\rho_{-1}\\
 \end{array} \right)\\
+ \left( \begin{array}{ccc}
\Omega_{+}\\
\gamma\\
 \end{array} \right)
\label{rho_equation}
\end{multline} 
The solution to the above equation could be found by first considering the first-order homogeneous  equation:
\begin{flalign}
\frac {d}{dt}
\left( \begin{array}{ccc}
\rho_{0}\\
\rho_{-1}\\
 \end{array} \right)
= \left( \begin{array}{ccc}
-\Omega_{-} - 2\Omega_{+} & -\Omega_{+}+\Omega_{-}\\
\Omega_{-} - \gamma & -\Omega_{-}-2\gamma\\
 \end{array} \right)
\left( \begin{array}{ccc}
\rho_{0}\\
\rho_{-1}\\
 \end{array} \right)
\end{flalign} 
Solving the eigenvalue equation 
\begin{flalign}
\left| \begin{array}{ccc}
-\Omega_{-} - 2\Omega_{+} -\lambda & -\Omega_{+}+\Omega_{-}\\
\Omega_{-} - \gamma & -\Omega_{-}-2\gamma -\lambda\\
 \end{array} \right|
= 0
\end{flalign} 
gives the two eigenvalues as
\begin{multline}
\lambda_{\pm} = -(\Omega_{-}+\Omega_{+}+\gamma )\\
 \pm \sqrt{\Omega_{+}^{2} +\Omega_{-}^{2}+ \gamma^{2}-\Omega_{+}\gamma - \Omega_{-}\gamma - \Omega_{-}\Omega_{+}}
\end{multline}
and the corresponding solution to the homogeneous differential equation is 
\begin{flalign}
c_{1}e^{\lambda_{+}t}\left( \begin{array}{ccc}
1\\
\frac{\Omega_{-}-\gamma}{\Omega_{-}+2\gamma+\lambda_{+}}\\
 \end{array} \right)
+ c_{2}e^{\lambda_{-}t}\left( \begin{array}{ccc}
1\\
\frac{\Omega_{-}-\gamma}{\Omega_{-}+2\gamma+\lambda_{-}}\\
\end{array} \right)
\end{flalign} 

Here $c_{1}$ and $c_{2}$ are constants to be evaluated based on the initial conditions. The particular solution of the inhomogeneous differential could be found by setting the populations $\rho_{0}$ and $\rho_{-1}$ as constants and the result is that $\rho_{0} = \rho_{-1} = 1/3$. Therefore the combined solution for equation \ref{rho_equation} is 

\begin{multline}
\left( \begin{array}{ccc}
\rho_{0}\\
\rho_{-1}\\
 \end{array} \right)
= c_{1}e^{\lambda_{+}t}\left( \begin{array}{ccc}
1\\
\frac{\Omega_{-}-\gamma}{\Omega_{-}+2\gamma+\lambda_{+}}\\
 \end{array} \right)\\
+ c_{2}e^{\lambda_{-}t}\left( \begin{array}{ccc}
1\\
\frac{\Omega_{-}-\gamma}{\Omega_{-}+2\gamma+\lambda_{-}}\\
\end{array} \right)
+ \frac{1}{3}\left( \begin{array}{ccc}
1\\
1\\
\end{array} \right)
\end{multline} 
The population in the +1 state, $\rho_{1}$, can be obtained with the expression $\rho_{1} = 1- \rho_{0}- \rho_{-1}$. Thus, 
\begin{flalign}
\rho_{1} = \frac{1}{3} - c_{1}e^{\lambda_{+}t}(2\Omega_{-}+\gamma +\lambda_{+} ) - c_{2}e^{\lambda_{-}t}(2\Omega_{-}+\gamma +\lambda_{-} ) 
\end{flalign}

The validity of the assumption that $\Omega_{-}=\Omega_{+}$ was verified using an NV spin state relaxation experiment where an NV initially polarized to the $\rho = \ket{0}\bra{0}$ state and read out with the $F_{1}$ sequence for nomalization. The constants $c_{1}$ and $c_{2}$ could be determined based on that fact that at time $\tau=0$, the populations should correspond to $\rho_{0}(t=0) = 1$ and $\rho_{\pm1}(t=0) = 0$.  These conditions lead to 
\begin{flalign}
c_{1} = -\frac{3\Omega_{-}+\lambda_{-} }{3(\Omega_{-}- \gamma)(\lambda_{-} - \lambda_{+})}
\end{flalign}
and
\begin{flalign}
c_{2} = \frac{3\Omega_{-}+\lambda_{+} }{3(\Omega_{-}- \gamma)(\lambda_{-} - \lambda_{+})}
\end{flalign}

The difference between the NV PL vs $\tau$ curves when the  $\pi$ pulse at the end of the dark time is tuned to the $\ket{1}$ state ($\pi_{0,+1}$) or the $\ket{-1}$ state ($\pi_{0,-1}$) can be used to probe whether $\Omega_{-}=\Omega_{+}$. The subtracted PL signal from such a pulse sequence is given by 
\begin{flalign}
\text{PL}_{\pi_{0,+1}-\pi_{0,-1}}  = -\frac{r(3\Omega_{-}+\lambda_{+})(3\Omega_{-}+\lambda_{-})}{3(\Omega_{-}-\gamma)(\lambda_{+}-\lambda_{-})}[e^{{\lambda_{+}t}} - e^{\lambda_{-}t}]
\label{PLdiffpipluspineg}
\end{flalign}
 where $r$ is the contrast between the 0 state and $\pm1$ states. In the equation for the PL difference between the two pulse sequences (equation \ref{PLdiffpipluspineg} ), the term $3\Omega_{-}+\lambda_{-}$ goes to zero in the event  $\Omega_{-}=\Omega_{+}$ bringing the PL difference to zero.
\begin{figure}
\includegraphics[width = 1.0\columnwidth]{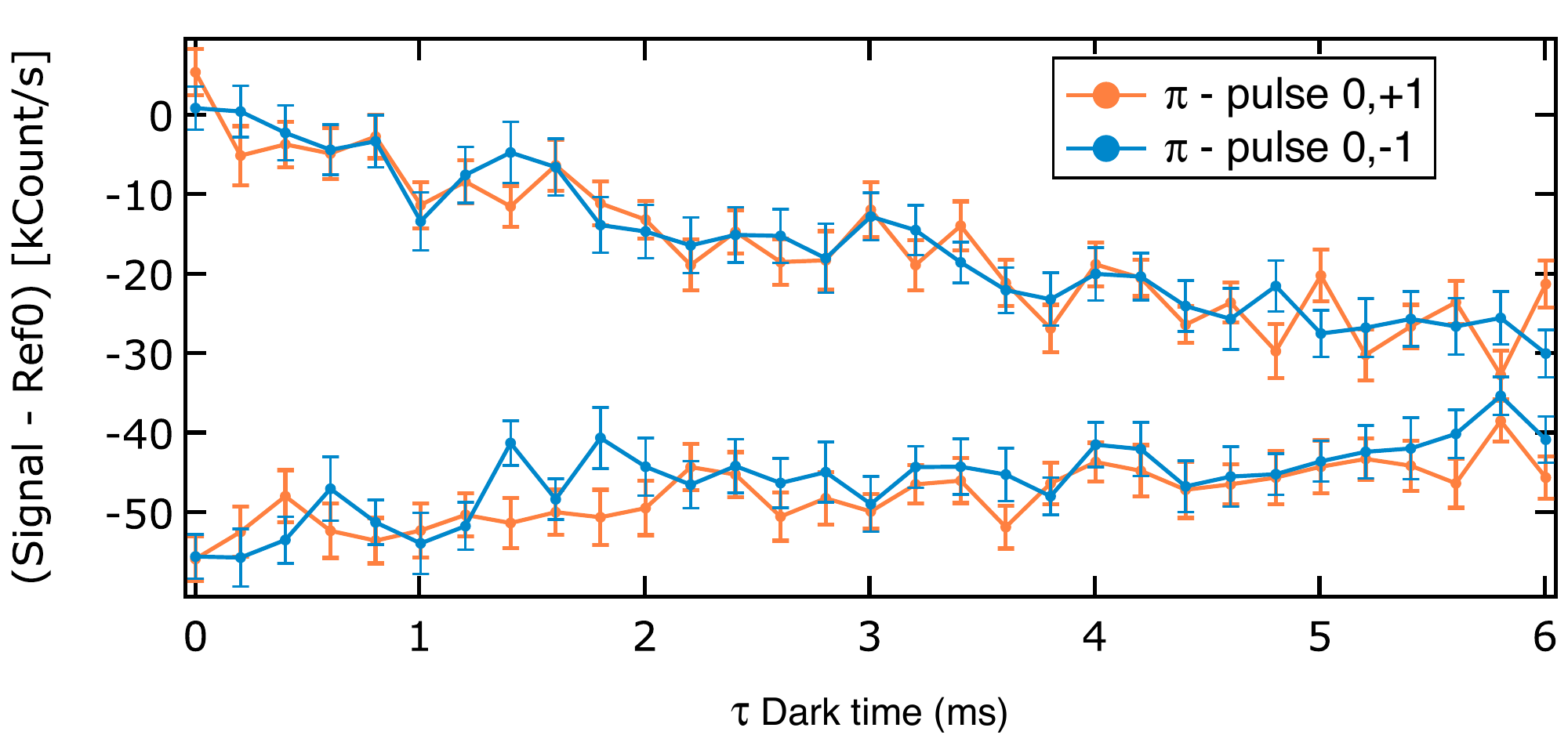}
 \caption{\label{fig:OmegaPlusMinus} Comparison of the $F_{1}$ results for NVA1 using $\pi$ pulses resonant with the $\omega_{+1,0}$ transition (orange) and the $\omega_{-1,0}$ transition (blue). The curves overlap and fit to the same relaxation rate, therefore we conclude that it is a good assumption to set $\Omega \equiv \Omega_{+}=\Omega_{-}$ to simplify the $\gamma$ analyses in this work.}
 \end{figure}

We compared measurements using the $F_{1}$ sequence with a $\pi_{0,+1}$ or $\pi_{0,-1}$ and found the $\Omega_{+} = \Omega_{-}$ in even the larger $\omega_{\pm1}$ regimes where $\omega_{1,0} \gg \omega_{-1,0}$. The results are plotted in Fig. \ref{fig:OmegaPlusMinus} for NVA1 at $\omega_{\pm1}$$=800$ MHz, a splitting large enough between $\omega_{+1,0}$ and $\omega_{-1,0}$ that some difference might be expected but was not observed. Subtracting the blue and orange curves gives zero on average as predicted by Eq. \ref{PLdiffpipluspineg}. Therefore, we use the simpler population dynamics model for relaxation measurements using one fitting parameter $\Omega$ for the $\left|\Delta m_{s}\right| = 1$ relaxation rates. 

\subsection{Pulse sequences and fitting for $\Omega$, $\gamma$}
\begin{figure}
\includegraphics[width = 1.0\columnwidth]{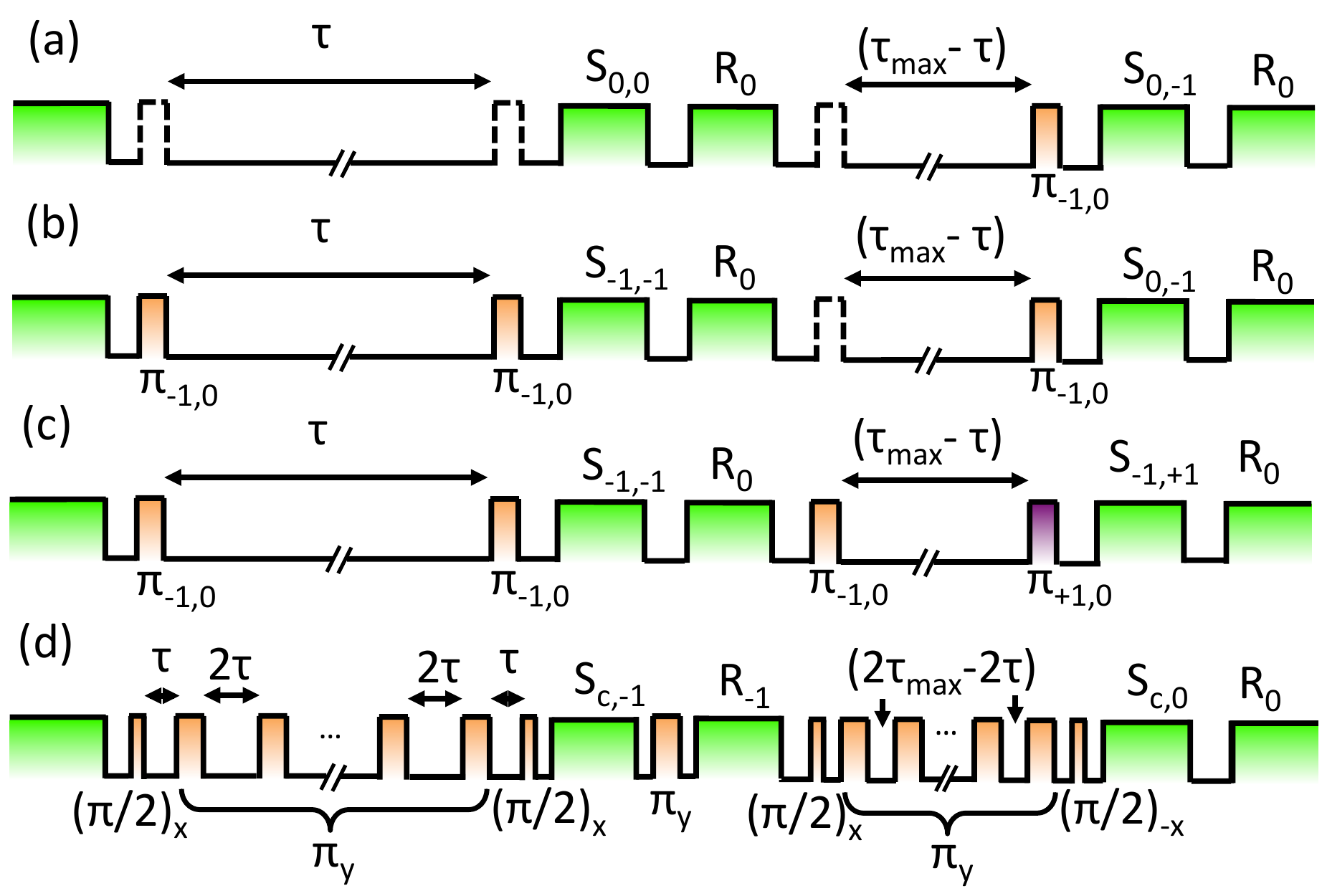}
 \caption{\label{fig:figSOMFullT1Sequence} Full experimental form of the pulse sequences for extracting $\Omega$ (a) and $\gamma$ (b,c) relaxation rates and (d) CPMG-enhanced $T_{2}$. Green pulses are for 532-nm initialization and readout, and orange (purple) pulses depict gating the in-series microwave switches with carrier on resonance with the $\omega_{0,-1}/2\pi$ ($\omega_{0,+1}/2\pi$) transition. Dashed-line boxes are blank delays when no pulse is done for $\ket{0}$ preparation or readout. The normalization procedure is described in the supplementary text. For each signal $S$ the first subscript is labeled to $0$, $-1$, or $c$ for the initialized state in $\ket{0}$, $\ket{-1}$, or SQ superposition, and the second subscript refers to the $m_{s}=0,-1,+1$ population that is read out, which is effectively a phase readout for the CPMG sequence. The $R$ pulse is for reference of the PL $a_{0}$ or $a_{-1}$ of each state $\ket{0}$ or $\ket{\pm1}$. The IQ phases $\pm x,y$ are indicated for the CPMG sequence.}
 \end{figure}

The full experimental form of the pulse sequences described in Fig. \ref{fig:Figure1} of the main text is shown in Fig. \ref{fig:figSOMFullT1Sequence}. The data for the $F_{1}$ sequence in Fig. \ref{fig:figSOMFullT1Sequence}(a) is normalized as
\begin{equation}
F_{1} = \left(S_{0,0} - R_{0}\right) - \left(S_{0,-1} - R_{0}\right).
\label{eq:SaNormalized}
\end{equation}
$S_{0,0}$ was then fit to Eq. \ref{eq:fit0t0} in the main text to yield parameters $r$ and $\Omega$. The data plotted as a relaxation signal is normalized by $r$. The $F_{2}$ sequence in Fig. \ref{fig:figSOMFullT1Sequence}(c) was then performed and normalized as
\begin{equation}
F_{2} = \left(S_{-1,-1} - R_{0}\right) - \left(S_{-1,+1} - R_{0}\right).
\label{eq:ScNormalized}
\end{equation}
An alternative to the $F_{2}$ signal to extract $\gamma$ is the $F_{3}$ signal in Fig. \ref{fig:figSOMFullT1Sequence}(b), giving
\begin{equation}
F_{3} = \left(S_{-1,-1} - R_{0}\right) - \left(S_{0,-1} - R_{0}\right).
\label{eq:SbNormalized}
\end{equation}
$F_{3}$ is fit to a sum of Eqs. \ref{eq:fit0t0} and \ref{eq:fitptp2Tone} $(F_{1}\left(\tau\right) + F_{2}\left(\tau\right))/2$ with $\Omega$ fixed to the $F_{1}$ result to yield parameters $r$ and $\gamma$. The data plotted as a relaxation signal is normalized by $r$. Joint fits of $F_{1}$ and $F_{3}$ to data gave the same results and uncertainties in $\Omega$ and $\gamma$ as obtaining first $\Omega$ from $F_{1}$ and then fixing $\Omega$ to fit $F_{3}$ for $\gamma$. $F_{3}$ is an alternative to the two-tone measurement $F_{2}$ (Fig. \ref{fig:figSOMFullT1Sequence}(c)) and $F_{1}$, however, an advantage of using $F_{2}$ are smaller error bars for $\gamma$ due to a single-exponential fit function rather than a biexponential fit function in $F_{3}$. $F_{2}$ also demonstrates most clearly how the pulse sequences give direct readout of the different spin populations.

\begin{figure}
\includegraphics[width = 1.0\columnwidth]{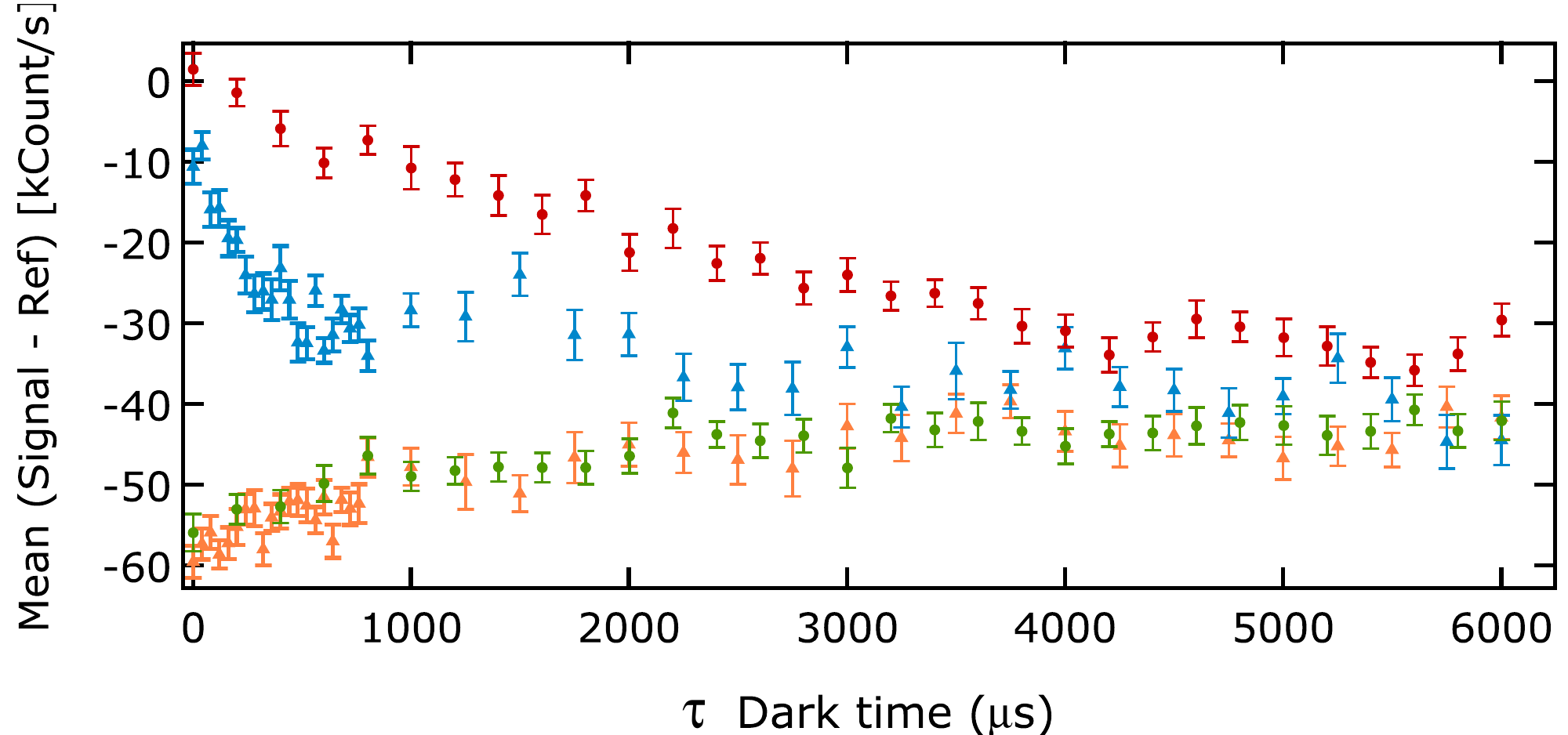}
 \caption{\label{fig:figSOMrawT36} Four data sets for NVA1 taken at 37.1 MHz using relaxation pulse sequences Fig. \ref{fig:figSOMFullT1Sequence}(a) (circles) and Fig. \ref{fig:figSOMFullT1Sequence}(b) (triangles) of Fig. \ref{fig:figSOMFullT1Sequence}. Red is $\left(S_{0,0}-R_{0}\right)$, green is $\left(S_{0,-1}-R_{0}\right)$, blue is $\left(S_{-1,-1}-R_{0}\right)$, and orange is $\left(S_{0,-1}-R_{0}\right)$. The $\pi-\tau-\pi$ data at short $\tau<1000$ \us{} (blue) is sampled at a smaller time spacing to accurately capture the fast $\gamma$ decay.}
 \end{figure}

The subtraction $S_{0,0}-R_{0}$ eliminates some laser and PL common-mode noise within a sequence shot. Two example data sets before final subtraction and fitting to $F_{1}$ and $F_{3}$ are shown in Fig. \ref{fig:figSOMrawT36}, where the biexponential (blue data) indicates a large $\gamma$ and small $\Omega$.

We use only ``symmetrized'' pulse sequences (same total shot time for each $\tau$ point) that keep the average laser and microwave power relatively constant for stable AOM operation and heating. The microwave signal was gated by two in-series fast switches and IQ modulation to ensure no spin rotations during the long dark times for $T_{1}$ measurements. The on-diamond microwave waveguide also enables stable Rabi frequencies over long time periods. Because the timescales of $\gamma$ and $\Omega$ sometimes differ by more than an order of magnitude the time point sampling must be chosen carefully, for example a larger number of time sampling points at small $\tau$ helps to resolve the initial fast DQ relaxation.

CPMG measurements (Fig \ref{fig:figSOMFullT1Sequence}(d)) were normalized on a coherence scale $[0,1]$ by measuring the reference PL in the $\ket{0}$ ($R_{0}$) and $\ket{-1}$ ($R_{-1}$) states during every shot. The two measurements with a $(\pi/2)_{\pm x}$ pulse at the end project the phase into a population of either state and these two coherence curves are subtracted and normalized by the contrast $R_{0} - R_{-1}$ to obtain the plotted result.  Common mode PL noise is primarily subtracted out by (signal $-$ reference) since the two $(\pi/2)_{\pm x}$ signals at a given $\tau$ were acquired at different times (seconds to minutes) due to the symmetrization of the sequence.

Error bars on the data points for relaxation population and CPMG coherence data versus time are standard errors over many repetitions at each individual $\tau$ point. Error bars in the extracted parameters $\gamma$, $\Omega$, and $T_{2}$ are fit standard errors, and uncertainties in $T_{1}$ were propagated from the errors in $\gamma$ and $\Omega$.

\subsection{Spin flips in spin $S=1$ ground state}
The three eigenstates of the $S_{z}$ operator $\ket{m_{s}}$ with $m_{s}=0,\pm1$ are alternately written in terms of the single-electron spin-$1/2$ states as $\ket{0} = \left(\ket{\uparrow}\ket{\downarrow} + \ket{\downarrow}\ket{\uparrow}\right)/\sqrt{2}$, $\ket{1} = \ket{\uparrow}\ket{\uparrow}$, and $\ket{-1} = \ket{\downarrow}\ket{\downarrow}$. In this form, the term single-quantum diretly refers to a single electron spin flip, and the double-quantum refers to both electrons flipping simultaneously, $\ket{\uparrow\uparrow}\leftrightarrow\ket{\downarrow\downarrow}$. Therefore, state $\ket{0}$ is immune to double-quantum spin relaxation.

\subsection{Limits to $T_{2}$ in $S=1$ ground state}
\begin{figure}
\includegraphics[width = 1.0\columnwidth]{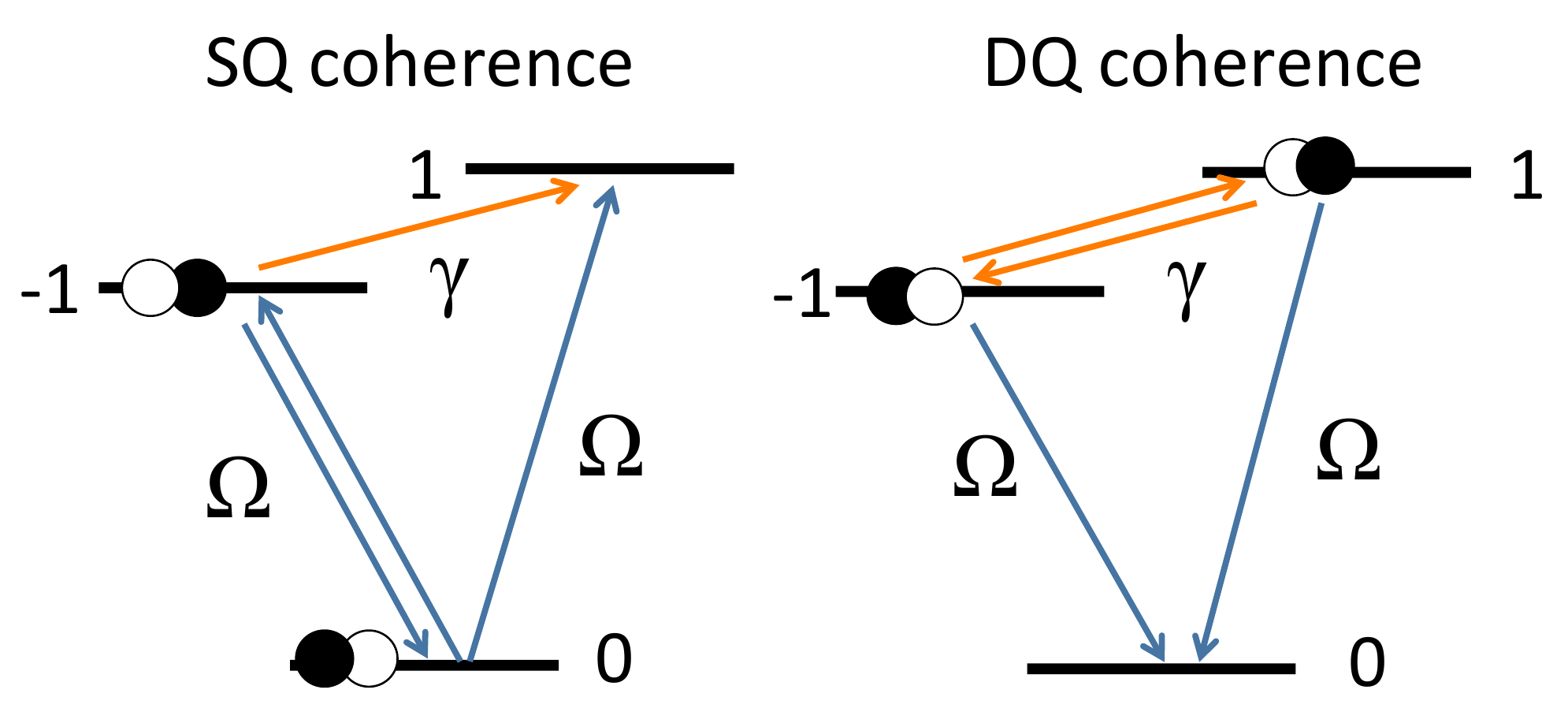}
 \caption{\label{fig:figSOMcoherenceSchematics} Diagrams of the NV ground state spin triplet with two-state coherences at finite magnetic field along the $z$ axis. A coherence is illustrated here between the two levels occupied with black and white discs. In the single-quantum (SQ) coherence there are three $\Omega$ relaxation events possible (blue arrows) and one $\gamma$ relaxation event (orange arrow) to leak population out of the superposition state. In the double-quantum (DQ) coherence case there are two relaxation events possible of each type. The total $1/T_{1}$ relaxation rate in each case is the sum of these four rates.}
 \end{figure}

With the relaxation rates between the three sublevels measured using the sequences presented in the main text Fig. \ref{fig:Figure1} and supplement Fig. \ref{fig:figSOMFullT1Sequence}, we can compute the fundamental limits to $T_{2}$ in the case of no dephasing. By $\left(T_2\right)^{-1}$ here we mean the total decay rate of the off-diagonal coherence term in the density matrix, $\rho_{0-1}$. An intuitive way to look at the $T_{1}$ quantities for the three-level NV ground state superpositions is in terms of the constituent relaxation rates, so $1/T_{1}^{\mathrm{SQ}} = 3\Omega+\gamma$ and $1/T_{1}^{\mathrm{DQ}} = 2\Omega+2\gamma$. That is, once a coherence is initialized it can decohere due to a quantum jump via any one of four channels, as illustrated in Fig. \ref{fig:figSOMcoherenceSchematics}.
The physical properties of the density matrix $\rho$ and its time evolution require the coherences (off-diagonal elements) to decay under specific constraints relating the dephasing and relaxation rates (Eqs. 20 and 41 in ref. \cite{threelevel2004}). We assume first that we perform a standard Hahn echo on the superposition of $\ket{0}$ and $\ket{-1}$. Let $\Gamma^{(ab)}_{d}$ denote the pure dephasing rate between $\ket{a}$ and $\ket{b}$. The pure dephasing rate we are interested in $\Gamma^{(-10)}_{2}$ must follow the constraints required for a three-level system

\begin{equation}
\Gamma_{d}^{(-10)} = \Gamma^{(-10)}_{2} - \frac{\Omega + \Omega + \Omega + \gamma}{2}
\label{eq:pureDephasingConstraint1}
\end{equation}
where we identify the total decoherence rate $\Gamma^{(-10)}_{2} = \Gamma_{2} = 1/T_{2}$ that we measure in a N-pulse single-quantum coherence sequence. The dephasing rate $\Gamma^{(-10)}_{d}$ cannot be negative, therefore the zero-dephasing limit of $\Gamma_{2}$ is
\begin{equation}
\Gamma_{2} = \frac{3\Omega + \gamma}{2}.
\label{eq:decoherenceLimit01}
\end{equation}
Thus, a ratio $\gamma/\Omega > 3$ means that SQ coherence is limited more by $\gamma$ than $\Omega$, which is what we observe at $\omega_{\pm1}/2\pi \lesssim 100$ MHz (Fig. \ref{fig:Figure3}(b)). We identify \Tonez{} in the main text as $T_{1}^{(0)} = 1/\left(3\Omega\right)$. This gives the final result
\begin{equation}
T_{2} \leq T_{2}^{\mathrm{SQmax}}= \frac{2}{3\Omega + \gamma} =  2\left(\frac{1}{T_{1}^{(0)}} + \frac{1}{T_{1}^{(+1,-1)}} \right)^{-1}.
\label{eq:T2limit}
\end{equation}
Table \ref{tab:Ttwomax} lists the measured $\Omega$, $\gamma$, \Tpm$=1/\gamma$, \Tonez, and CPMG-N $T_{2}$ along with the theoretical upper bound \Ttwomax{}. In our experiments at best the NV $T_{2}$ reached just over half of this maximum: $T_{2}\gtrsim0.5$ \Ttwomax $=T_{1}$. From the spectral analysis we can infer that the existence of finite-frequency noise, even if decaying with frequency, will keep the dephasing term non-zero even for very large $N$. We infer that technical challenges, a combination of pulse errors and finite $\pi$-pulse times $t \sim \tau$, are responsible for the inability to space ideal pulses close enough together to completely eliminate dephasing from high-frequency noise. For the high-field case shown in Fig. \ref{fig:Figure2}(b) of the main text we found the $T_{2}\left(N\right)$ to begin saturating around $N=1024$ pulses. For NVA7, applying $N=2048$ appeared to cause a reduction in $T_{2}\left(N\right)$, and it appears that the main limitations to increasing it further were technical in combination with inherent surface noise sources.

\begin{table*}[t]
\begin{tabular}{|c|r|r|r|r|r|r|r||r|r|r|} 
\hline
NV &    $\omega_{\pm1}$$/2\pi$    &  $\Omega$ [kHz]  &   $\gamma$ [kHz]  & $1/\left(3\Omega\right)$ [ms]   &  $1/\gamma$ [ms] &  \Tone{} [ms] & $T_{2}\left(N\right)$ [ms]  &  \Ttwomax{} = 2\Tone [ms] & $T_{2}/$\Ttwomax{} ($\%$) \\ \hline
NVA1 & 20 & 0.11(1) & 1.7(3) & 3.1(3) & 0.6(1) &    0.50(8) &  NA &   1.0(2) &   NA\\
NVA1 & 28 & 0.084(8) & 1.14(9) & 4.0(4) & 0.88(7) &    0.72(5) &  NA &   1.44(9) &   NA\\
NVA1 & 29.2 & 0.14(1) & 1.2(2) & 2.4(2) & 0.8(1) &    0.61(8) &  NA &   1.2(2) &   NA\\
NVA1 & 37.1 & 0.115(4) & 2.4(6) & 2.9(1) & 0.4(1) &    0.37(9) &  0.41(4) & 0.7(2) &   60(20)  \\
NVA1 & 39.2 & 0.142(10) & 0.6(2) & 2.4(2) & 1.6(6) &    1.0(2) &  NA &   1.9(3) &   NA\\
NVA1 & 51 & 0.22(1) & 0.52(9) & 1.55(7) & 1.9(3) &    0.85(7) &  NA & 1.7(1) &   NA  \\
NVA1 & 69.9 & 0.16(1) & 0.33(5) & 2.1(1) & 3.1(5) &    1.23(9) &  NA & 2.5(2) &   NA  \\
NVA1 & 120 & 0.194(9) & 0.24(4) & 1.72(8) & 4.2(7) &    1.22(7) &  NA &   2.4(1) &   NA\\
NVA1 & 240.6 & 0.095(5) & 0.17(4) & 3.5(2) & 6(1) &    2.2(2) &  NA & 4.4(4) &   NA  \\
NVA1 & 774.5 & 0.235(9) & 0.15(3) & 1.42(5) & 7(1) &    1.17(6) &  NA &   2.3(1) &   NA\\
NVA1 & 1612 & 0.111(8) & 0.17(2) & 3.0(2) & 5.8(7) &    2.0(1) &  NA &   4.0(3) &   NA\\ \hline
NVB1 & 32.5 & 0.100(6) & 0.45(4) & 3.3(2) & 2.2(2) &    1.32(7) &  NA &   2.6(1) &   NA\\ \hline
NVB2* & 32.5 & 0.066(8) & 0.11(2) & 5.0(6) & 9(2) &    3.2(3) &  NA &   6.5(6) &   NA\\ \hline
NVA2 & 30.8 & 0.31(2) & 8(2) & 1.08(7) & 0.12(3) &    0.11(2) &  0.13(2) & 0.21(4) &   60(10)  \\ \hline
NVA5* & 28 & 0.13(2) & 0.17(2) & 2.6(4) & 6.0(7) &    1.8(2) &  NA &   3.6(3) &   NA\\
NVA5* & 37 & 0.15(1) & 0.23(4) & 2.2(1) & 4.4(8) &    1.5(1) &  NA &   2.9(2) &   NA\\
NVA5* & 80 & 0.14(2) & 0.13(2) & 2.3(3) & 8(1) &    1.8(2) &  NA &   3.6(3) &   NA\\
NVA5* & 240 & 0.16(2) & 0.19(6) & 2.1(3) & 5(2) &    1.5(2) &  NA &   3.0(4) &   NA\\ \hline
NVA7 & 43 & 0.09(2) & 0.7(2) & 3.8(8) & 1.4(4) &    1.0(2) &  NA &   2.1(4) &   NA\\
NVA7 & 59 & 0.062(6) & 0.5(1) & 5.4(5) & 2.2(4) &    1.6(3) &  NA &   3.1(6) &   NA\\
NVA7 & 73.97 & 0.072(7) & 0.30(7) & 4.6(5) & 3.4(8) &    2.0(3) &  NA &   3.9(6) &   NA\\
NVA7 & 104 & 0.077(8) & 0.22(4) & 4.3(4) & 4.6(8) &    2.2(2) &  NA &   4.5(5) &   NA\\
NVA7 & 137.7 & 0.066(6) & 0.30(8) & 5.1(5) & 3.3(9) &    2.0(3) &  NA &   4.0(6) &   NA\\
NVA7 & 214 & 0.058(6) & 0.4(1) & 5.8(6) & 2.3(6) &    1.7(3) &  NA &   3.3(6) &   NA\\
NVA7 & 465.7 & 0.070(7) & 0.27(5) & 4.8(5) & 3.7(7) &    2.1(2) &  NA &   4.2(4) &   NA\\
NVA7 & 797 & 0.067(5) & 0.26(4) & 5.0(4) & 3.8(6) &    2.2(2) &  NA &   4.3(4) &   NA\\
NVA7 & 1431 & 0.054(6) & 0.27(6) & 6.2(7) & 3.7(8) &    2.3(4) &  1.05(4) & 4.6(7) &   23(4)  \\ \hline
NVA8 & 26.95 & 0.066(6) & 0.7(2) & 5.1(5) & 1.3(4) &    1.1(2) &  NA &   2.1(4) &   NA\\
NVA8 & 37.4 & 0.061(5) & 0.6(1) & 5.5(4) & 1.6(3) &    1.2(2) &  NA &   2.4(4) &   NA\\
NVA8 & 63 & 0.061(6) & 0.45(8) & 5.4(5) & 2.2(4) &    1.6(2) &  NA &   3.2(4) &   NA\\
NVA8 & 119.85 & 0.061(6) & 0.33(7) & 5.5(5) & 3.0(6) &    2.0(3) &  NA &   3.9(6) &   NA\\
NVA8 & 798.4 & 0.056(6) & 0.17(3) & 5.9(6) & 6(1) &    3.0(3) &  NA &   5.9(6) &   NA\\
NVA8 & 1375.9 & 0.056(6) & 0.24(4) & 5.9(6) & 4.2(7) &    2.5(3) &  1.28(6) & 4.9(6) &   26(3) \\ \hline

\hline
\end{tabular}
\caption{Measured relaxation rates $\Omega$, $\gamma$, and maximum coherence time $T_{2}\left(N\right)$ for the NVs in this work. The NV names with a * are deep NVs and do not show an increasing $\gamma$ at low $\omega_{\pm1}$. The second column from the right shows the computed maximum expected spin coherence time from Eq. \ref{eq:T2limit}, and the rightmost column shows the ratio of experimental to theoretical maximum results. Standard errors from fitting routines are given parentheses for the least significant digit. ``NA'' refers to $\omega_{\pm1}$ values at which the maximum $T_{2}$ was not measured. The large difference between $1/\left(3\Omega\right)$ and the full $T_{1}$, in all cases, emphasizes the significant effect of $\gamma$ on the relevant relaxation rate. 
\label{tab:Ttwomax}}

\end{table*}

We also consider the case of a DQ coherence where a superposition of $\ket{1}$ and $\ket{-1}$ is prepared, as has been relevant in experiments to enhance sensitivity to nuclear spins \cite{Kim2015,Mamin2014}. In this case we are interested in the total decay rate $\Gamma^{(-11)}_{2}$ of the $\rho_{-11}$ term and so

\begin{equation}
\Gamma_{d}^{(-11)} = \Gamma^{(-11)}_{2} - \frac{\Omega +\gamma + \Omega + \gamma}{2},
\label{eq:pureDephasingConstraintDQ}
\end{equation}
which leads to a symmetric result in the contributions of the relaxation rates. The resulting zero-dephasing limit to the DQ coherence time $T_{2}^{\mathrm{DQ}} = 1/\Gamma^{(-11)}_{2}$ is

\begin{equation}
T_{2}^{\mathrm{DQ}} \leq T_{2}^{\mathrm{DQmax}}= \frac{2}{2\Omega + 2\gamma} = 2 \left(\frac{2}{3T_{1}^{(0)}} + \frac{2}{T_{1}^{(+1,-1)}} \right)^{-1}
\label{eq:T2limitDQ}
\end{equation}
where here we retain the definitions of \Tonez{} $=\left(3\Omega\right)^{-1}$ and \Tpm{} $=\left(\gamma\right)^{-1}$ that were stated for the SQ coherence case. Thus, a ratio $\gamma/\Omega > 1$ means that DQ coherence is limited more by $\gamma$ than $\Omega$.

The ratio of DQ and SQ coherence time limits $T_{2}^{\mathrm{DQmax}}/T_{2}^{\mathrm{SQmax}}$ can be written in terms of $\gamma$ and $\Omega$.

\begin{equation}
T_{2}^{\mathrm{DQmax}}/T_{2}^{\mathrm{SQmax}} =  \frac{\gamma + 3\Omega}{2\gamma + 2\Omega}
\label{eq:T2DQSQratio}
\end{equation}

\begin{figure}
\includegraphics[width = 1.0\columnwidth]{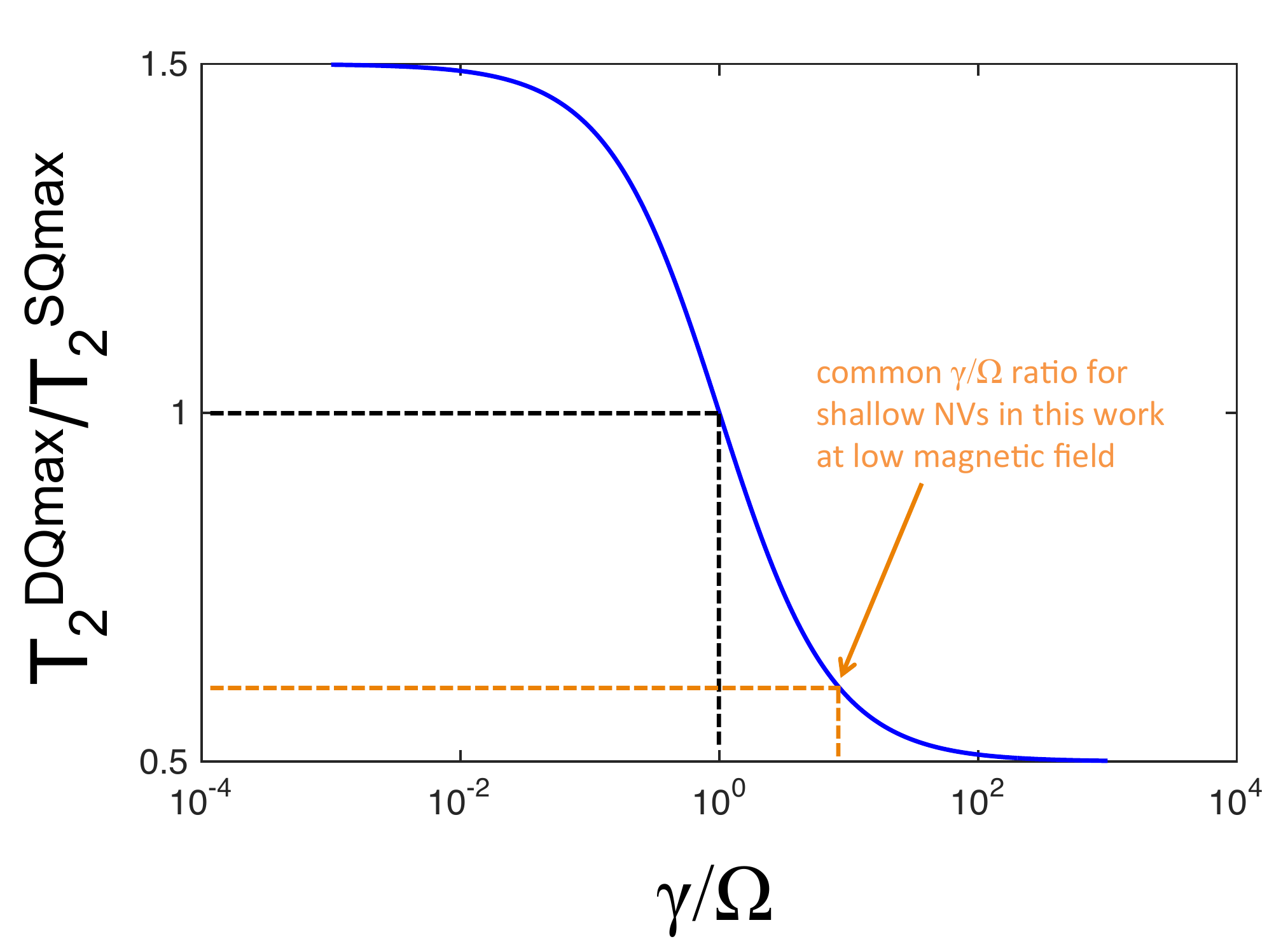}
 \caption{\label{fig:computedRatioDQSQmax} Ratio Eq. \ref{eq:T2DQSQratio} of theoretical $T_{2}$ limits (i.e., zero dephasing) for the double-quantum to single-quantum cases, $T_{2}^{\mathrm{DQmax}}/T_{2}^{\mathrm{SQmax}}$ as a function of the SQ and DQ relaxation rates $\Omega$ and $\gamma$, respectively. This ratio can vary from 0.5 to 1.5 depending on which relaxation channel is dominant. The case most relevant to the main text experiments is $\gamma/\Omega > 1$ (bottom left quadrant of image) where DQ coherences are expected to be more limited by relaxation compared to SQ coherences.}
 \end{figure}

We plot this quantity as a function of $\gamma/\Omega$ in Fig. \ref{fig:computedRatioDQSQmax}, and we find that $T_{2}^{\mathrm{DQmax}}/T_{2}^{\mathrm{SQmax}} < 1 $ when $\gamma > \Omega$ and  $T_{2}^{\mathrm{DQmax}}/T_{2}^{\mathrm{SQmax}} > 1 $ when $\gamma < \Omega$. This is an intuitively simple result that implies that the DQ relaxation channel will limit the coherence of the DQ coherences more than it limits the SQ coherences. For example, in the case of the NV we measured with $\gamma = 8(2)$ kHz and $\Omega = 0.31(2)$ kHz, the result is $T_{2}^{\mathrm{DQmax}}/T_{2}^{\mathrm{SQmax}} \approx 0.5$. When $\gamma/\Omega = 1$, the decoherence rate limits are equal in the two cases because effectively the three levels are all on the same footing, in a phenomenological way, despite having different spin projections. In a multipulse experiment that aims to measure both $T_{2}^{\mathrm{DQ}}$ and $T_{2}^{\mathrm{SQ}}$, the ratio $T_{2}^{\mathrm{DQ}}/T_{2}^{\mathrm{SQ}}$ may not match the theoretical $T_{2}^{\mathrm{DQmax}}/T_{2}^{\mathrm{SQmax}}$ because the dephasing will be different for SQ and DQ coherences. For example, the pure DQ dephasing rate is more sensitive to magnetic $B_{z}$ noise, but insensitive to electric $E_{z}$ noise \cite{Dolde2011, Kim2015}.

The definition of \Tpm$=\left(\gamma\right)^{-1}$ is based is based on the SQ coherence case because that quantity appears in Eq. \ref{eq:T2limit} for SQ coherence time limits. However, looking at Eq.  \ref{eq:T2limitDQ} one could also define time constant from the view of DQ coherence: $\left(2\gamma\right)^{-1}$ captures the bidirectional chance of relaxation between the $\ket{\pm1}$ states when in superposition. This definition does not change the overall limit $T_{2}^{\mathrm{DQmax}}$. If we define these time constants with respect to the coherences they most affect then $T_{1}^{\mathrm{(0,SQ)}} \equiv T_{1}^{(0)} = \left(3\Omega\right)^{-1}$ and $T_{1}^{\mathrm{(\pm1,DQ)}} \equiv \frac{1}{2}T_{1}^{(+1,-1)} =\left(2\gamma\right)^{-1}$. Using these definitions then Eqs. \ref{eq:T2limit} and \ref{eq:T2limitDQ} can be written as
\begin{equation}
T_{2} \leq T_{2}^{\mathrm{SQmax}}= \frac{2}{3\Omega + \gamma} =  2\left(\frac{1}{T_{1}^{\mathrm{(0,SQ)}}} + \frac{1}{2}\frac{1}{T_{1}^{\mathrm{(\pm1,DQ)}}} \right)^{-1}
\label{eq:T2limitSQ_redfine}
\end{equation}
and
\begin{equation}
T_{2}^{\mathrm{DQ}} \leq T_{2}^{\mathrm{DQmax}}= \frac{2}{2\Omega + 2\gamma} =  2\left(\frac{2}{3}\frac{1}{T_{1}^{\mathrm{(0,SQ)}}} + \frac{1}{T_{1}^{\mathrm{(\pm1,DQ)}}} \right)^{-1}.
\label{eq:T2limitDQ_redfine}
\end{equation}
Eqs. \ref{eq:T2limitSQ_redfine} and \ref{eq:T2limitDQ_redfine} perhaps more equitably show the influence of each of these time constants on SQ versus DQ coherences. 

Pulse sequences employing dual-frequency ``DQ swap'' pulses \cite{Mamin2014} could be useful tools for studying the $\gamma$ transition rate as well. Even in the present work focused on SQ coherence times, we consider the possibility that at low applied $B_{z}$ our $\pi_{0,-1}$-pulses also create small coherences within the $\ket{+1}$ and $\ket{-1}$ subspace due to a small Rabi detuning of 10s of MHz. The ratio of DQ to SQ coherence times is $\approx 0.5$ for $\gamma \gg \Omega$, and therefore any unintentional DQ coherences formed by Rabi driving at at low magnetic fields would decay at a faster rate, if limited by relaxation, than would the intentional SQ coherences. Furthermore, in the absence of nuclear-bath-related anomalous decoherence effects \cite{Huang2011} (as we use \C{12}-enriched diamond films) the DQ coherences also generally dephase $2-4\times$ faster from magnetic noise \cite{Mamin2014,Kim2015}. This DQ-to-SQ maximum coherence time ratio that we derive is a prediction that could be evaluated in future experiments in the context of $\Omega$ and $\gamma$ measurements. 

Although prior work on NV dephasing at finite $B_{z}$ shows that the DQ spin coherence time is not affected to first order by electric fields \cite{Kim2015} (since it shifts the $\ket{\pm1}$ energy levels equally), we predict from the above analysis that electric fields in fact should greatly influence the DQ spin decoherence rate through the DQ relaxation channel. Also, near $B_{z}=0$ the spin eigenstates become DQ superpositions of $\ket{m_{s}=\pm1}$, and one may find a crossover between dominance of dephasing \cite{Dolde2011,electric2016} and DQ relaxation from electric fields.

\subsection{Single-quantum relaxation data, $\Omega$}
\begin{figure}
\includegraphics[width = 1.0\columnwidth]{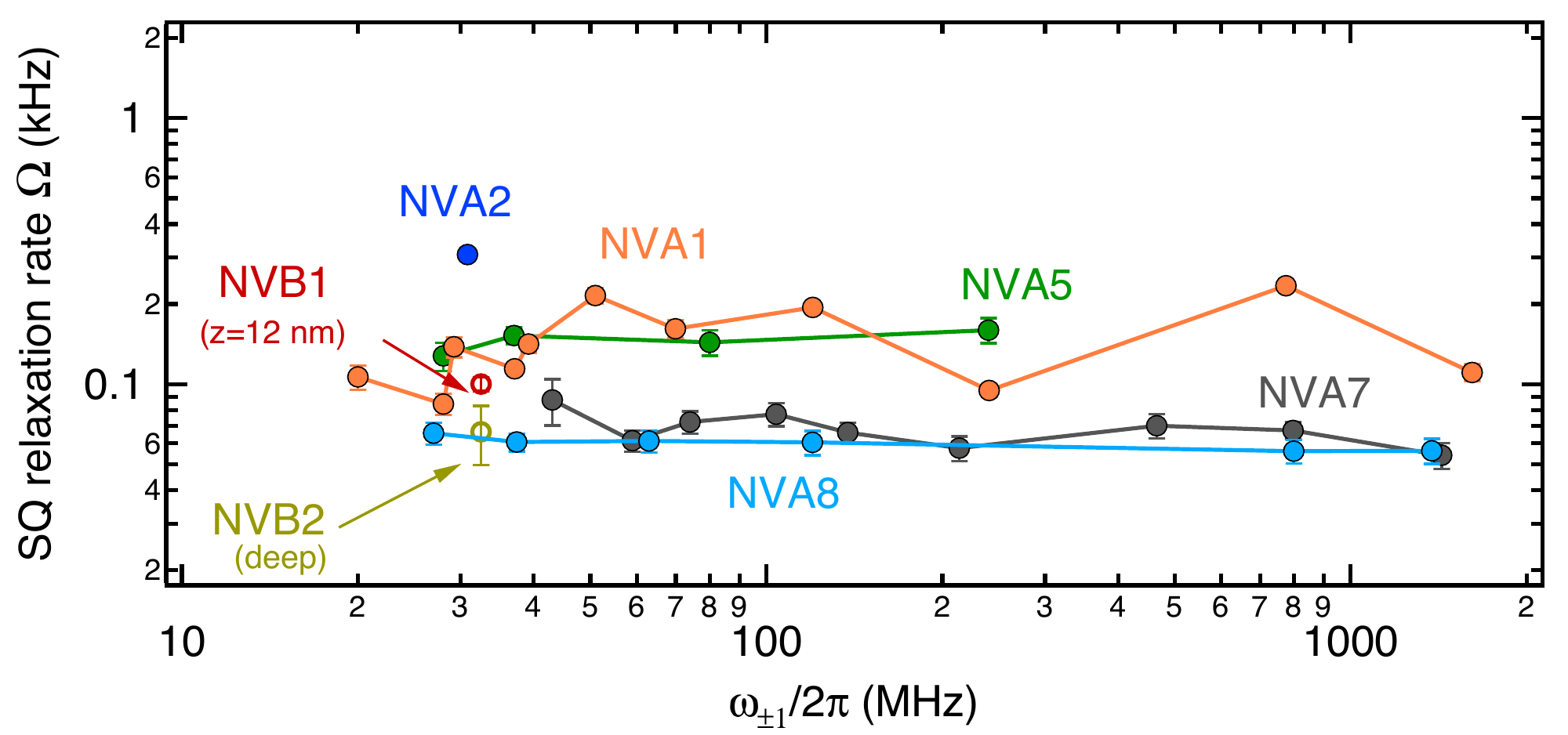}
 \caption{\label{fig:SOMOmegaVsSplitting} Single-quantum relaxation rate $\Omega$ measured as a function of the $\ket{-1}\leftrightarrow\ket{1}$ transition frequency $\omega_{\pm1}$ plotted for several NV centers in samples A and B. The data is used to compute the $\gamma/\Omega$ ratios plotted in the main text Fig. \ref{fig:Figure3}(b). $\Omega$ shows no dependence on the $\omega_{\pm1}$ splitting; over this range the $\omega_{0,-1}/2\pi$ transition frequency, which determines the magnetic noise $\Omega$ is sensitive to, varies over only a small fractional range of about 2060 MHz - 2860 MHz.}
 \end{figure}

Figure \ref{fig:SOMOmegaVsSplitting} is a plot of the single-quantum relaxation rates of the data shown in the main text Fig. \ref{fig:Figure3}. The $\Omega$ data show a generally flat behavior with the $\omega_{\pm1}$ splitting and the magnitudes even at large $\omega_{\pm1}$ are typically smaller than the $\gamma$ data. The flat behavior is likely due to the fractionally small change in $\omega_{0,-1}$, which is approximately given by $\omega_{0,-1} = 2\pi D_{\mathrm{gs}}-\omega_{\pm1}/2$. The surface magnetic noise spectrum affecting $\Omega$ is either flat or negligible over this frequency range of about 2060 MHz - 2860 MHz. We determined that the magnitude of the magnetic noise spectra we measured with dephasing spectroscopy is too small to affect $\Omega$ if it continues to fall off as $1/f^{2}$, reaching NV-magnetic coupling strengths of around 1 Hz or less. Therefore, as noted in an earlier study \cite{Rosskopf2014}, the \Tonez{} for near-surface NVs may be limited by a magnetic noise source with a cutoff at much higher frequencies on the order of gigahertz.

Of the NVs studied in most detail (A1, A7, A8) NVA1 showed the fastest SQ relaxation and larger variation in $\Omega$ (up to a factor of 2). This is interesting because our dephasing and relaxation spectroscopy showed that NVA1 had a higher level of what we identified as magnetic noise (red dashed Lorentzian curve in Fig. \ref{fig:Figure4}(a)). We have already stated that this same magnetic noise that is responsible for dephasing cannot also affect $\Omega$ given the single correlation time because $S\left(\omega\right)$ becomes too small at $\omega=\omega_{0,-1}$. However, our and others' prior studies suggest that both \Tonez{} \cite{Myers2014,Rosskopf2014} and $T_{2}$ \cite{Myers2014,Romach2015} have surface-distance dependence and perhaps there are at least two different regimes for surface-related magnetic noise correlation rates, a $\sim0.1-1$ \us{} rate and a much faster sub-nanosecond. We discuss more about future experiments to address this in a later section on spin-locking and $T_{1\rho}$.

\subsection{Relaxation for shallow versus deep NVs}
\begin{figure}
\includegraphics[width = 1.0\columnwidth]{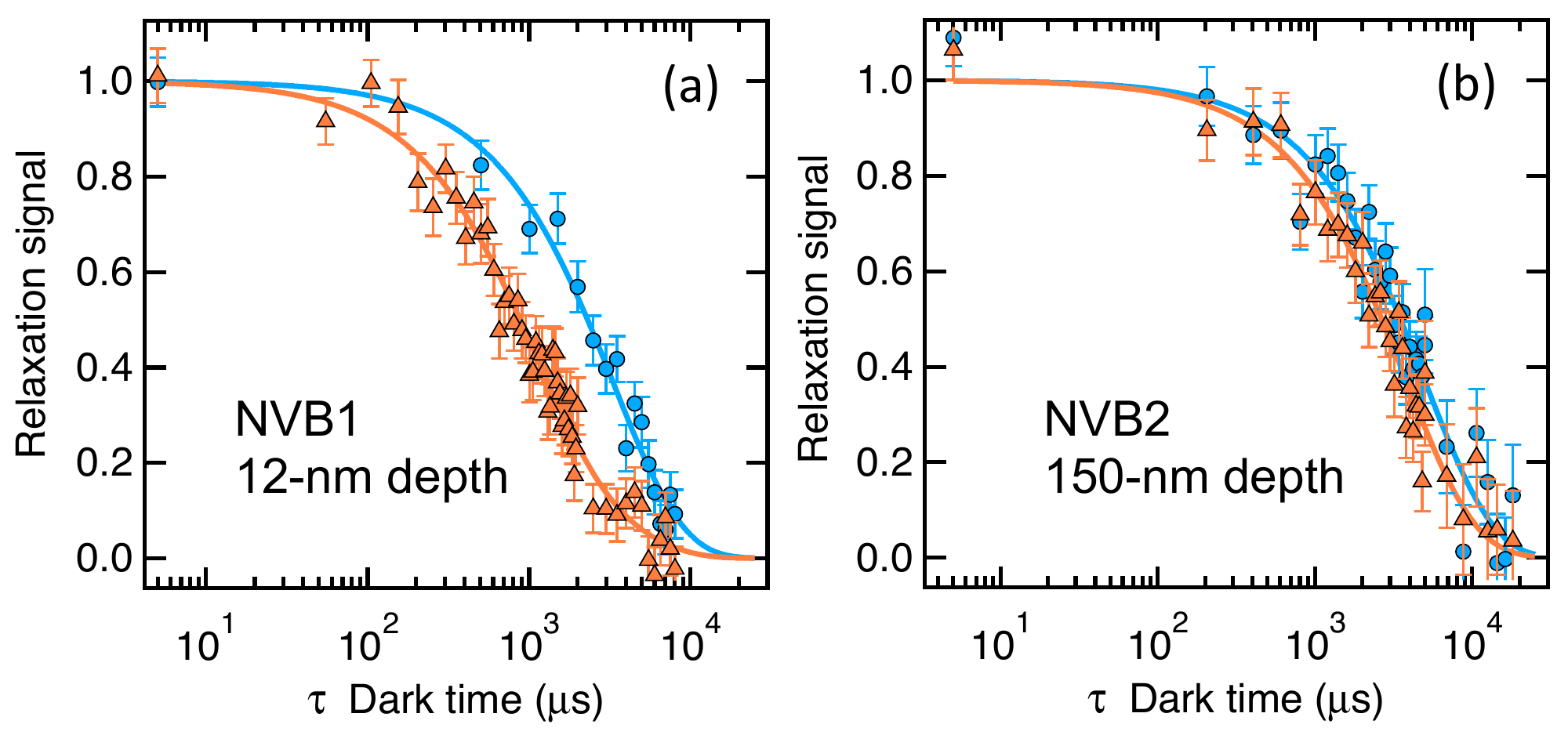}
 \caption{\label{fig:k7k26Comparison} Comparison of the relaxation rate measurements at $\omega_{\pm1}$$/2\pi=32.5$ MHz for a shallow NV (a) and deeper NV (b) both in a \nit{15} delta-doped film Sample B. In the 12-nm-deep NV the double-quantum relaxation sequence data (orange triangles) decays faster than does the single-quantum data (blue circles), giving $\gamma = 0.45(4)$ kHz compared to $\Omega = 0.10(1)$ kHz. In the deeper NVB2 the extracted rate is $\gamma = 0.11(2)$ kHz, which is more comparable to $\Omega = 0.066(8)$ kHz. The larger $\gamma$ rate for NVs at few-nanometer depths is evidence for an effect of surface-related electric field noise at frequency $\omega_{\pm1}$.}
\end{figure}

Figure \ref{fig:k7k26Comparison} shows $F_{1}$ (blue circles) and $F_{3}$ (orange triangles) relaxation signals for two nitrogen delta-doped NVs in sample B, NVB1 at depth 12 nm and NVB2 at 150 nm; depths were independently measured using magnetic resonance depth imaging and checked with surface-proton spectroscopy, as we previously reported \cite{Myers2014}. In the case of deep NVB2 we found $\gamma/\Omega = 1.7$, while for NVB1 we measured $\gamma/\Omega = 4.5$. In sample A the main-text conclusion of surface noise origin of the $\gamma$ relaxation is also supported by the observation that NVA5 showed a relatively small $\gamma$ at $\omega_{\pm1}$$=37$ MHz in conjunction with its identification as a deep NV; it's Hahn echo showed coupling to a \C{13} spin bath outside the \C{12} film and a relatively long Hahn echo $T_{2}=147(6)$ \us{} \cite{Ohno2012,Myers2014}. Other NVs in sample A, identified as shallow, did not exhibit \C{13} oscillations in the echo data and had shorter bare Hahn echo $T_{2} < 100$ \us{}. 

In the discussion of Fig. \ref{fig:Figure3} of the main text we introduce the form of $\gamma$ dependence on frequency $f=\omega_{\pm1}/2\pi$ as $\gamma\left(f\right) = 1/f^{\alpha} + \gamma_{\infty}$. Our claim that $\gamma_{\infty}$ is primarily due to bulk effects is based on examining the mean $\gamma_{\infty}$ value for NVs identified as shallow (A1, A7, A8) and those identified as deep (A5, B2). For the shallow NVs $\langle \gamma_{\infty} \rangle = 0.21(4)$ and for the deep NVs $\langle \gamma_{\infty} \rangle = 0.14(5)$. These values are comparable enough that bulk effects likely contribute to most of the frequency-independent $\gamma_{\infty}$ rate for shallow NVs. The question of whether $\gamma_{\infty}$ is caused by spin-lattice relaxation, electric field noise internal to the diamond, or both is an open question, and a study of $\gamma$ dependence on both temperature and $f$ may be helpful.

\subsection{Additional data comparing $T_{2}$ and $T_{1}$}
\begin{figure}
\includegraphics[width = 1.0\columnwidth]{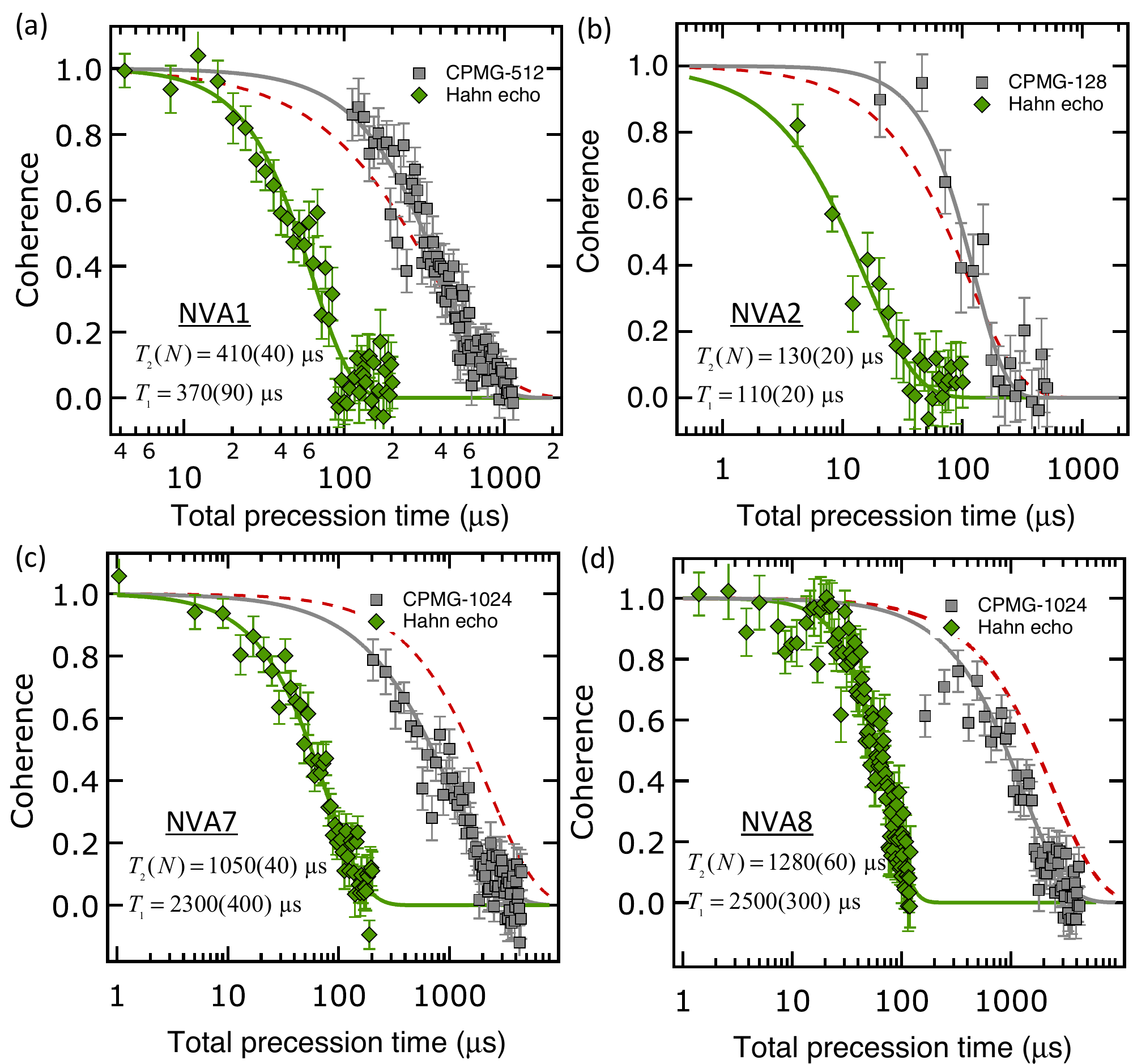}
 \caption{\label{fig:SOMFourNVT1T2} Enhancement of single-quantum coherence time using CPMG-$N$ for shallow implanted NVs under conditions of (a,b) large $\gamma$ at small $\omega_{\pm1}$ and (c,d) small $\gamma$ at large $\omega_{\pm1}$. Data shown are Hahn echo (green diamonds) and CPMG-$N$ (gray squares) where $N$ is the total number of $\pi$ pulses, and solid lines are fits to $\exp\left[-(T/T_{2})^{n}\right]$ with free exponent $n$. Dashed red lines are reference plots of $\exp\left(-T/T_{1}\right)$ using the $T_{1} = \left(3\Omega + \gamma\right)^{-1}$. (a) NVA1 with splitting $\omega_{\pm1}$$/2\pi=37.1$ MHz, where the CPMG-512 yields a $T_{2}(512) = 1.2T_{1}$.  (b) NVA2 with $\omega_{\pm1}$$/2\pi=30.6$ MHz, where CPMG-128 yields $T_{2}(128) =1.2T_{1}$. (c) NVA7 with $\omega_{\pm1}$$/2\pi=1431$ MHz where the CPMG-1024 yields $T_{2}(1024) = 0.45T_{1}$. (d) NVA8 with $\omega_{\pm1}$$/2\pi=1376$ MHz where the CPMG-1024 yields $T_{2}(1024) = 0.52  T_{1}$.}
 \end{figure}

In Fig. \ref{fig:SOMFourNVT1T2} we show supplemental data to the main text Fig. \ref{fig:Figure2} on comparing CPMG-based $T_{2}$ enhancement to the $T_{1}$ computed from $\gamma$ and $\Omega$ for four NV centers, two taken at small $\omega_{\pm1}$ (a,b) and two taken at large $\omega_{\pm1}$ (c,d).

\begin{figure}
\includegraphics[width = 1.0\columnwidth]{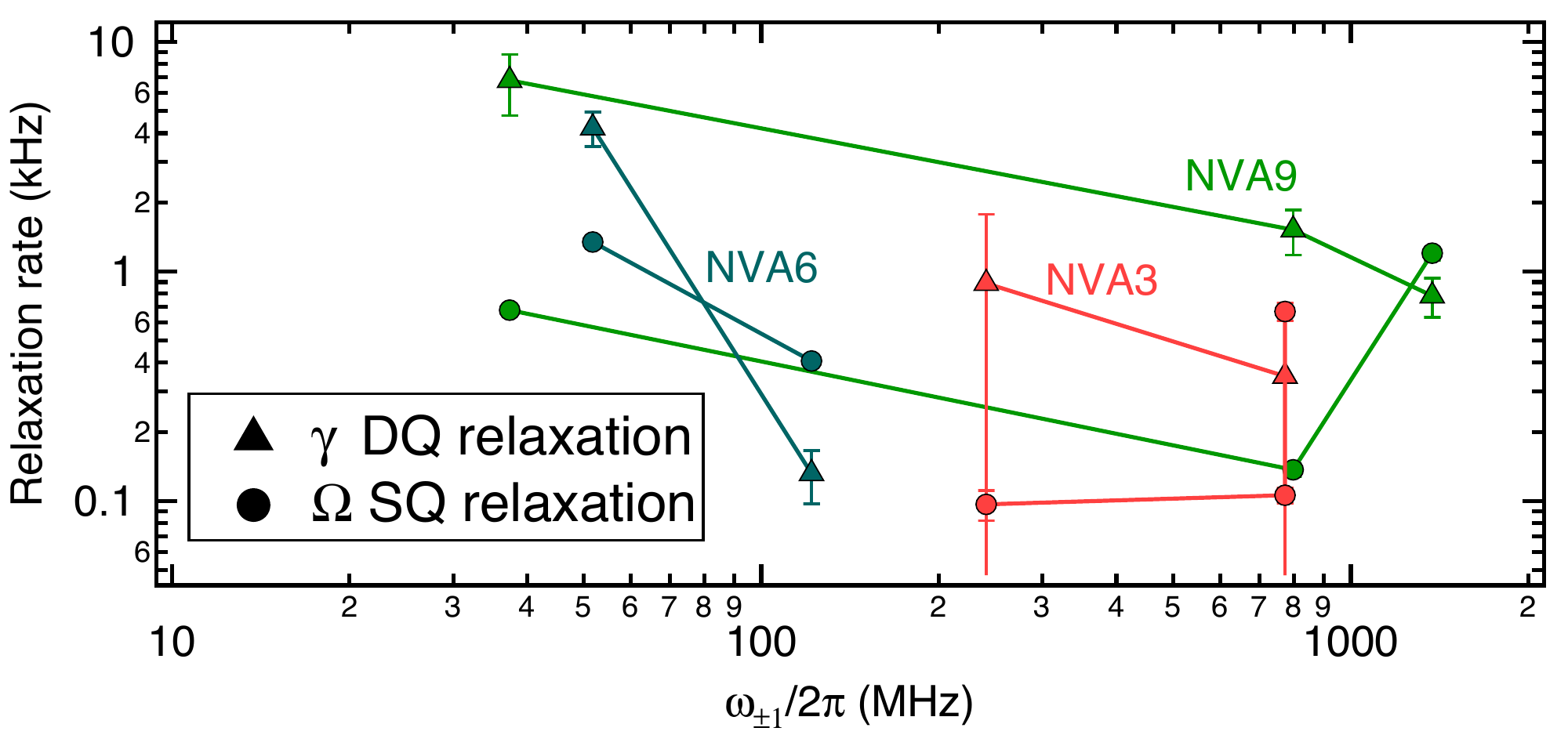}
 \caption{\label{fig:SOMUnstableNVdata} SQ and DQ relaxation rates $\Omega$ and $\gamma$ for NVs that showed unstable $\Omega$ over hourly time periods of multiple measurements. For example, NVA3 (red circles) exhibited $\Omega$ that changed from one measurement to the next even at the same $\omega_{\pm1}$ value. In general, however, $\gamma$ still appeared to increase at lower $\omega_{\pm1}$ values, most noticeably in NVA9. NVA9 is likely quite near to the surface because it shows a relatively short Hahn echo coherence time $T_{2}=9.5(9)$ \us{}.}
 \end{figure}

\begin{figure}
\includegraphics[width = 0.5\columnwidth]{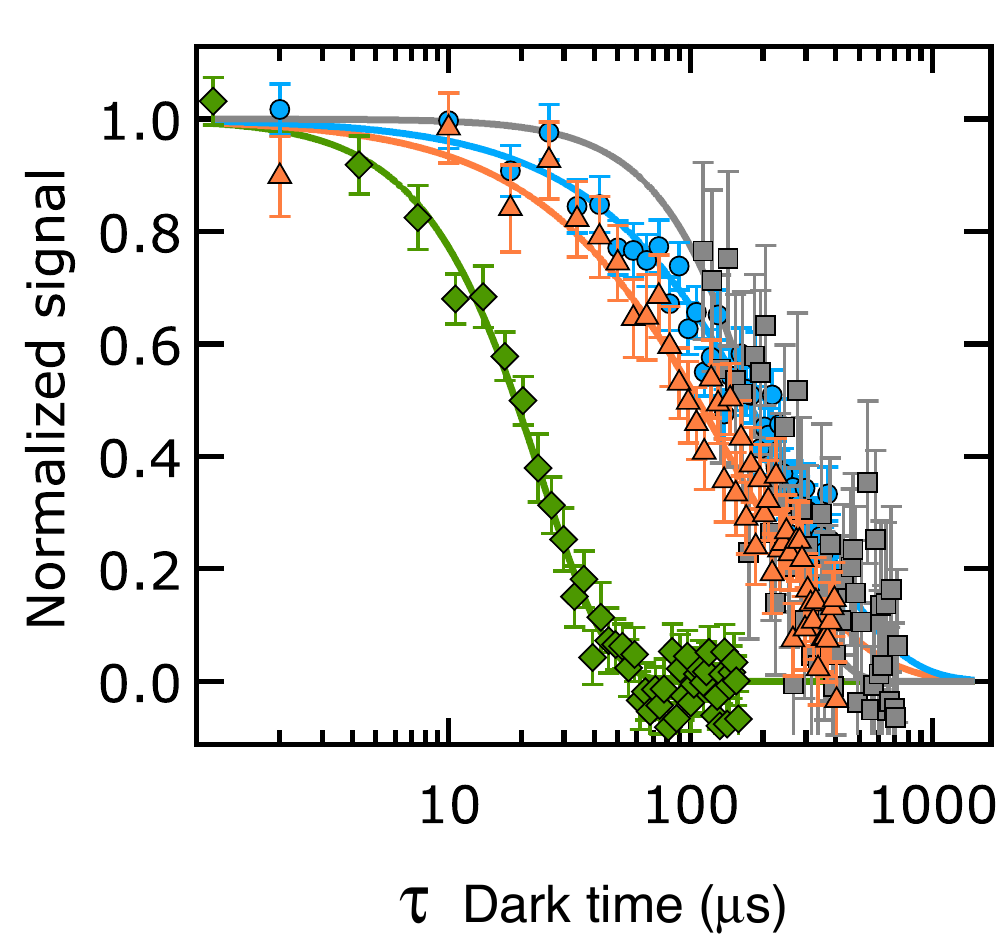}
 \caption{\label{fig:SOMNVA6CPMG} An example of an NV that showed an unstable and increasing $\Omega$ SQ relaxation rate over time. Relaxation data at $\omega_{\pm1}$$/2\pi=57.1$ MHz is shown for signals $F_{1}$ (blue circles) and $F_{3}$ (orange triangles) in comparison to $T_{2}=24(1)$ \us{} with Hahn echo (green diamonds) and $T_{2}\left(N\right)=220(70)$ \us{} with CPMG-$N$ (gray squares) for $N=512$ pulses. Total $T_{1}=120(10)$ \us{} is so short that $T_{2,N}$ could be extended to about 90$\%$ of $T_{2}^{\mathrm{SQmax}} = 240(20)$ \us{}, which is made evident in this plot, although note this comparison is only for qualitative reference as coherence and relaxation curves should not be directly compared.}
 \end{figure}

In sample A we have focused measurements on NVs that showed a consistent $\Omega$ over time. Some NV centers in the nanopillars exhibited $\Omega$ values that increased or decreased by up to an order of magnitude and sometimes between two values measurement to measurement. For these shown in Fig. \ref{fig:SOMUnstableNVdata} we judged that it was not reliable to continue to measure pairs of $\gamma$ and $\Omega$ versus $\omega_{\pm1}$ for a larger range of values. NVA6 for example had a particularly unstable $\Omega$ value (not all data points shown) that tended to decrease over time. In fact, the total $T_{1}$ for NVA6 was so short that we were able to decouple its echo $T_{2}=24(1)$ \us{} up to $T_{2}\left(N=512\right) = 216(70)$ at $\omega_{\pm1}$$=57.1$ MHz. This $T_{2}\left(N\right)$ approaches $90\%$ of $T_{2}^{\mathrm{SQmax}} = 242$ \us{} since the relaxation of both SQ and DQ channels was fast, where $T_{1}^{(0)} = 250$ \us{}. Relaxation and CPMG data for this NVA6 are shown in Fig. \ref{fig:SOMNVA6CPMG}.

\subsection{Spectral deconvolution of CPMG data: Method and limitations}
We outline here a numerical method for deconvolution of the multipulse coherence data from the CPMG-$N$ filter functions. A single stretched exponential with coherence time $T_{2}\left(N\right)$ to fit decoherence data is only an approximation in the presence of finite-frequency electromagnetic fluctuations along the $z$ axis; that is the pure-dephasing contribution to coherence $C_{\rm{SQ}}\left(t\right)$ is generally non-exponential and depends on the frequency spectrum of these fluctuations \cite{deSousa2009,Bar-Gill2012,Romach2015}. The coherence defined on the interval $[0,1]$ for total free precession time $T$ is given by \cite{dasSarma2008,deSousa2009}

\begin{equation}{\label{eq:SpectralCoherence}}
C(T) = \exp{\left[-\chi(T)\right]}.
\end{equation}
where the functional depends on the noise spectrum \Sw{}
\begin{equation}{\label{eq:SpectralCoherence}}
\chi(T) = \frac{1}{\pi}\int_{0}^{\infty} d\omega \Swe{} \frac{F_{N}\left(\omega T\right)}{\omega^2}.
\end{equation}

The total time $T$ that we plot on the $x$ axis for all coherence data is $T= 2\tau_{\mathrm{max}}N + N\tau_{\pi}$ because $2\tau$ is the delay between $\pi$-pulses of duration $\tau_{\pi}$. The $N$-pulse filter function is 
\begin{equation}{\label{eq:SpectralFilterFunctionN}}
F_{N}\left(\omega T \right) = 8 \sin^4\left(\frac{\omega T}{4 N}\right) \frac{ \sin^2 \left(\frac{\omega T}{2}\right)}{\cos^2 \left(\frac{\omega T}{2 N}\right)},
\end{equation}
which has a peak maximum at $\omega = \pi N /T$.

The problem is to extract \Sw{} given the known filter function and coherence data. This is a challenging task in general because there are a finite number of $C(T)$ points. Moreover the contributions to this noise spectrum at a given $\omega_{i}$ depend on unknown values of \Sw{} at other frequencies due to convolution with $F_{N}$, which has a finite peak linewidth and harmonics.

An approximate method is to begin at the high frequency values of \Sw{} by looking at $C(T_{i})$ for the shortest $T_{i}$ points. Particularly for large $N$ the filter function is sharply peaked and can be approximated as a delta function in the form $F_{N}(\omega T)/\omega^2 = T_{i} \delta(\omega-\omega_{i})$. Therefore, $\chi(T) \approx T_{i} \tilde{S}\left(\omega_{i}\right)/\pi$ at this point $\omega_{i}=\pi N /T_{i}$ and the noise spectral density by this method is
\begin{equation}{\label{eq:SpectralDeltaSpectrum}}
\tilde{S}_{\delta}\left(\omega_{i}\right) = \pi\frac{\chi\left(T_{i}\right)}{T_{i}}
\end{equation}
This approximation $\tilde{S}_{\delta}\left(\omega_{i}\right)$ can first be computed for all $T_{i}$ points in the data set as a first pass. 


Assuming that the noise spectrum falls off as some power law of frequency, then the higher frequency $(\omega > \omega_{i})$ unknown parts of the spectrum won't contribute as largely to the full $\tilde{S}\left(\omega_{i}\right)$ as the $\tilde{S}_{\delta}\left(\omega_{i}\right)$ term will. We also assume that all harmonics in $F_{N}(\omega T)$ are at higher frequencies than the peak frequency $\omega_{i}$. More completely, the coherence signal could be described as

\begin{equation}{\label{eq:SpectralFullChi1}}
\chi\left(T_{i}\right) \approx\frac{1}{\pi}  \left[\int_{0}^{\omega_i-\epsilon} +\int_{\omega_i-\epsilon}^{\omega_i+\epsilon} + \int_{\omega_{i}+\epsilon}^{\infty} \right]d\omega\Swe{}\frac{F_{N}\left(\omega T_{i}\right)}{\omega^2},
\end{equation}
for $\epsilon\rightarrow0$. The first term we take as negligible since, especially for larger $N$, $F_{N}\left(\omega T_{i}\right)$ has a small value and no harmonics for $\omega < \omega_{i}$. The second term can be approximated as the delta function result in Eq. \ref{eq:SpectralDeltaSpectrum}. The final term is treated more carefully to subtract out the effect of higher frequency contributions to each $\tilde{S}\left(\omega_{i}\right)$, giving

\begin{equation}{\label{eq:SpectralFullChi2}}
\chi\left(T_{i}\right) \approx \frac{1}{\pi}\tilde{S}\left(\omega_{i}\right) T_{i} + \frac{1}{\pi}\int_{\omega_{i}}^{\infty}\Swe{}\frac{F_{N}\left(\omega T_{i}\right)}{\omega^2}.
\end{equation}
This can be solved to find $\tilde{S}\left(\omega_{i}\right)$ as
\begin{equation}{\label{eq:SpectralFullSolved}}
 \tilde{S}\left(\omega_{i}\right) \approx \frac{\pi}{T_{i}}\chi\left(T_{i}\right)-\frac{1}{T_{i}}\int_{\omega_{i}}^{\infty}\Swe{}\frac{F_{N}\left(\omega T_{i}\right)}{\omega^2}
\end{equation}
where again $\omega_{i} = \pi N/ T_{i}$. The calculation of the noise spectrum assumes that \Sw{} falls off as an arbitrary power law in $\omega$ and the steps are 

1) Scale data to coherence $\in [0,1]$ using PL references of $\ket{0}$ and $\ket{-1}$ from a differential measurement. Exclude any data points $C\left(T\right)$ that are saturated near either 1 or 0 coherence as these do not contain useful spectral information.

2) For all $i=1...n$ compute $\chi(T_{i})=-\rm{ln}\left[C(T_{i})\right]$. 

3) For all $i=1...n$ compute the frequencies $\omega_{i}$ and the first term of Eq. \ref{eq:SpectralFullSolved}.

4) Starting from $\omega_{i=1}$ as the highest frequency data point recursively compute the second term in Eq. \ref{eq:SpectralFullSolved} by a numerical integration. This was done by first interpolating the \Sw{} result of previous steps over an upsampling rate of $R \propto N$ to accommodate for the narrowing of the harmonics with larger $N$. The reason for this interpolation is that if a higher harmonic of $F_{N}\left(\omega T_{i}\right)$ coincides with an $\omega_{j}$ (where all  $\omega_{j}> \omega_{i}$) then its width will be much smaller than the value of $\Delta \omega = \omega_{j}-\omega_{j+1}$ and result in a greatly overestimated value of the subtraction term. In other words, the interpolation point spacing should be fine enough to ``resolve'' the sharp features of the filter function. For each $\omega_{i}$ of the $i=1....Rn$ this subtraction term uses the full analytical form of $F_{N}$ in Eq. \ref{eq:SpectralFilterFunctionN} in the numerical integration. The routine was performed on each CPMG-$N$ data set in isolation, so there are several data points at particular frequencies. The highest-frequency points of each set were discarded in the final plot as they were only used for interpolation and subsequent subtraction since the spectrum to the right of them was fully unknown.

We also emphasize that the method of spectroscopic deconvolution of the coherence signal using a multipulse filter function \cite{Bar-Gill2012,Romach2015,Kim2015} becomes incorrect when $T_{2} \sim T_{1}$ because no longer is the assumption valid that low-frequency pure-dephasing noise determines the shape of the CPMG data curve $C\left(T\right)$. The valid use of this dynamical decoupling deconvolution requires that the decay in $C\left(T\right)$ data is due solely to dephasing effects because the derivation of Eq. \ref{eq:SpectralCoherence} relies upon the assumption that the transverse noise terms, those that cause relaxation, can be set to zero \cite{deSousa2009}. It is possible that neglecting to consider the full $T_{1}$, which is always shorter than \Tonez{}, could lead one to assume dephasing deconvolution is valid when it would not be. Therefore, to judge this method's range of applicability it is important to measure the full $T_{1}$ from both $\gamma$ and $\Omega$. This is the reason we employ the technique only at high magnetic fields ($B_{z}\gtrsim 100$ G) where \Tpm{} is sufficiently long compared to $T_{2}$ for at least some useful range of $N$. For example, at large $N$ the $\tau$-points used for low-frequency $S(f)$ points are particularly susceptible to influence from relaxation, and the data's high-frequency components will look like a ``wall'' of white noise that is not an accurate reflection of dephasing effects. For this reason, in our deconvolution results we place more confidence in the low- to mid- frequency parts of the CPMG spectrum where we found electric fields to contribute most significantly, as the higher-frequency Lorentzian may tend to be overestimated due to the onset of relaxation effects. A more-complete non-analytical model of surface-induced decoherence that incorporates relaxation should be developed to provide further insight.

The corollary to this relaxation effect, however, is that if one is studying how $T_{2}$ increases with $N$ beginning with small $N$ and the $T_{2}\left(N\right)$ appears to saturate this does not necessarily mean that relaxation timescales are yet significant. Rather this apparent saturation could be a frequency window of actual white-like dephasing noise that is followed by a decay in the noise spectrum where $T_{2}\left(N\right)$ rises once more pulses are used. The simplest example is the flat low-frequency part of a Lorentzian spectrum that would be accessed by small $N$. This slow start in the ``decoupling efficiency'' at small $N$ can for example be seen in the experiment and modeling of the $\lambda=\log_{N}\left[T_{2}\left(N\right)/T_{2}\left(N=1\right)\right]$ parameter plotted in Fig. 4 of our prior work on shallow-NV dynamical decoupling \cite{Myers2014}. 

A possible future experiment may be to employ ``DQ swap'' multipulse techniques discussed in \cite{Mamin2014} at large magnetic fields (100s of Gauss) to demonstrate a dephasing spectroscopy using the DQ qubit coherence. DQ swap sequences are more challenging to execute without accumulation of pulse errors than SQ dephasing CPMG sequences because two calibrated microwave tones are required. However, such DQ coherence data is insensitive to $E_{\parallel}$, so a deconvolution should yield a magnetic noise spectrum.

\subsection{Spectroscopy with double-quantum relaxometry}
We derive the relationship between the DQ relaxation rate $\gamma$ and the electric field noise spectrum transverse to the NV axis. The result is analogous to the case of a $S_{x}$ spin-$1/2$ operator for a qubit \cite{spectrometer2003}, however the noise operator connecting the $\ket{1}$ and $\ket{-1}$ is the $S_{\pm}^{2}$ spin-$1$ operator, so we overview the calculation to find the correct multiplicative factors relating $\gamma$ to the power spectral density of electric field noise.

The relevant Hamiltonian for the $\ket{1}$ and $\ket{-1}$ sub-basis in energy units is only the Zeeman term since $DS_{z}^2$ gives equal energy for the $\ket{\pm1}$ states:
\begin{equation}
H_{0} = g\mu_{B} B_{z} S_{z}
\label{eq:H0spectroscopy}
\end{equation}
where $S_{z}$ is still the standard $S=1$ spin operator, and this Hamiltonian yields the expected energy difference $\hbar\omega_{\pm1} = 2g\mu_{B}B_{z}$. From the full ground state spin Hamiltonian stated in the main text the time-dependent perturbation for the transverse electric field noise is
\begin{equation}
V(t) = \frac{-d_{\perp}E_{\perp}\left(t\right)}{2}\left[e^{-i\phi(t)}S_{+}^{2} + e^{i\phi(t)}S_{-}^{2}\right]
\label{eq:Vspectroscopy}
\end{equation}
where $\phi\left(t\right) = \tan^{-1}E_{y}\left(t\right)/E_{x}\left(t\right)$, $S_{\pm}=S_{x} \pm iS_{y}$ are the $S=1$ raising and lowering operators, and $d_{\perp}$ is the transverse electric dipole coupling of the NV. The time-dependent parts will be specified later and here are general.

As in a DQ relaxation measurement we assume the initial state is $\ket{\psi\left(0\right)} = \ket{-1}$ after a green-laser pulse and $\pi_{-1}$ pulse. We can assume that $\gamma \gg \Omega$ such that all the population in $\ket{0}$ remains zero, though this is only valid at small $\omega_{\pm1}$. Even so, the $S_{+}^{2}$ term in the Hamiltonian is the only one that gives nonzero first-order coupling of the $\ket{1}$ and $\ket{-1}$ states, and second-order magnetic terms are suppressed in their contribution to $\gamma$ (see later supplement section on DQ magnetic driving). If $\Omega \sim \gamma$ then first-order magnetic terms ($S_{x},S_{y}$) do contribute to the $\ket{1}$ state amplitude, but only directly through the $\Omega$ rate. Therefore, since our goal is to compute $\gamma$ and we focus on the $S_{+}^{2}$ term only, then the final derivative in the following steps (on Eq. \ref{eq:averagePopulationTDPT}) can be identified directly with $\gamma$, even at large $\omega_{\pm1}$. Then first-order time-dependent perturbation theory in the interaction picture yields an amplitude for the $\ket{1}$ state

\begin{equation}
\alpha_{1} = 0 - \frac{i}{\hbar}\int_{0}^{t}d\tau \bra{1}\left(\frac{-d_{\perp}}{2}E_{\perp}\left(\tau\right)e^{-i\phi(\tau)}S_{+}^{2}\left(\tau\right) \right)\ket{-1}
\label{eq:firstOrderTDPT1}
\end{equation}
where the $\bra{1}S_{-}^{2}\left(\tau\right)\ket{-1}$ term has vanished, as justified below. The time-dependent quadratic raising operator in the interaction picture is given by
\begin{equation}
S_{+}^{2}\left(\tau\right) = e^{i g\mu_{B}B_{z}S_{z}\tau/\hbar}\left(S_{+}^{2}\right)e^{-i g\mu_{B}B_{z}S_{z}\tau/\hbar} = S_{+}^{2}e^{i\omega_{\pm1}\tau}
\label{eq:interactionSPlus}
\end{equation}
which simplifies in the same way as in a $\sigma_{x}$ perturbation operator in a two-level system case. With $\bra{1}S_{+}^{2}\ket{-1} = 2$ the we find the amplitude is
\begin{equation}
\alpha_{1} =  \frac{i d_{\perp}}{\hbar}\int_{0}^{t}d\tau E_{\perp}\left(\tau\right)e^{-i\phi(\tau)} e^{i\omega_{\pm1}\tau}
\label{eq:firstOrderTDPT2}
\end{equation}
The population of the $\ket{1}$ state in time is
\begin{multline}
\rho_{1}\left(t\right)=\left|\alpha_{1}\right|^{2} =\left(\frac{d_{\perp}}{\hbar}\right)^{2}\int_{0}^{t}\int_{0}^{t}\\
d\tau_{1}d\tau_{2} E_{\perp}\left(\tau_{1}\right)E_{\perp}\left(\tau_{2}\right)e^{-i\left(\phi(\tau_{2})-\phi(\tau_{1})\right)}  e^{-i\omega_{\pm1}\left(\tau_{1}-\tau_{2}\right)}
\label{eq:firstOrderTDPT3}
\end{multline}
As in the case for a spin-$1/2$ coupling operator \cite{spectrometer2003} a change of variables can be done with $\tau=\tau_{1}-\tau_{2}$ and $T=\left(\tau_{1}+\tau_{2}\right)/2$. The integration limits on the new $\int d\tau$ integral can be taken to $\pm \infty$ in the case that we only look at timescales such that $t \gg \tau_{c}$ \cite{spectrometer2003}, where $\tau_{c}$ is the correlation time of the noise. This is a valid assumption because the noise sources we find from analyzing CPMG and DQ relaxation data have correlation times of 100 ns to a few microseconds. To measure $\gamma$ we use dark times (after initialization into $\ket{-1}$) of several to tens of microseconds to milliseconds, so $t \gg \tau_{c}$. We suppose a stationary noise process that gives translation invariance in time, and we are interested in the average population for many iterations of the $E_{\perp}\left(t\right)e^{i\phi(t)}$ noise trajectory since we execute 10s of thousands of pulse sequence shots to measure the populations. Under these assumptions the average population of the $\ket{1}$ state becomes

\begin{equation}
\bar{\rho}_{1}\left(t\right)= t\left(\frac{d_{\perp}}{h} \right)^{2} S_{E_{\perp}}\left(-\omega_{\pm1}\right)
\label{eq:averagePopulationTDPT}
\end{equation}
where the noise spectral density for the perpendicular electric field has been identified as
\begin{equation}
S_{E_{\perp}}\left(\omega\right) = \int^{\infty}_{-\infty} d\tau e^{i\omega \tau} \left\langle E_{\perp}\left(\tau\right)e^{i\phi(\tau)}E_{\perp}\left(0\right)e^{-i\phi(0)}\right\rangle
\label{eq:spectralDensityTDPT}
\end{equation}
The noise-induced transition rate from the initial $\ket{-1}$ to final $\ket{1}$ is the time derivative of Eq. \ref{eq:averagePopulationTDPT} which gives the final result in units of Hz
\begin{equation}
\gamma = \left( \frac{d_{\perp}}{h}\right)^{2} S_{E_{\perp}}\left(-\omega_{\pm1}\right) \equiv S_{\gamma}\left(-\omega_{\pm1}\right) 
\label{eq:gammaTDPT}
\end{equation}
where the latter definition shows equivalence between the $\gamma$ data shown in the main text and a coupling noise power spectral density $S_{\gamma}\left(\omega\right)$ of units Hz$^{2}/$Hz or simply Hz. Therefore the DQ relaxation rate depends only on the transverse coupling and the noise power spectral density at the frequency splitting of the $\ket{\pm1}$ levels.
Because we consider the electric field trajectory as a classical random variable, the spectrum is the same on the positive and negative frequency side $S_{E_{\perp}}\left(-\omega_{\pm1}\right) = S_{E_{\perp}}\left(\omega_{\pm1}\right)$ \cite{spectrometer2003}.

The more-complete consideration of three different $\omega_{\pm1}$ resonances due to the NV-\nit{14} hyperfine interaction can be treated in the same way, except the initial population is $1/3$ in each of the three $m_{I}=0,\pm1$ nuclear spin states (assuming zero polarization of the nuclear spin at low magnetic field). The NV electronic transitions between $m_{s} = \pm 1$ are spread out over three lines due to the hyperfine interaction with the host \nit{14} nuclear spin. For relaxation spectroscopy using these DQ transitions then the filter function is effectively $\approx 2A_{\parallel|} \approx 4.4$ MHz wide \cite{Doherty2013}. Because the smallest frequency we probe is noise at $\omega_{\pm1}$$/2\pi \sim 20$ MHz, the convolution effect becomes more important at these low frequencies, particularly as $\gamma$ changes fastest in that region. However, the filter function is still 1-4 orders of magnitude narrower than the frequencies 20-1612 MHz probed. With the starting factor of $1/3$ assuming no nuclear polarization, three delta-function-like resonances would give a relaxation rate
\begin{multline}
\gamma_{N14} = \frac{1}{3}\left( \frac{d_{\perp}}{2}\right)^{2} [S_{E_{\perp}}\left(-\omega_{\pm1}\right)\\
+S_{E_{\perp}}\left(-\omega_{\pm1}+A_{\parallel}\right)+S_{E_{\perp}}\left(-\omega_{\pm1}-A_{\parallel}\right)]
\label{eq:gammaTDPThyperfine}
\end{multline}
In principle such a deconvolution could be more effective if one probes $\gamma\left(\omega\right)$ at steps of $\Delta\omega =  A_{\parallel}$ so that the contributions of neighboring points can be measured and known for subtraction (without interpolation) to find the spectrum $S$ at a single frequency point. We do not attempt this deconvolution because the main conclusions of our paper do not rely upon this detail: 1) that $S_{\gamma}\left(\omega\right)$ increases at low frequencies and 2) that $S_{\gamma}$ can be combined with dephasing data to learn about the spectral contributions of electric field noise. If the deconvolution were done we would expect that the lower-$\omega$ contribution of the three terms in Eq. \ref{eq:gammaTDPThyperfine} would contribute slightly more than the others since the noise is of $1/f^{\alpha}$ type, so this would have a net effect of reducing the actual $S_{\gamma}\left(\omega\right)$ at low $\omega$. If anything this makes the noise decay appear more gradual, further supporting the identification of the electric noise as the lower-frequency-cutoff noise affecting dephasing.

The quantity $c\left(\tau\right)\equiv\left\langle E_{\perp}\left(\tau\right)e^{i\phi(\tau)}E_{\perp}\left(0\right)e^{-i\phi(0)}\right\rangle$ in Eq. \ref{eq:spectralDensityTDPT} is the classical correlation function of the transverse electric field fluctuations. The phase gives the radial direction of the transverse field on the plane normal to the NV $z$ axis, $\tan\phi = E_{y}/E_{x}$. Therefore, the function can also be written
\begin{equation}
S\left(\tau\right) = \left\langle\left[E_{x}\left(\tau\right) + iE_{y}\left(\tau\right)\right]\left[E_{x}\left(0\right) -i E_{y}\left(0\right)\right]\right\rangle
\label{eq:correlationExEy}
\end{equation}
which has two autocorrelation terms and two $x,y$ cross-correlation terms. There are certainly non-zero correlations between $E_{x}$ and $E_{y}$ from, for example, a surface distribution of electric dipoles fluctuating simply because the two transverse components $E_{x}$ and $E_{y}$ are ultimately generated from the same set of dipoles that have a specific configuration of orientations at time $\tau$. However, because we have assumed time-translation invariance the terms are equivalent, that is $\langle E_{x}\left(\tau\right)E_{y}\left(0\right)\rangle = \langle E_{x}\left(0\right)E_{y}\left(\tau\right)\rangle$. Therefore the cross-terms cancel out because they have opposite signs in the full correlation function Eq. \ref{eq:correlationExEy}. 

The result is that the $x$ and $y$ transverse components of the field relative to the NV axis are treated as if they are uncorrelated from one another. For a simple stationary Gauss-Markov process, where the mean and variance do not depend on past history, the correlation function looks like
\begin{equation}
S\left(\tau\right) = \langle E_{\perp}^{2} \rangle e^{-\left|\tau\right|/\tau_{c}}
\label{eq:correlationEperp}
\end{equation}
where $\langle E_{\perp}^{2} \rangle = \langle E_{x}^{2} \rangle+\langle E_{y}^{2} \rangle$, again with $x$ and $y$ here defined perpendicular to the NV axis as throughout this section. The inverse Fourier transform of this function gives a Lorentzian power spectral density \cite{deSousa2009}
\begin{equation}
S_{E_{\perp}}\left(\omega\right) = \int_{-\infty}^{\infty}\langle E_{\perp}^{2} \rangle e^{-\left|\tau\right|/\tau_{c}}e^{i\omega t} dt = \frac{\langle E_{\perp}^{2} \rangle \tau_{e}}{\pi\left(1+\omega^{2}\tau_{e}^{2}\right)}
\label{eq:FourierS}
\end{equation}
which appears like white noise for $\omega\tau_{e} \ll 1$, that is, for frequencies far below the cutoff. This function returns the total electric field noise power when integrated over all frequencies
\begin{equation}
\int_{-\infty}^{\infty}\frac{\langle E_{\perp}^{2} \rangle \tau_{e} d\omega}{\pi\left(1+\omega^{2}\tau_{e}^{2}\right)} = \int_{-\infty}^{\infty}\frac{\langle E_{\perp}^{2} \rangle \tau_{e} 2\pi df}{\pi\left(1+(2\pi f)^{2}\tau_{e}^{2}\right)}  = \langle E_{\perp}^{2} \rangle
\label{eq:integratePower}
\end{equation}

\subsection{Modeling the combined $\gamma$ and CPMG data}
The relaxation and dephasing noise spectra obtained from the above data analysis methods are each due to distinct noise sources and either transverse (relaxation) or parallel (dephasing) fields, relative to the NV axis. Therefore, the $y$ axes of the extracted noise spectra are not equivalent in the two cases, even though the units of Hz$^{2}/$Hz are the same. To compare them directly, we scale each to find an effective $E_{\perp}$ electric field noise spectrum; in other words we multiply the dephasing data by a geometric factor and NV-field coupling to reflect the $E_{\perp}$ fields that would be associated with $E_{\parallel}$ fields that actually cause first-order dephasing. Part of this scaled spectrum is not due to electric fields, rather there is a magnetic component, which is the reason for the term ``effective''. With this electric field noise spectrum for relaxation and dephasing we can then plot the two on the same $y$ axis and fit to a joint model of electric and magnetic field noise, as we next explain in detail. At the end, we scale the model parameters back to coupling units to present the data in main text Fig. \ref{fig:Figure4}.

We extracted an effective $E_{\perp}$ noise spectrum (units V$^{2}\cdot$m$^{-2}/$Hz) from each dephasing spectroscopy and DQ relaxometry data set in the following scaling
\begin{equation}
S_{E_{\perp}}^{\mathrm{cpmg}}\left(f\right) = 2\frac{S_{\mathrm{cpmg}}\left(f\right)}{d_{\parallel}^{2}/h^{2}}, S_{E_{\perp}}^{\gamma}\left(f\right) = \frac{S_{\gamma}\left(f\right)}{d_{\perp}^{2}/h^{2}}
\label{eq:cpmgGammaSEperp}
\end{equation}
where the factor of $2$ in the CPMG expression comes from considering the NV orientation in a surface electric field model of $\langle E_{\parallel}^{2}\rangle$ and $\langle E_{\perp}^{2}\rangle$. The geometry is described in the sections below about electric field noise sources. 

We describe the details of the fitting procedure shown in Figure \ref{fig:Figure4} of the main text. The fit function for the combined data set has three terms written as
\begin{multline}
S\left(f\right) =  \frac{\tau_{e}\langle E_{\perp,e}^{2}\rangle }{\pi\left[1+ \left(2\pi f\tau_{e}\right)^{\alpha}\right]} \\
+ \frac{\gamma_{\infty}}{\left(d_{\perp}/h\right)^{2}} + \frac{\tau_{m}\langle E_{\perp,m}^{2} \rangle }{\pi\left[1+ \left(2\pi f\tau_{m}\right)^{2}\right]} \frac{1}{1+\exp{\left[q \left(f-10^{7} \mathrm{Hz}\right)\right]}}
\label{eq:logisticLorentzianFit}
\end{multline}
where the five fit parameters are $\gamma_{\infty}$, $\langle E_{\perp,e}^{2}\rangle$, $\tau_{e}$, $\langle E_{\perp,m}^{2}\rangle$, and $\tau_{m}$. Table \ref{tab:spectra} summarizes the fit parameter results converted into NV-noise coupling units for comparison to previous work on CPMG-based spectroscopy \cite{Myers2014,Romach2015}.

A sixth parameter exponent $\alpha$ in the first term would typically be equal to 2 for a Lorentzian, however, we let it vary and resulting fits gave $1 < \alpha < 2$. At high frequencies where $2\pi f \tau_{e} \gg 1$ this quantity in the denominator dominates and looks like the canonical $A/f^{\alpha}$ noise. We kept the 1 from the pure-Lorentzian form in the denominator to allow for the concept of a total sum of many constituent Lorentzians in the total electric noise spectrum, which has frequency regimes of effective $\alpha \sim 0$, $ 1<\alpha < 2$, and $\alpha = 2$. This is the exponent shown in the end column of Table \ref{tab:spectra}. 

The sigmoid function in Eq. \ref{eq:logisticLorentzianFit} ($q$ arbitrarily large for step function) that multiplies the second Lorentzian is simply to give a cutoff frequency (here 10 MHz) above which the Lorentzian has zero effect; that is magnetic noise affects the CPMG dephasing data but not the $\gamma$ relaxation data. This is similar to calculating a joint fit with two separate functions that have shared parameters, where the CPMG data is fit to the full form of Eq.  \ref{eq:logisticLorentzianFit} and the relaxation data is fit only to the two terms on the bottom line.

\begin{table*}[t]
\begin{tabular}{|c|c|c|c|c|c|c|r|} 
\hline
NV &    $\omega_{\pm1}$$/2\pi$  &  $\sqrt{\langle E_{\perp,e}^{2}\rangle}$  $(b_{e})$ & $\tau_{e}$ &  $\gamma_{\infty}/\left(d_{\perp}/h\right)^{2} (\gamma_{\infty})$   & $\sqrt{\langle E_{\perp,m}^{2}\rangle}$ $(b_{m})$  &  $\tau_{m}$ & E-noise $\alpha$ \\ \hline
NVA1 & 784 MHz & $6.9\times10^{7}$ V$/$m & 1.2 \us{}  & 5800 V$^{2}$m$^{-2}/$Hz & $25\times 10^{7}$ V$/$m & $130$ ns & 2 \\
& &(170 kHz) & & (0.17 kHz) & (620 kHz) & & \\ \hline
NVA1 & 1612 MHz & $6.9\times10^{7}$ V$/$m & 1.0 \us{}  & 5800 V$^{2}$m$^{-2}/$Hz & $25\times 10^{7}$ V$/$m & $140$ ns & 2.1 \\
& &(170 kHz) & & (0.17 kHz) & (620 kHz) & & \\ \hline
NVA8 & 1431 MHz & $3.4\times10^{7}$ V$/$m & 32 \us{}  & 7000 V$^{2}$m$^{-2}/$Hz & $4.6\times 10^{7}$ V$/$m & $440$ ns & 1.5 \\
& &(85 kHz) & & (0.20 kHz) & (110 kHz) & &  \\ \hline
NVA7 & 797 MHz & $2.6\times10^{7}$ V$/$m & 8.1 \us{}  & 8600 V$^{2}$m$^{-2}/$Hz & $6.5\times 10^{7}$ V$/$m & $360$ ns & 1.6 \\
& &(65 kHz) & & (0.25 kHz) & (160 kHz) & &  \\ \hline

\hline
\end{tabular}
\caption{Summary of fit results to combined dephasing and DQ relaxation noise spectra using the simple double-Lorentzian model where the $b_{e} = \sqrt{\langle E_{\perp,e}^{2}\rangle} \sqrt{1/2} \left(d_{\parallel}/h\right)$ is the total NV coupling rate, for dephasing, to the ``electric'' Lorentzian noise. $\tau_{e}$ is the correlation time of this noise source. $b_{m}=\sqrt{\langle E_{\perp,m}^{2}\rangle} \sqrt{1/2} \left(d_{\parallel}/h\right)$ and $\tau_{m}$ are the corresponding parameters for the ``magnetic'' noise source that is relevant only to dephasing. $\gamma_{\infty}$ is the saturation level of the DQ relaxation rate at high frequency. The $\omega_{\pm1}$$/2\pi$ stated in this table refers to the value at which the CPMG spectroscopy measurements were done, which was always in the limit that $\gamma \rightarrow \gamma_{\infty}$. The electric field noise exponent $\alpha$ in the last column refers to the $1/f^{\alpha}$ component of the blue dashed curves in the fitted spectra. We find $\alpha$ is between 1 and 2 as would be expected for a sum of many Lorentzians, for example in the case of electric dipoles with various correlation times even for a single type of adatom on the diamond surface. The NVA1 measurements repeated at two large $\omega_{\pm1}$ shown yielded a spectrum fit to the same parameter values within the significant figures.
\label{tab:spectra}}
\end{table*}

\begin{figure}
\includegraphics[width = 1.0\columnwidth]{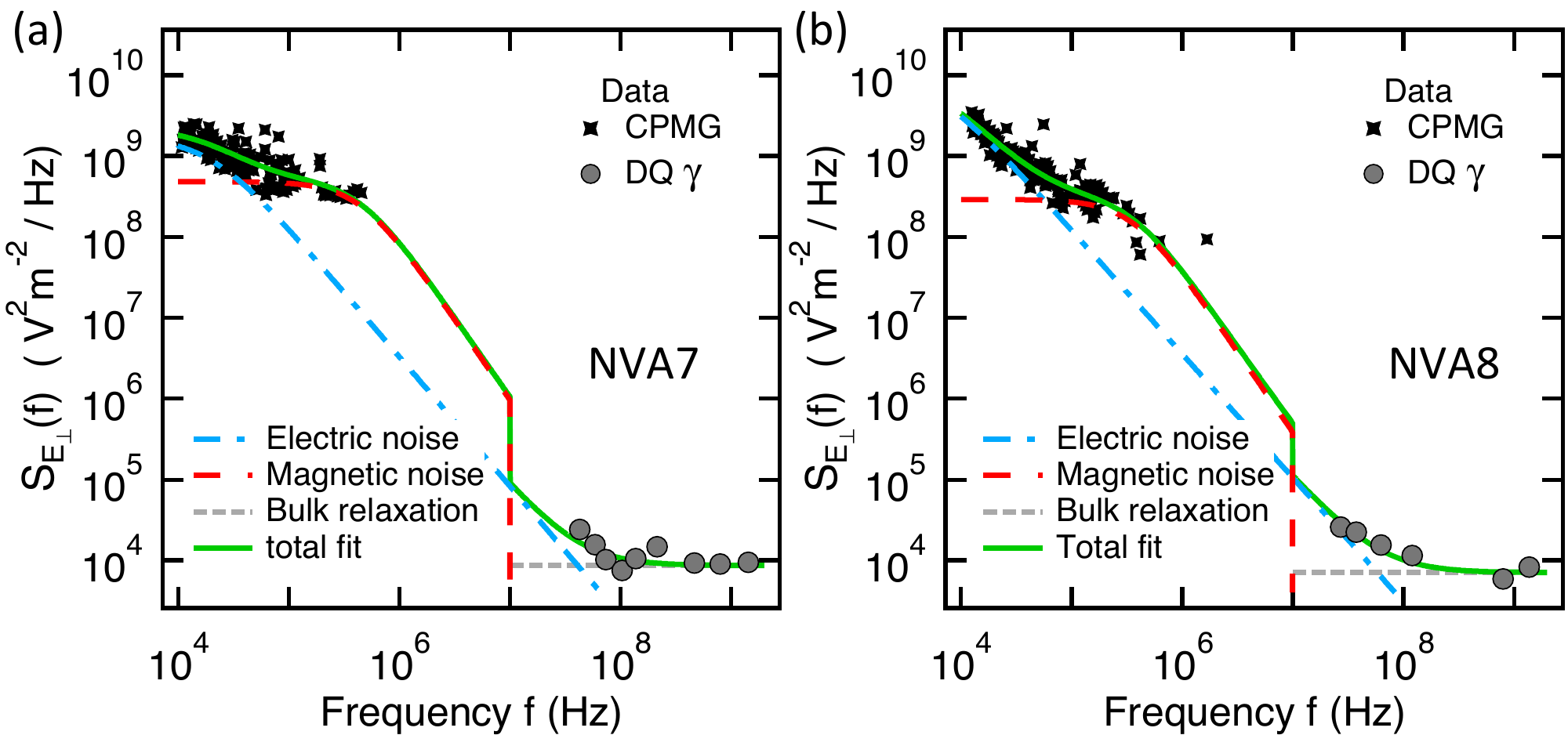}
 \caption{\label{fig:SOMA7A8spectra} Transverse electric field noise spectra fits for NVA7 (a) and NVA8 (b), showing the deduced contributions from electric, magnetic, and bulk noise sources for dephasing and DQ relaxation. For both NVs the electric field component fit well to a $1/f^{\alpha}$ function with $\alpha \approx 1.5-1.6$. Such an exponent $1 < \alpha <2$ can can arise from a sum of many Lorentzian power spectra with a distribution of different correlation times, for example from charge traps or electric dipoles. The power spectrum parameters are listed in Table \ref{tab:spectra} for both electric field and coupling units.}
 \end{figure}

\begin{figure}
\includegraphics[width = 1.0\columnwidth]{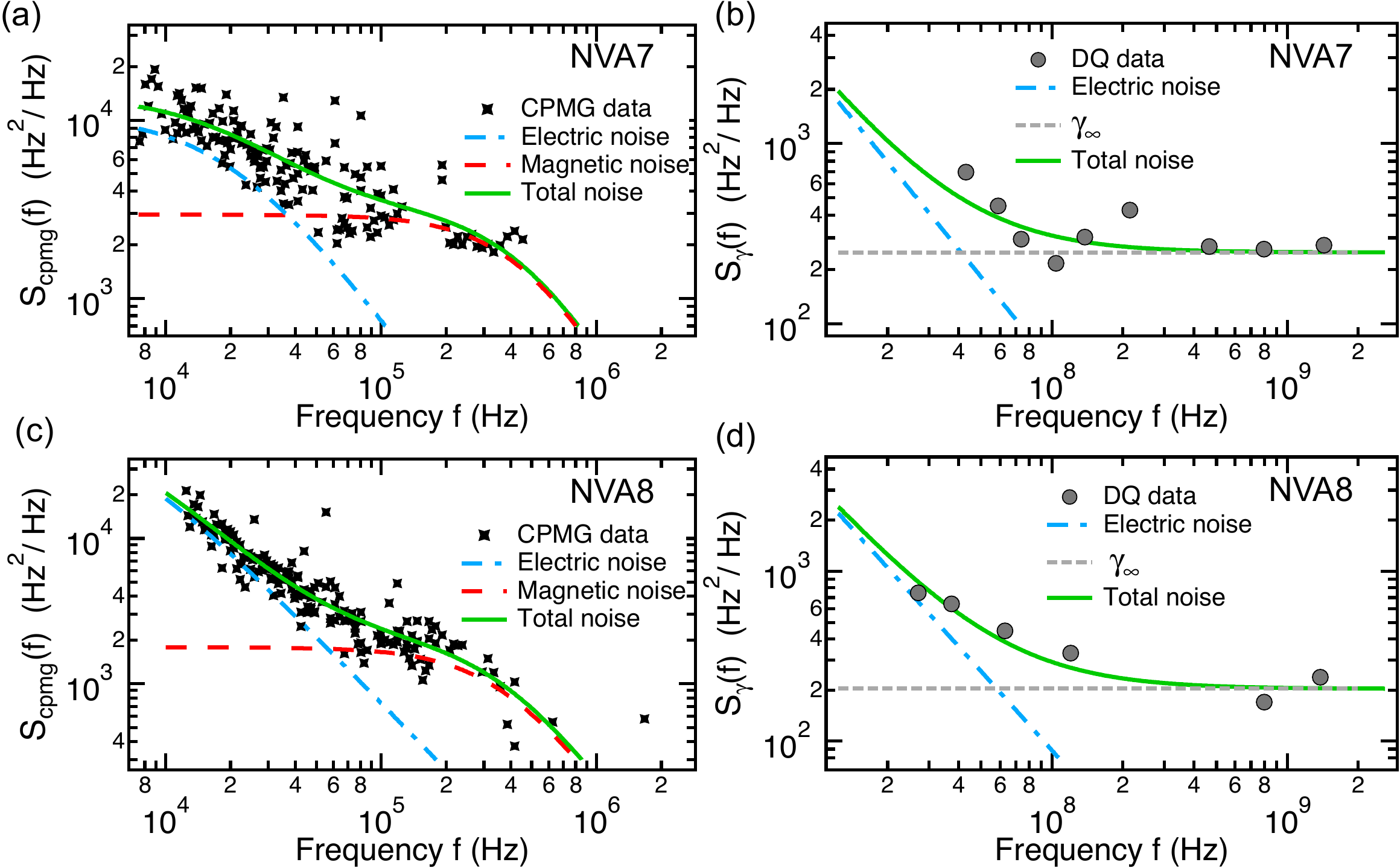}
 \caption{\label{fig:SOMA7A8spectraCoupling} Coupling noise power spectra fits for NVA7 (a,b) and NVA8 (c,d), showing the deduced contributions from electric, magnetic, and bulk noise sources for dephasing and DQ relaxation. The power spectrum parameters are listed in Table \ref{tab:spectra} for both electric field and coupling units.}
 \end{figure}

In Fig. \ref{fig:SOMA7A8spectra}  we plot spectrum data and fits for the two NVs that show higher electric field noise at low frequencies, NVA7 (a) and NVA8 (b), the latter shown in the main text Fig. \ref{fig:Figure4}(b). The parameters of the fits are included in Table \ref{tab:spectra}, and the final scaling to dephasing and relaxation coupling is plotted in Fig. \ref{fig:SOMA7A8spectraCoupling} using the inverse procedure of Eq. \ref{eq:cpmgGammaSEperp}. One could alternatively write two separate fit functions for the dephasing and DQ relaxation data and fit their parameters jointly, however, the method of the effective transverse field is helpful to display the data as one complete noise spectrum. 

\begin{figure}
\includegraphics[width = 1.0\columnwidth]{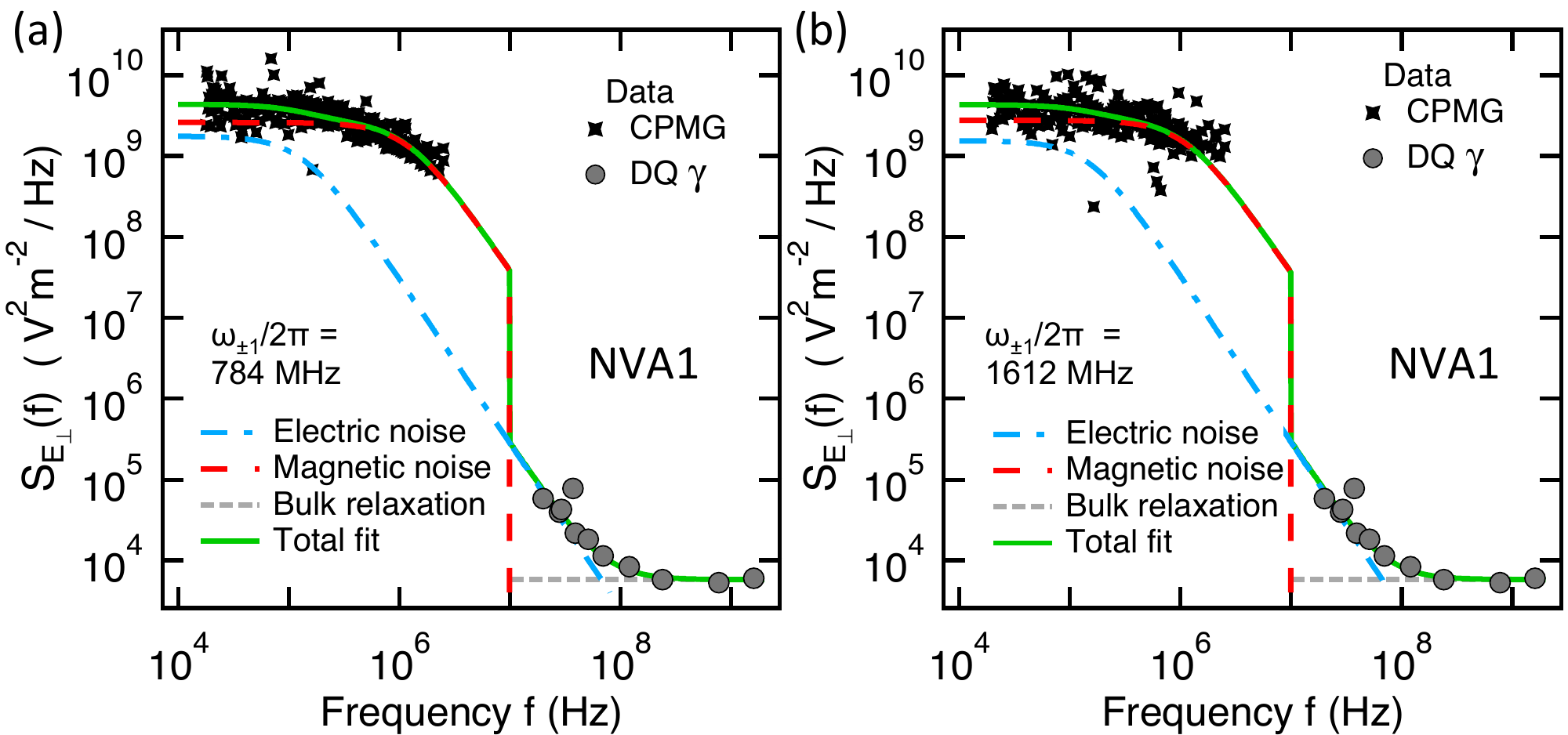}
 \caption{\label{fig:SOMNVA1spectraVsSplitting} Transverse electric field noise spectra fits for NVA1 at $\omega_{\pm1}$$/2\pi=784$ MHz (a) and $\omega_{\pm1}$$/2\pi=1612$ MHz (b), showing that for sufficiently large applied magnetic fields the extracted dephasing spectra are described well by the same noise model parameters as shown in Table \ref{tab:spectra}.}
 \end{figure}

\begin{figure}
\includegraphics[width = 1.0\columnwidth]{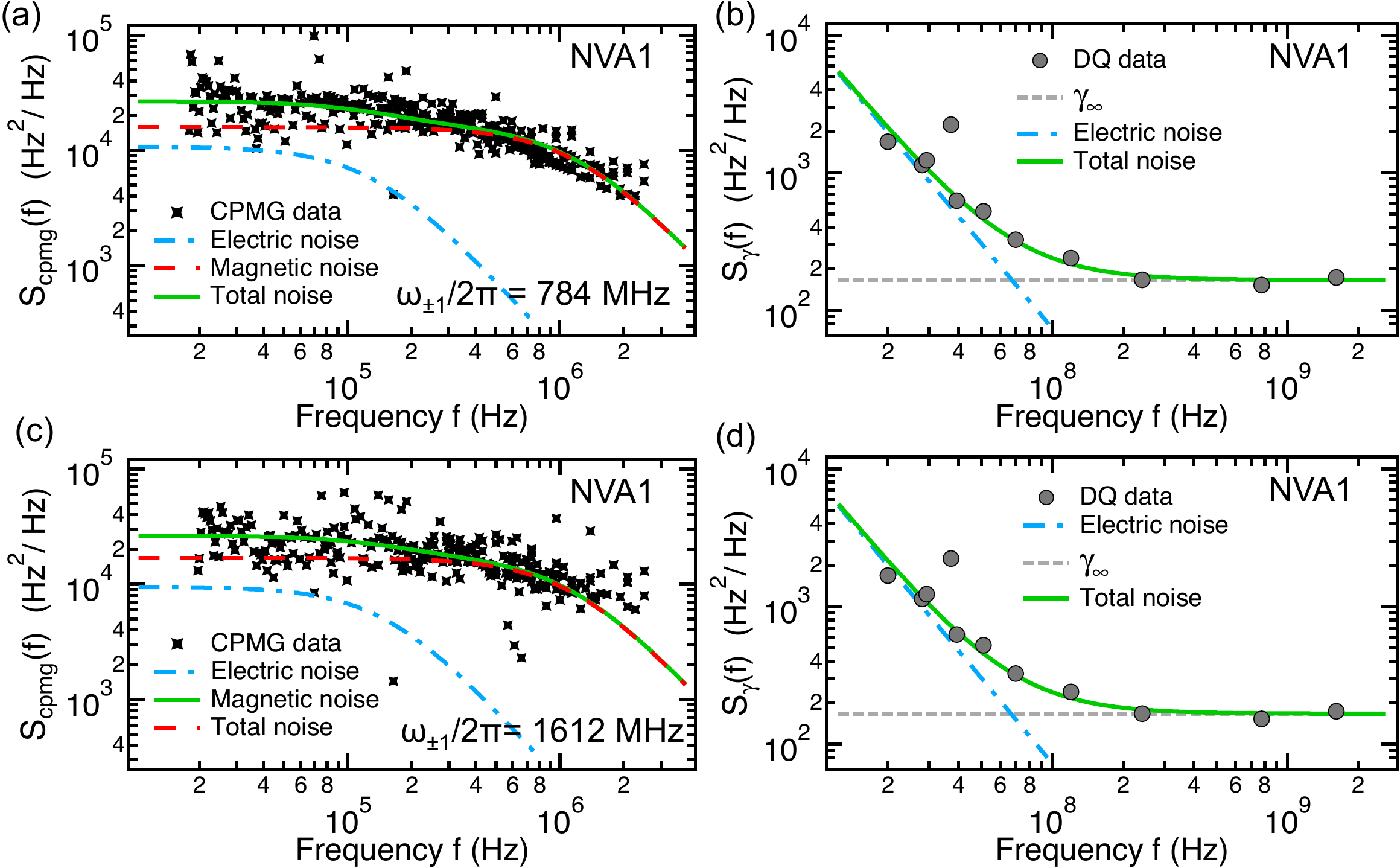}
 \caption{\label{fig:SOMNVA1spectraCoupling} Coupling noise power data and fits for NVA1 at $\omega_{\pm1}$$/2\pi=784$ MHz (a,b) and $\omega_{\pm1}$$/2\pi=1612$ MHz (c,d), showing that for sufficiently large applied magnetic fields the extracted dephasing spectra are described well by the same noise model parameters (within fitting uncertainty) as shown in Table \ref{tab:spectra}. The data in (b) and (d) are the same for this comparison, since DQ measurements are taken by tuning $\omega_{\pm1}$, but the fit results of electric noise (blue dash-dot line) and total noise (green solid line) are independently generated since the dephasing data are distinct in (a) and (c).}
 \end{figure}

In Fig. \ref{fig:SOMNVA1spectraVsSplitting} we plot effective transverse electric field noise spectra data and fits for the NVA1 at two different applied magnetic fields. Fig. \ref{fig:SOMNVA1spectraCoupling} show the same data sets for final scaling back to dephasing and relaxation coupling units. This comparison shows that there is negligible dependence on the magnetic field between 140 G and 290 G because the model fit parameters are nearly identical in the two $\omega_{\pm1}$ cases. We avoided performing dephasing spectroscopy at low magnetic fields since $\gamma$ becomes large at small $\omega_{\pm1}$.

Finally, we predict based on our model of dephasing and relaxation that there is quite possibly no fundamental reason that the $T_{2}$ cannot be increased to beyond $T_{1}$, even at high fields where dephasing is dominant, because the noise spectrum continues to fall above the MHz regime. It is evident from the combined data that there is not significant white electric or phononic noise level due to the surface otherwise $\gamma$ would be limited to a faster value than observed. Using present hardware, even increasing the number of pulses $N$ past a certain point will help very little due to the limited pulse delay resolution, phase cycling rate, and finite pulse time. In other words, as $N$ is increased in our example we probed a maximum CPMG frequency of $f_{\mathrm{max}} \gtrsim 2$ MHz on the noise spectrum. Once $N$ is large enough to reach $f_{\mathrm{max}}$ then the portion of the accessible spectrum becomes shorter from the low-frequency side as $N$ is increased, thus limiting the ability to even accurately measure the $T_{2}$ with CPMG.

\subsection{Stationary versus non-stationary electric noise source}
In the preceding analysis we have made the common assumption of an Ornstein-Uhlenbeck process, which is the unique first-order stationary Gauss-Markov process \cite{Uhlenbeck1930}. The correlation time $\tau_{e}$ of the electric fluctuations affecting $\gamma$ have been assumed to be much shorter than the single-shot measurement time (milliseconds) as well as the full measurement time (hours). However, we also observed two long-lived ``anomalous'' data points for $\gamma$ both shown in the main text: one being that $\gamma =2.4(6)$ kHz at $\omega_{\pm1}$$/2\pi=37.1$ MHz deviated more than a factor of two larger than the $\gamma=0.63(16)$ at $\omega_{\pm1}$$/2\pi=39.2$ MHz for NVA1, and the other anomalous point being for a relatively low value of $\gamma$ at $\omega_{\pm1}$$/2\pi=104$ MHz for NVA7. It is possible that a long-lived noise source (minutes or hours) was present at the time of the anomalous point, for example if an electric-dipole-containing molecule had adsorbed in a surface position very close to the NV for some time. Another possibility is the change of charge states of surrounding vacancy defects. It has been shown that the charge state of the NV center (NV$^{-}$ or NV$^{0}$) can be switched for example optically \cite{Manson2005} or electrically \cite{Grotz2012}. However, this particular NV is verified to be a single NV within the diffraction-limited spot size of $\sim 500$ nm, so it is likely not electric noise from another NV in this case. Note that these $\gamma\left(\omega_{\pm1}\right)$ measurements were not taken in a monotonically increasing or decreasing $\omega_{\pm1}$, rather we took them out of order pseudo-randomly, and so $\gamma$ is not for example slowly increasing in time but is quite stable.

Another interesting route of further research in DQ spin relaxation would involve looking at changes in $\gamma$ with manipulation of the charge state of a small ensemble of NV centers or as a function of laser illumination or static electric fields on surface electrodes. We used constant low-power 532-nm illumination for all measurements in this work of $60$ $\mu$W into the back of the objective, which still yields high PL counts due to the enhanced collection efficiency of the diamond nanopillars.

\subsection{Model of fluctuating surface electric dipoles}
We derive a model of the power spectral density of fluctuating electric fields at the NV center based on the assumption of electric dipoles on the surface. First we derive the electric field at the NV of depth $z$ for a single dipole at some distance $d$ from the surface and then take the case $d\rightarrow 0$. 

We suppose an interface with surface normal $\hat{z}$ between air and diamond with permittivities $\epsilon_{0}$ and $\epsilon_{d}\epsilon_{0}$, respectively. From the image-dipole method of calculating the potential of an electric dipole composted of charges $q$ and $-q$ near a dielectric interface, we obtain a potential inside the $\epsilon_{d}$ region $(z<0)$ of
\begin{equation}
\Phi^{(z)}\left(\textbf{r}\right)  = \frac{p}{4\pi\epsilon_{0}} \left(\frac{2}{1+\epsilon_{d}}\right)\frac{z}{\left(x^{2}+y^{2}+z^{z}\right)^{3/2}}
\label{eq:imageDipole1}
\end{equation}
for a dipole oriented parallel to the surface normal, $\textbf{p} = p \hat{z}$. For the case of a dipole oriented perpendicular to the surface normal we obtain
\begin{equation}
\Phi^{(x)}\left(\textbf{r}\right) = \frac{p}{4\pi\epsilon_{0}} \left(\frac{2}{1+\epsilon_{d}}\right)\frac{x}{\left(x^{2}+y^{2}+z^{z}\right)^{3/2}}
\label{eq:imageDipole1}
\end{equation}
Let $E_{0} \equiv \frac{p}{4\pi\epsilon_{0}} \left(\frac{2}{1+\epsilon_{d}}\right)$. We find the mean-square electric field from a fluctuating dipole sampling all directions by considering the $x,y,z$ contributions to the electric field from each component $p_{x},p_{y},p_{z}$, a total of nine components:
\begin{multline}
E_{x}^{(p_{z})} = \frac{\partial \Phi^{(z)}}{\partial x} = -E_{0}\frac{3xz}{\left(x^{2}+y^{2}+z^{2}\right)^{5/2} } \\
E_{y}^{(p_{z})} = \frac{\partial \Phi^{(z)}}{\partial y} = -E_{0}\frac{3yz}{\left(x^{2}+y^{2}+z^{2}\right)^{5/2} } \\
E_{z}^{(p_{z})} = \frac{\partial \Phi^{(z)}}{\partial z} = E_{0}\frac{x^{2}+y^{2}-2z^{2}}{\left(x^{2}+y^{2}+z^{2}\right)^{5/2} }\\
E_{x}^{(p_{x})} = \frac{\partial \Phi^{(x)}}{\partial x} = E_{0}\frac{-2x^{2} + y^{2} + z^{2}}{\left(x^{2}+y^{2}+z^{2}\right)^{5/2} } \\
E_{y}^{(p_{x})} = \frac{\partial \Phi^{(x)}}{\partial y} = -E_{0}\frac{3xy}{\left(x^{2}+y^{2}+z^{2}\right)^{5/2} } \\
E_{z}^{(p_{x})} = \frac{\partial \Phi^{(x)}}{\partial z} = -E_{0}\frac{3xz}{\left(x^{2}+y^{2}+z^{2}\right)^{5/2} }
\label{eq:dipolePhiToE}
\end{multline}
where the components $E_{i}^{(p_{y})}$ are identical to the $E_{i}^{(p_{x})}$ with $x$ replaced by $y$ since the orientation of the dipole perpendicular to the surface normal is arbitrary.

The $x,y,x$ coordinates are defined relative to the surface, and the NV is crystallographically aligned at an angle of $\theta_{\mathrm{NV}}\approx 54.74^{\circ}$ to the surface normal. We take the NV to lie in the $xz$ plane, and the goal is to compute the mean-square electric fields $\left\langle E_{\perp}^{2} \right\rangle$ and $\left\langle E_{\parallel}^{2} \right\rangle$ that appear in the NV Hamiltonian Eq. \ref{eq:hamiltonian}. These fields for the single electric dipole are given by
\begin{equation}
\left\langle E_{\perp}^{2} \right\rangle_{\mathrm{dipole}} = \left\langle E_{x}^{2} \right\rangle\cos^{2}\theta_{\mathrm{NV}} + \left\langle E_{z}^{2} \right\rangle\sin^{2}\theta_{\mathrm{NV}} + \left\langle E_{y}^{2} \right\rangle
\label{eq:EperpSquared}
\end{equation}
\begin{equation}
\left\langle E_{\parallel}^{2} \right\rangle_{\mathrm{dipole}} = \left\langle E_{x}^{2} \right\rangle\sin^{2}\theta_{\mathrm{NV}} + \left\langle E_{z}^{2} \right\rangle\cos^{2}\theta_{\mathrm{NV}}
\label{eq:EparSquared}
\end{equation}
Each $\left\langle E_{i}^{2} \right\rangle$ with $i=x,y,z$ is computed from the mean-square sum of the three components in Eq. \ref{eq:dipolePhiToE}. This sum introduces a factor of $1/6$ from taking the rms per each fluctuating $p_{i}$ component and the total mean, for example $\left\langle E_{x}^2\right\rangle = \left[\left(E_{x}^{(p_{x})}/\sqrt{2}\right)^2 + \left(E_{x}^{(p_{y})}/\sqrt{2}\right)^2 + \left(E_{x}^{(p_{z})}/\sqrt{2}\right)^2 \right]/3$.

For the parallel and perpendicular mean-square fields from a uniform sheet of fluctuating electric dipoles, the terms of Eqs. \ref{eq:EperpSquared} and \ref{eq:EparSquared} are computed from an integral over the surface area such that a small number of dipoles in a given area is $dN = \sigma_{p}rdrd\phi$ where $\sigma_{p}$ is the areal number density of dipoles and $\phi$, $r=x^{2}+y^{2}$ are polar coordinates. For example in non-reduced form
\begin{multline}
\left\langle E_{x}^{2} \right\rangle_{\mathrm{surf}} = \frac{\sigma_{E}E_{0}^2}{6}\int_{0}^{2\pi}\int_{0}^{\infty}r dr d\phi \left[\left( \frac{3rz\cos\phi}{\left(r^{2}+z^{2}\right)^{5/2}} \right)^{2}\right. \\
+\left( \frac{-2r^{2}\cos^{2}\phi + r^{2}\sin^{2}\phi + z^{2}}{\left(r^{2}+z^{2}\right)^{5/2}} \right)^{2} \\
+\left.\left( \frac{-2r^{2}\sin^{2}\phi + r^{2}\cos^{2}\phi + z^{2}}{\left(r^{2}+z^{2}\right)^{5/2}} \right)^{2}\right]
\label{eq:EperpSurfIntegral}
\end{multline}
The final results for surface-induced electric fields at the NV of depth $z$ are
\begin{equation}
\left\langle E_{\perp}^{2} \right\rangle_{\mathrm{surf}} = \left(\frac{p}{4\pi \epsilon_{0}}\right)^{2} \left(\frac{2}{1+\epsilon_{d}}\right)^{2} \frac{3\pi}{8} \frac{\sigma_{p}}{z^{4}}
\label{eq:EperpSurf}
\end{equation}
\begin{equation}
\left\langle E_{\parallel}^{2} \right\rangle_{\mathrm{surf}} = \left(\frac{p}{4\pi \epsilon_{0}}\right)^{2} \left(\frac{2}{1+\epsilon_{d}}\right)^{2} \frac{3\pi}{16} \frac{\sigma_{p}}{z^{4}}
\label{eq:EparSurf}
\end{equation}
The ratio of $\langle E_{\perp}^{2}\rangle_{\mathrm{surf}}/\langle E_{\parallel}^{2} \rangle_{\mathrm{surf}}=2$ between these two results is a direct result of the orientation of the NV described in Eqs. \ref{eq:EperpSquared} and \ref{eq:EparSquared} and is the reason we scale the noise spectral density for the CPMG data with $2$ in the main text Eq. \ref{eq:cpmgGammaSEperp}. In fact this ratio $\langle E_{\perp}^{2}\rangle_{\mathrm{surf}}/\langle E_{\parallel}^{2} \rangle_{\mathrm{surf}}=2$ does not change if the particular surface electric field model has an altered ratio of $\langle E_{z}^{2}\rangle_{\mathrm{surf}}/\langle E_{x}^{2} \rangle_{\mathrm{surf}}$ and $\langle E_{x}^{2}\rangle_{\mathrm{surf}}=\langle E_{y}^{2} \rangle_{\mathrm{surf}}$; the second equality would be the case for a fairly uniform surface noise source. This invariant ratio is a consequence of the NV angle given by $\cos^{-1}\left(\theta_{\mathrm{NV}}\right)=\sqrt{1/3}$, or $\theta_{\mathrm{NV}}\approx 54.74^{\circ}$. That is, from Eqs. \ref{eq:EperpSquared} and \ref{eq:EparSquared} the ratio
\begin{equation}
\left\langle E_{\perp}^{2} \right\rangle/\left\langle E_{\parallel}^{2} \right\rangle= \frac{\left(\cos^{2}\theta+1\right) + k \sin^{2}\theta}{\sin^{2}\theta + k \cos^{2}\theta}
\label{eq:Eperp2Epar2Ratio}
\end{equation}
becomes equal to 2 for any $k = \left\langle E_{z}^{2} \right\rangle/\left\langle E_{x}^{2} \right\rangle$ if $\theta = \theta_{\mathrm{NV}}$ and if on average $\left\langle E_{y}^{2} \right\rangle=\left\langle E_{x}^{2} \right\rangle$.

Because the electric field power spectral densities are proportional to $\langle E_{\perp}^{2}\rangle_{\mathrm{surf}}$ ($\langle E_{\parallel}^{2}\rangle_{\mathrm{surf}}$), they will have the depth dependence of $1/z^{4}$. Surface-induced anomalous heating of trapped ions due to non-Johnson-noise-related electric fields from electrode surfaces has been an important subject of study for several decades. At least two theories predict the distance dependence of noise power to be $1/z^{4}$, though they differ in the dependence on frequency. A model (A) of adatoms diffusing on the surface predicts $S \sim 1/f^{3/2}$ and a phonon-induced electric dipole fluctuator model (B) predicts a transition from white noise, to $1/f$, to $1/f^2$ as would occur from a sum of many Lorentzians with different cutoff frequencies \cite{Enoise2013}. The differences between those models and the present diamond situation are 1) The nearly atomic-scale distances we consider are much smaller, nanometers rather than 10s to 100s of micrometers, and 2) The NV is actually inside the dielectric material rather than in vacuum above a surface, so in model (B) it could potentially feel the phononic noise directly in addition to feeling electric field noise. We do not attempt to distinguish between these two models since even model (B) would exhibit $\sim1/f^{3/2}$ in some regimes due to the transition region from flat to $1/f^{2}$. The averaging we have done in summing up the components of these fields accounts for the various orientations of the dipole $\textbf{p}$ that are sampled as it fluctuates measurement-to-measurement. These fluctuations will have a power spectrum $S_{p}\left(f\right)$ that in general depends on the type of adatom species, its trapping potential on the surface, and its vibronic spectrum as a function of temperature \cite{Safavi2011}. Since each possible vibronic transition may have different frequency cutoffs then there may be a sum of Lorentzian-like components in the total noise spectral density. For simplicity we assume the single-Lorentzian case and a typical molecular dipole magnitude of 2 Debye in order to compute an order of magnitude for the possible surface areal density of dipoles. We fit this power spectrum model to the combined CPMG and $\gamma$ data assuming an NV depth of $7$ nm consistent with the N implantation parameters and known $T_{2}$'s from prior work \cite{Myers2014,Romach2015}. To produce the $\sqrt{\langle E_{\perp}^{2}\rangle_{\mathrm{surf}}} \sim 10^{7}$ V$/$m observed in the spectra, the extracted $\sigma_{p}$ is orders of magnitude larger than would be physically possible. This either means that the magnitude of $p$ at the surface must be much greater than 1 D or that the surface electric dipole model is not an accurate description of the surface electric field noise. In contrast to $\sim1$-D electric dipoles, only a small number of elementary electric charges are required to produce such a $\sim 10^{7}$ V$/$m magnitude of electric field at nanometric distances. A simple uniform charge sheet model, as studied in the next section, has an incorrect depth dependence of $S(f) \propto 1/z^{2}$, however accounting for discrete charges and only short-range NV-charge interactions can make the depth dependence look like $\alpha=3-4$. It is also likely that for NVs within nanometers of electric dipole phenomena that quadrupolar fluctuations \cite{iontrap2016} may become important as well, and these terms will fall off more rapidly with depth. 

We compare our electric field noise spectral density to those extracted from experiments on motional heating rates of trapped ions due to the trap's metal electrodes. For an ion-electrode distance of 75 \um{} and trap frequency 1 MHz, experiments at room temperature report $S_{E_{\perp}}^{\rm{ion}} \approx 2\times10^{-11}$ V$^{2}$m$^{-2}/$Hz \cite{ion2008}. We compare our value of $S_{E_{\perp}}(f=1$ MHz$) \approx 3.5\times 10^{6}$ V$^{2}$m$^{-2}/$Hz for NVA8 (Fig. \ref{fig:Figure4}(c)) by scaling $S_{E_{\perp}}\left(f\right)$ to account for the $\sim7$-nm NV-surface separation and diamond dielectric half-space, and we obtain $S_{E_{\perp}}\left(f\right) \approx 3 \times 10^{-10}$ V$^{2}$m$^{-2}/$Hz. This estimate is of similar magnitude as the $S_{E_{\perp}}^{\rm{ion}}$ that is often attributed to ``patch potential'' electric fields. We note finally that, while ion trap studies suggest a wide range of $S_{E_{\perp}}^{\rm{ion}}$ values experimentally and theoretically \cite{ion2007,ionclean2012,Safavi2011}, the $1/d^{4}$ model appears widely accepted for various possible models \cite{Enoise2013}. Our comparison here is important not only to point out the possible universality of surface-related electric field noise, but also to suggest that the magnitudes seen in non-diamond surfaces are of the order that can be probed using the demonstrated sensitivity of the NV scanning probe.

\subsection{Model of fluctuating surface charges}
Perhaps the simplest surface-based model of electric field noise is due to a uniform density of charge traps that periodically become occupied or unoccupied. In the McWhorter charge trap model for surfaces \cite{McWhorter1957} the $1/f$ noise regime arises due to a sum of many trap relaxation phenomena with a uniform distribution of frequencies $\lambda \in \left[\lambda_{1},\lambda_{2}\right]$ and $\lambda_{1} \ll f \ll \lambda_{2}$. In the regime that $f \gg \lambda_{2} \gg \lambda_{1}$ the spectrum resembles a $\sim 1/f^{2}$ Lorentzian, and therefore the crossover in noise may occur in the transition regime of $f \sim \lambda_{2}$. 

We consider a simplistic model of a ``sheet'' of surface charge traps that cause a fluctuating electric field at the NV at depth $z$. Although a static infinite sheet of charge shows no distance dependence of the electric field, the total mean square fluctuations of the constituent charges will have a distance dependence. Using the method of image charges at a dielectric-air interface, the electric potential inside the diamond $(z > 0)$ at NV position $r,z$ due to a charge $q$ at distance $d$ from the interface is
\begin{multline}
\Phi\left(z > 0\right) = \frac{1}{4\pi \epsilon_{0}\epsilon_{d}} \frac{q}{\sqrt{r^{2}+(d+z)^{2}}} +\\ 
\frac{q}{4\pi \epsilon_{0}\epsilon_{d}}\left(\frac{\epsilon_{d}-1}{\epsilon_{d}+1}\right)\frac{1}{\sqrt{r^{2} + (d+z)^{2}}} 
\label{eq:potentialImageCharge}
\end{multline}
where $\epsilon_{d}=5.7$ is the relative permittivity of diamond and $r^{2}=x^{2}+y^{2}$. For a charge $q=-e$ on the surface $d\rightarrow 0$ the three electric field components $E_{x_{i}} = d\Phi/ dx_{i}$ are
\begin{equation}
E_{x_{i}}\left(r,z\right) = A_{0}\frac{x_{i}}{\left(r^{2} + z^{2}\right)^{3/2}}e
\label{eq:surfaceChargeField}
\end{equation}
where $A_{0} = \frac{1}{4\pi \epsilon_{0}}\frac{2}{1+\epsilon_{d}}$. We take the NV center axis to be in the $xz$ plane an angle $\theta_{\mathrm{NV}}$ from the z-axis normal as in the electric dipole noise calculation. Also like the dipole calculation, the contributions of mean square fluctuations on the perpendicular and parallel NV axes are given by Eqs. \ref{eq:EperpSquared} and \ref{eq:EparSquared}, now for charges
\begin{equation}
\left\langle E_{\perp}^{2} \right\rangle_{\mathrm{charge}} = \left\langle E_{x}^{2} \right\rangle\cos^{2}\theta_{\mathrm{NV}} + \left\langle E_{z}^{2} \right\rangle\sin^{2}\theta_{\mathrm{NV}} + \left\langle E_{y}^{2} \right\rangle
\label{eq:EperpSquaredCharge}
\end{equation}
\begin{equation}
\left\langle E_{\parallel}^{2} \right\rangle_{\mathrm{charge}} = \left\langle E_{x}^{2} \right\rangle\sin^{2}\theta_{\mathrm{NV}} + \left\langle E_{z}^{2} \right\rangle\cos^{2}\theta_{\mathrm{NV}}.
\label{eq:EparSquaredCharge}
\end{equation}
The calculation of the mean-square electric field for a charge trap could be computed by considering that a charge trap is either occupied by an electron with electric field given by Eq. \ref{eq:surfaceChargeField}, or it is unoccupied with field $E_{x_{i}}=0$. This assumption for rms gives $\left\langle E_{x}^{2} \right\rangle = E_{x}^{2} / 3$.

If a uniform sheet of surface charge is assumed then the number of charges in a small area is $dN = \sigma_{e}rdrd\phi$ for surface charge number density $\sigma_{e}$, and the surface area integrals ($r\rightarrow\infty$) yield
\begin{equation}
\left\langle E_{x}^{2} \right\rangle_{\mathrm{surf}} = \left\langle E_{y}^{2} \right\rangle_{\mathrm{surf}} = \frac{\pi A_{0}^{2}e^{2}\sigma_{e}}{12z^{2}}
\label{eq:surfaceChargeIntegrals}
\end{equation}
\begin{equation}
\left\langle E_{z}^{2} \right\rangle_{\mathrm{surf}} = \frac{\pi A_{0}^{2}e^{2}\sigma_{e}}{6z^{2}}
\label{eq:surfaceChargeIntegrals}
\end{equation}
Combining these results with Eqs. \ref{eq:EperpSquaredCharge} and \ref{eq:EparSquaredCharge} gives the results relevant to the axes in the NV Hamiltonian
\begin{equation}
\left\langle E_{\perp}^{2} \right\rangle_{\mathrm{surf}} = \frac{1}{\left(4\pi \epsilon_{0}\right)^{2}}\left(\frac{2}{1+\epsilon_{d}}\right)^{2} \frac{2\pi e^{2} \sigma_{e}}{9z^{2}}
\label{eq:surfaceChargeIntegrals}
\end{equation}
\begin{equation}
\left\langle E_{\parallel}^{2} \right\rangle_{\mathrm{surf}} = \frac{1}{\left(4\pi \epsilon_{0}\right)^{2}}\left(\frac{2}{1+\epsilon_{d}}\right)^{2} \frac{\pi e^{2} \sigma_{e}}{9z^{2}}
\label{eq:surfaceChargeIntegrals}
\end{equation}
which again is consistent with $\left\langle E_{\perp}^{2} \right\rangle_{\mathrm{surf}}/\left\langle E_{\parallel}^{2} \right\rangle_{\mathrm{surf}} = 2$ due to the magic NV angle. For an example of $\left\langle E_{\perp}^{2} \right\rangle_{\mathrm{surf}}=(7\times 10^7$ V$/$m$)^2$ and assuming $z=7$ nm for the NV depth this yields a surface charge density $\sigma_{e} \approx 1.9$ nm$^{-2}$.

If the surface integral is not taken out to $r\rightarrow \infty$, but rather some nanometric distance on the order of the NV depth, then $\left\langle E_{\perp}^{2} \right\rangle_{\mathrm{surf}} \propto 1/z^{\alpha}$ with $\alpha \sim 2-4$. This could arise for example due to electric-field screening effects at the nontrivial semiconductor interface \cite{Luth2010}. In such a case, surface charges could potentially account for the $1/z^{3.6(4)}$ dependence of the surface noise power observed in prior studies of coherence versus NV depth \cite{Myers2014,Romach2015}.

The depth dependence of the Lorentzian that has a higher frequency cutoff, which we now suggest to be the magnetic noise, also needs further depth-correlated study, though \cite{Romach2015} results have pointed to a roughly $S\left(f\right) \propto 1/z^{1.8}$ behavior. Rosskopf et. al. have studied high-frequency magnetic noise through \Tonez{} and $T_{1,\rho}$ measurements, finding a much shorter $\tau_{e}=0.2$ ns \cite{Rosskopf2014}, though depth dependence must still be characterized. We have discussed in a section above that this noise with ultra-short correlation time can explain faster $\Omega$ near the surface but not dephasing due to the necessarily low noise power at low frequencies. As there is a gap in frequency space in the characterization of magnetic noise, it may be that there is a broad distribution of magnetic noise correlation times from 100s of nanoseconds to less than 1 nanosecond. 

\subsection{Surface-related strain noise}
\begin{figure}
\includegraphics[width = 1.0\columnwidth]{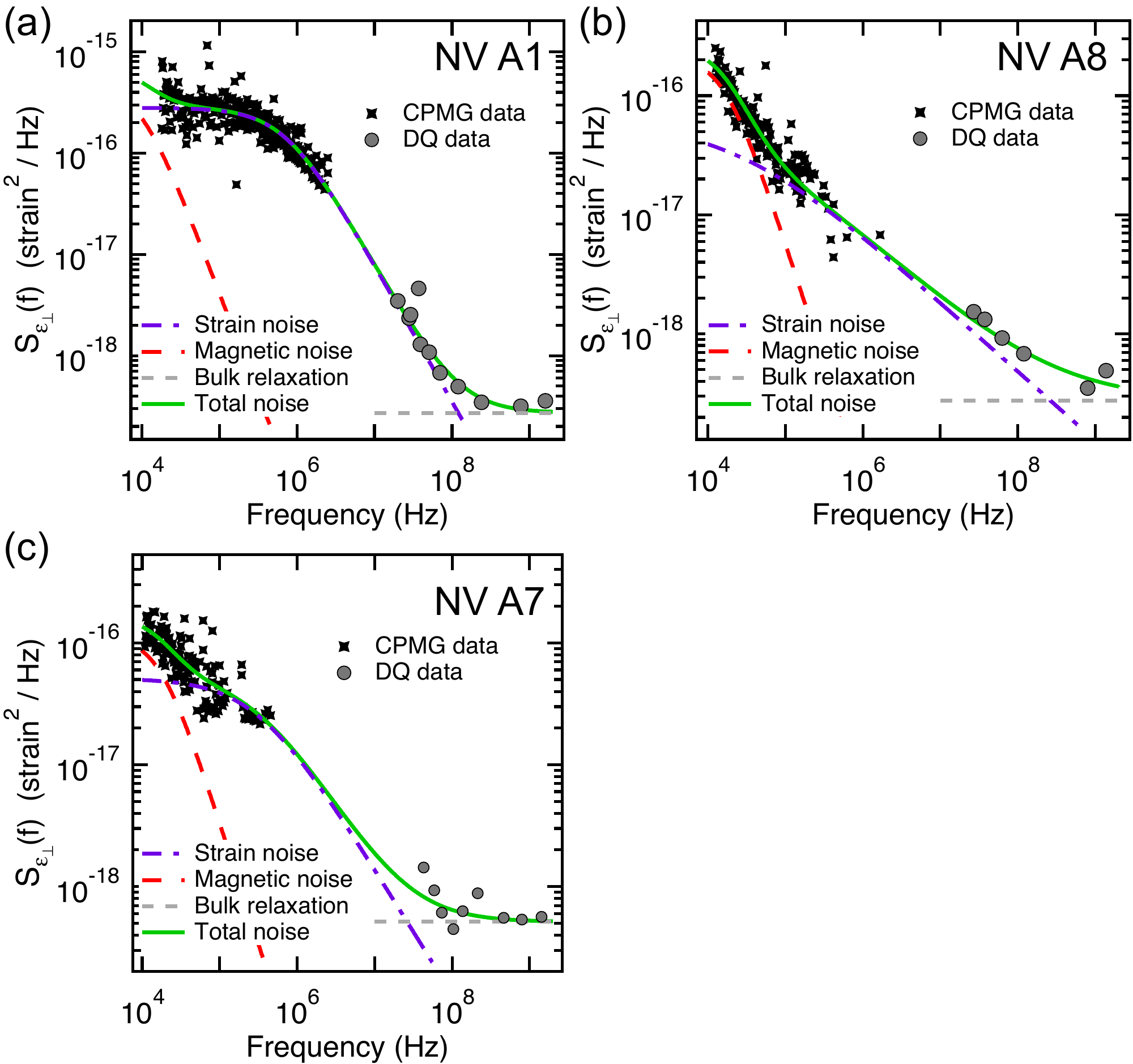}
 \caption{\label{fig:FigureSOMStrainSpectra} Noise spectra modeling for NVs A1, A8, and A7 with strain assumption rather than electric field. Strain coupling constants are more comparable, yielding a different relationship between the SQ dephasing and DQ relaxation spectra. The qualitative result, different from the electric case, is that strain noise is predicted as the higher frequency contribution and the remaining low frequency part is then magnetic. The exponents of the strain curves (purple dash-dot) are smaller, $\alpha = 1.4, 0.6, 1.0$ for A1, A8, and A7, respectively. This constraint means that the quality of the strain fit to the DQ data is lower in comparison to the electric case.}
 \end{figure}
In the main text we have chosen to present the DQ relaxation in terms of electric noise. Here we consider the possibility of surface-modified phonon noise in in the place of surface electric field noise as the cause of faster DQ relaxation for small $\omega_{\pm1}$. This case is more difficult to analyze because the strain coupling coefficients, while measured in the literature, are more complicated if fully considered in the tensor formalism \cite{Doherty2012, Lee2016}. Although stress and strain are tensors, the NV susceptibility to strain is adequately described by the transverse and axial terms in the Hamiltonian Eq. \ref{eq:hamiltonian}, where here we write the strain terms separately
\begin{multline}
H_{\mathrm{NV}} = \left(h\Dgs + d_{\parallel}E_{\parallel} + \tilde{d}_{\parallel}\xi_{\parallel}\right)S^2_{z} + g\mu_{B} \textbf{B}\cdot\textbf{S}\\
- \left(\frac{d_{\perp}E_{\perp}}{2}+ \frac{\tilde{d}_{\perp}\xi_{\perp}}{2}\right)\left(S^{2}_{+} +S^{2}_{-}\right)
\label{eq:hamiltonianStrain}
\end{multline}
with strain coupling parameters $\tilde{d}_{\parallel}/h \approx 13$ GHz/strain and $\tilde{d}_{\perp}/h \approx 22$ GHz/strain \cite{Ovartchaiyapong2014}. The coupling ratio $\tilde{d}_{\perp} /\tilde{d}_{\parallel} \approx 1.7$ is much smaller than that of the electric field coupling parameters, $d_{\perp} /d_{\parallel} \approx 49$. Therefore, when we alternatively scale the dephasing and DQ spectroscopy data of coupling power spectral densities by these strain couplings the two spectra have a different relative separation compared with those for the pure-electric case in main text Fig. \ref{fig:Figure4}. We make the scaling 
\begin{equation}
S_{\varepsilon_{\perp}}^{\mathrm{cpmg}}\left(f\right) = 2\frac{S_{\mathrm{cpmg}}\left(f\right)}{\tilde{d}_{\parallel}^{2}/h^{2}}; \mathrm{  } S_{\varepsilon_{\perp}}^{\gamma}\left(f\right) = \frac{S_{\gamma}\left(f\right)}{\tilde{d}_{\perp}^{2}/h^{2}}.
\label{eq:scaleSpectraStrain}
\end{equation}
from Hz$^{2}/$Hz units to (strain)$^{2}/$Hz. Also the factor of 2 in Eq. \ref{eq:scaleSpectraStrain} has been borrowed from the simpler electric field case in Eq. \ref{eq:scaleSpectra}, however, we do not know the form of surface-related strain fields (e.g., from defects, terminating atoms and dimers, etc...), so the relation between $\varepsilon_{\parallel}$ and $\varepsilon_{\perp}$ is likely off by at least a constant of order 1. In Fig. \ref{fig:FigureSOMStrainSpectra} we plot the analyses of dephasing and DQ relaxation spectra with the same data as for the electric field case but now with the strain coupling re-scaling.

The most obvious qualitative result for the three NVs studied is that strain noise (purple dash-dot line), that which affects both spectroscopy data sets, is now the high frequency component of the double-Lorentzian, while the magnetic component (red dashed line) is then assigned to the low frequency part. Second, the exponents of the strain noise curve are smaller than for the electric case, $\alpha = 1.4, 0.6, 1.0$ for A1, A8, and A7, respectively, which are farther from resembling Lorentzian spectra. In fact, the DQ data is less well fit for these exponents constrained to small values, which makes a case for electric fields as a more likely explanation. For example, in Fig. \ref{fig:FigureSOMStrainSpectra}(a) the total noise curve (green line) noticeable overestimates all of the data points around $10^8$ Hz in comparison to the electric case in Fig. \ref{fig:SOMNVA1spectraVsSplitting}(a). 

Another number to consider from the strain spectra model is the total amount of rms strain, or area under the strain noise curve $\sqrt{\langle \varepsilon_{\perp}^{2} \rangle}$. This value is $6\times10^{-5}$ strain, $5\times10^{-6}$ strain, and $2\times10^{-5}$ strain for NVs A1, A8, and A7, respectively. One question is whether this level of total dynamic strain is physically reasonable for NVs approximately 7 nanometers or farther from the diamond surface. 

As we discuss in the main text, the dephasing studies of ref. \cite{Kim2015} using dielectric liquids on diamond also point strongly to a source of electric field noise at the diamond surface, and the existence of parallel electric fields deduced in that work necessarily imply the existence of transverse electric fields at the same NVs. This apparent electric noise is reported in \cite{Kim2015} to persist over a variety of surface treatments, including successive steps of oxygen annealing and boiling in perchloric, nitric, and sulfuric acids; therefore, it is likely that it exists as well in our diamond films prepared from similar starting material and same CVD growth parameters. Because terminating atoms of the surface have resonant modes that can cause strain in the top layers of diamond \cite{HREELS1995}, albeit mostly THz modes, it is perhaps plausible that the applied viscous liquids serve to change the frequency spectrum of the surface phonons. However, liquids with the higher dielectric constants, such as D-glycerol, showed greater enhancement of coherence $T_{2}$ times \cite{Kim2015}, making electric screening a more likely explanation. Furthermore, our recent scanning probe measurements with other external surfaces brought within nanometers of the diamond suggest that the noise affecting $\gamma$ may be mitigated, which is more consistent with the electric field hypothesis than with strain. These surface-noise measurements will be explored in a future paper.

\subsection{Rabi driving at low $B_{0}$}
Several of the low-field measurements are performed with $\ket{\pm1}$ splittings of $\omega_{\pm1}$$/2\pi\approx37.1$ MHz. Choosing a $\ket{0}\leftrightarrow\ket{-1}$ Rabi frequency is a trade-off between covering the hyperfine bandwidth and avoiding population transfer to the $\ket{1}$ state. For a resonant $\pi_{0,-1}$-pulse, the detuning of the $\omega_{1,0}$ transition is $\omega_{\pm1}$, which gives an ``unintentional'' maximum Rabi contrast of $\epsilon_{+1,\mathrm{max}} =\Omega_{R}^{2}/\left( \Omega_{R}^{2} + \omega_{\pm1}^2\right) \approx 5\%$ for $\Omega_{R}=8.33$ MHz (60-ns $\pi$-pulse). Over the duration of the $\pi$-pulse the final population leaked into $\ket{1}$ is then

\begin{equation}
\epsilon_{+1} = \frac{\Omega_{R}^{2}}{ \Omega_{R}^{2} + \omega_{\pm1}^2}\sin^{2}\left(\pi t\sqrt{\Omega_{R}^2+\omega_{\pm1}^2} \right)\approx 0.029
\label{eq:rabiLowLeak}
\end{equation}
This $3\%$ is a relatively low population compared to our PL signal noise, which is about $10\%$. In general this population leak in the $F_{3}$ or $F_{2}$ measurement will alter the initial conditions after the first $\pi_{-1}$-pulse to $\rho_{-1} = 1-\epsilon_{+1}$ and $\rho_{1}=\epsilon_{+1}$. Likewise, at the end of the dark time $\tau$, there will be an imperfect population transfer between $\ket{-1}$ and $\ket{0}$. Longer $\pi$ pulse durations can be used to reduce the bandwidth and prevent significant population transfer. These considerations do not affect the fundamental decay rates $\gamma$ and $\Omega$ between the levels, and low-$B_{z}$ effects are an interesting avenue for future relaxation studies on both SQ and DQ qubit coherence.

\subsection{Effects of static $B_{\perp}$ and magnetic field noise on $\gamma$}
The component of the applied static magnetic field transverse to the NV axis will change the eigenvectors $\ket{0}$, $\ket{\pm1}$ into a general mixed form. For relatively small $B_{z} < 200$ G the spin mixing is not severe. However, for a misalignment angle $\theta$ approaching $80-90^{\circ}$ most of the total field is $B_{\perp}$, and the mixing will be more significant even at small $B_{0}=\sqrt{B_{\perp}^2+B_{z}^2}$. The implication is that there will be a modified set of spin relaxation transition rates that differ from the pure $\Omega_{+}$, $\Omega_{-}$, and $\gamma$.

In the presence of a weak transverse magnetic field compared to $\Dgs$ and $B_{z}=0$ $(\theta=90^{\circ})$ the eigenstates are approximately \cite{Dolde2011,n14eseem2014}
\begin{equation}
\ket{\psi^{(0)}_{z}} \approx \ket{0} 
\label{eq:weakBperpMixingz}
\end{equation}
\begin{equation}
\ket{\psi^{(0)}_{x}} \approx \frac{1}{\sqrt{2}}\left(\ket{-1} - \ket{1}\right) 
\label{eq:weakBperpMixingx}
\end{equation}
\begin{equation}
\ket{\psi^{(0)}_{y}} \approx \frac{i}{\sqrt{2}}\left(\ket{-1} + \ket{1}\right) 
\label{eq:weakBperpMixingy}
\end{equation}

The measurements of NVA2 were taken in a non-zero small $B_{z}$ and transverse magnetic field of $B_{\perp}=44.8$ G and $\theta\approx 83^{\circ}$, which results it a splitting of the resonance lines $\omega_{xy} = 30.6$ MHz. Since $B_{z} = 5.4$ G in this case, diagonalizing $H = hDS_{z}^2 + g\mu_{B}\textbf{B}\cdot\textbf{S}$ shows that the mixing between all three $\ket{m_{s}}$ states is very weak (populations better than 99$\%$ pure), meaning that we can still consider to good approximation that it is the as-defined $\gamma$ that we are measuring with the $F_{1}$ and $F_{3}$ or $F_{2}$ methods. For all other NVs the $B_{z} \gg B_{\perp}$ condition was fulfilled even more strongly, so the $\ket{m_{s}}$ eigenstates are always valid.

In the hypothetical case that $\theta=90^{\circ}$ ($B_{z}\approx0$), there is still very little mixing of $\ket{1}$ and $\ket{-1}$ with $\ket{0}$ as shown by Eqs. \ref{eq:weakBperpMixingz}-\ref{eq:weakBperpMixingy}. This means that the DQ relaxation channel will act between $\ket{\psi^{(0)}_{x}}$ and $\ket{\psi^{(0)}_{y}}$, although the splitting $\omega_{x,y}$ will be much smaller and subject to other factors like static strain and static electric field in comparison to the $\omega_{\pm1}$ values considered in the present work \cite{Dolde2011,electric2016}. This may be an interesting versatility to the DQ relaxation method for spectroscopy. When $\textbf{B}$ is perfectly aligned to the NV center axis one can probe solely electric field and strain noise, and but tuning $\theta$ to the other extreme one can also become sensitive to magnetic noise using the same pulse sequence.  

Finite $B_{\perp}$ can also make the DQ transition of a $S=1$ defect system susceptible to coherent driving by magnetic fields, though it is a second-order effect that requires significantly more power \cite{MacQuarrie2013,Klimov2014}. We consider for our experiment the effect of incoherent magnetic noise on $\gamma$ to further check whether the observed frequency dependence can really be attributed to electric field noise. Our measurements were performed with misalignment angles $\theta = 0-8^{\circ}$ so there is a small contribution of $B_{\perp}$ to the total field.

As an experimental check of the potential for magnetic driving, we tuned the applied magnetic field such that $\omega_{\pm1}/2\pi = 1376$ MHz with $\theta\approx 8.7^{\circ}$ and $\gamma$ of NVA8 was mostly saturated at $\gamma=\gamma_{\infty}$. The corresponding SQ transition is at $\omega_{0,-1}/2\pi = 2188.65$ MHz. We set up a second microwave source to output a tone at $f_{2}=1376$ MHz with its phase modulated by a noise source. We combined this non-gated microwave channel with the double-gated coherent drive at $f_{1}=2188.65$ MHz. We coherently drove the $\omega_{0,-1}$ transition in order to perform the $F_{3}$ measurement while the noisy $f_{2}$ tone was continuously applied. We used the Rabi frequency due to $f_{1}$ for calibration to estimate that we could increase the measured $\gamma$ to 1.9 kHz using a magnetic-field amplitude of $B_{1x} =1$ G at $f_{2}$. A noise spectrum amplitude of 1 G$/$Hz$^{1/2}$ is many orders of magnitude larger than the amplitude of the magnetic $(b_{m})$ Lorentzian with parameters in Table \ref{tab:spectra} at frequencies of $\omega=\omega_{\pm1}=20-1375$ MHz and accounting for linewidth. Therefore, this magnetic noise from the surface is far too small to have an effect on $\gamma$ comparable to the applied test 1-G noisy driving field. We conclude that magnetic noise in the diamond or at the surface is not responsible for the observations of $\gamma$ in this work.

In a related argument, from second-order perturbation theory, with perturbation as a small $g\mu_{B}B_{\perp}$, the ratio of the DQ to SQ magnetic Rabi frequencies is approximately given by \cite{MacQuarrie2013}
\begin{equation}
R=\frac{\Omega_{R,\pm1}}{\Omega_{R,-10}} = \frac{\sqrt{2}\left(g\mu_{B}/\hbar\right)B_{\perp}D_{\mathrm{gs}}}{D_{\mathrm{gs}}^{2} - \left(g\mu_{B}/\hbar\right)^{2}B_{\parallel}^2}
\label{eq:RatioRabi}
\end{equation}
For the numbers $D_{\mathrm{gs}}=2870.5$ MHz, $B_{\perp}=37.7$ G, and $B_{\parallel}=246.0$ G we get $R_{\Omega} = 0.052$. We can also think of this ratio $R$ as relating the bare coupling of the NV magnetic fields $\gamma_{\mathrm{NV}}=2\pi \times 2.8$ MHz/G to an effective reduced coupling $\gamma_{\mathrm{NV}}^{(\mathrm{eff})} = R\gamma_{\mathrm{NV}}$. In this way the determination of the effect of magnetic noise on the DQ relaxation is in mathematical analogy to the disparate couplings to transverse and parallel electric fields $d_{\parallel}$ and $d_{\perp}$ shown in Eq. \ref{eq:cpmgGammaSEperp}. That is, when the effective $\ket{\pm1}$ coupling to magnetic fields $\gamma_{\mathrm{NV}}^{(\mathrm{eff})}$ comes into the noise spectral density it must be squared, meaning that the DQ relaxation rate $\gamma$ actually depends on the even smaller factor $R^{2}=0.0027$ in relation to $\langle B^{2}\rangle$ noise. For sufficiently misaligned magnetic fields then the inequality $B_{\parallel} \gg B_{\perp}$ does not hold and Eq. \ref{eq:RatioRabi} is not correct. We performed noise-spectroscopy measurements only under the conditions of $B_{\parallel} \gg B_{\perp}$, so we do not consider the other cases here.

\subsection{Implications for $T_{1\rho}$ and continuous dynamical decoupling}
Future measurements comparing continuous-DD $T_{1,\rho}$ to the more complete definition of $T_{1}$ could shed light on the dephasing limitations we find at high applied magnetic field. The consideration of the $\gamma$ relaxation channel has not been considered in prior work on SQ and rotating-frame relaxation of shallow NV centers \cite{Rosskopf2014}. The transition matrix used for analyzing spin-locking $T_{1,\rho}$ measurements \cite{Rosskopf2014} has treated these as zero-valued matrix elements though they will depend on $\gamma$, and at small $\omega_{\pm1}$ or very shallow NVs it may dominate the observed rotating-frame decay. Future $T_{1,\rho}$ experiments tuning both the Rabi frequency and $\omega_{\pm1}$ could help continue to elucidate relative contributions of magnetic and electric noise in the 1 MHz - 30 MHz regime. Likewise surface electric fields may play an even more important role in continuous dynamical decoupling using coherent electrical or mechanical driving in the $\ket{\pm1}$ manifold \cite{MacQuarrie2015}.

\section{}
\subsection{}
\subsubsection{}

\end{document}